\documentclass[12pt,preprint]{aastex}

\def\ref{\par\noindent\hangindent=1in\hangafter=1}

\newcommand{\Av}{\mbox{$A_V$}}		% Av (extinction)
\newcommand{\ci}{\mbox{$C$}}		% Concentration index
\newcommand{\etal}{\hbox{et al.}}			
\newcommand{\Halpha}{H$\alpha$}
\newcommand{\kms}{km~s$^{-1}$}		% km/sec
\newcommand{\Mv}{\mbox{$M_V$}}		% Mv (absolute visual mag)
\newcommand{\msun}{\mbox{$M_{\odot}$}}	% Solar mass
\newcommand{\vi}{\mbox{$V\!-\!I$}}	% Color index V-I
\def\H0{$H_0$~= 75 \kms\ Mpc$^{-1}$}

\def\ref{\par\noindent\hangindent 30pt}
\def\farcs{\hbox{$.\!\!^{\prime\prime}$}}

\def\v16{$\Delta V_{1-6}$}
\def\lea{\mathrel{<\kern-1.0em\lower0.9ex\hbox{$\sim$}}}
\def\gea{\mathrel{>\kern-1.0em\lower0.9ex\hbox{$\sim$}}}
\newcommand{\Ha}{H$\alpha$}

\newcommand{\Msun}{\mbox{$M_{\odot}$}}

\newcommand\cola {\null}
\newcommand\colb {&}
\newcommand\colc {&}
\newcommand\cold {&}
\newcommand\cole {&}
\newcommand\colf {&}
\newcommand\colg {&}
\newcommand\colh {&}
\newcommand\coli {&}
\newcommand\colj {&}
\newcommand\colk {&}
\newcommand\coll {&}
\newcommand\colm {&}
\newcommand\coln {&}
\newcommand\eol{\\}
\newcommand\colzero {\null}

\begin{document}

\title{THE ANTENNAE GALAXIES (NGC 4038/4039) REVISITED: ACS AND NICMOS OBSERVATIONS OF A PROTOTYPICAL MERGER\altaffilmark{1} }

\author{\sc Bradley C.\ Whitmore,\altaffilmark{2} 
Rupali Chandar,\altaffilmark{3}
Fran\c cois Schweizer,\altaffilmark{4}\\
Barry Rothberg,\altaffilmark{2,5}
Claus Leitherer,\altaffilmark{2}
Marcia Rieke,\altaffilmark{6}
George Rieke,\altaffilmark{6}\\
W.\ P.\ Blair,\altaffilmark{7}
S. Mengel,\altaffilmark{8}
\& A.~Alonso-Herrero\/\altaffilmark{9}}

\altaffiltext{1}{Based on observations with the NASA/ESA {\it Hubble
Space Telescope\/}, obtained at the Space Telescope Science Institute,
which is operated by the Association of Universities for Research in
Astronomy, Inc.\ under NASA contract NAS5-26555. 
Also based on data obtained from the Hubble Legacy Archive, which is a
collaboration between the Space Telescope Science Institute
(STScI/NASA), the Space Telescope European Coordinating Facility
(ST-ECF/ESA) and the Canadian Astronomy Data Centre (CADC/NRC/CSA).
 }
\altaffiltext{2}{Space Telescope Science Institute, 3700 San Martin
Drive, Baltimore, Maryland 21218; whitmore@ stsci.edu, etc.}
\altaffiltext{3}{University of Toledo, Department of Physics \& Astronomy, 
Toledo, OH 43606}
\altaffiltext{4}{Carnegie Observatories, 813 Santa Barbara Street, Pasadena,
CA 91101}
\altaffiltext{5}{Naval Research Laboratory, Remote Sensing Division, Code 7211, 4555 Overlook Ave., Southwest, Washington, D.C. 20375}
\altaffiltext{6}{The University of Arizona, Department of Astronomy,
933 N. Cherry Ave., Tucson, AZ 85721-0065}
\altaffiltext{7}{Center for Astrophysical Sciences, The Johns Hopkins University, 3400 N.\ Charles Street, Baltimore, MD 21218}
\altaffiltext{8}{European Southern Observatory, Germany; s.mengel@gmx.net}
\altaffiltext{9}{Instituto de Estructura de la Materia, CSIC, Serrano 121, 28006 Madrid, Spain}

\begin{abstract}

The Advanced Camera for Surveys (ACS) and the Near Infrared Camera and Multi-Object Spectrometer  (NICMOS) have been used to obtain new 
{\em Hubble Space Telescope} images of NGC~4038/4039 (``The Antennae'').
These new observations allow us to better differentiate compact star clusters from individual stars, based on both size and color. We use this ability to extend the cluster luminosity function by approximately two magnitudes over our previous WFPC2 results, and find that it continues as a single power law, $dN/dL \propto L^{\alpha}$ with $\alpha = -2.13 \pm 0.07$, down to the observational limit of $\Mv\approx -7$. Similarly, the mass function is a single power law $dN/dM \propto M^{\beta}$ with $\beta = -2.10 \pm 0.20$
for clusters with ages  $<\!3\times10^8$~yr, corresponding to lower mass limits that range from $10^4$ to $10^5\,\msun$, depending on the age range of the subsample. Hence the power law indices for the luminosity and mass functions are essentially the same. The luminosity function for intermediate-age clusters (i.e., $\sim$100--300 Myr old objects found in the loops, tails, and outer areas) shows no bend or turnover down to $\Mv\approx -6$, consistent with relaxation-driven cluster disruption models which predict the turnover should not be observed until $\Mv \approx -4$. An analysis of individual $\sim$0.5-kpc sized areas over diverse environments shows good agreement between values of $\alpha$ and $\beta$, similar to the results for the total population of clusters in the system. There is tentative evidence that the values of both $\alpha$ and $\beta$ are flatter for the youngest clusters in some areas, but it is possible that this is caused by observational biases. Several of the areas studied show evidence for age gradients, with somewhat older clusters appearing to have triggered the formation of younger clusters. The area around Knot~B is a particularly interesting example, with an $\sim$10--50~Myr old cluster of estimated mass $\sim\!10^6\,\Msun$ having apparently triggered the formation of several younger, more massive (up to $5\times10^6\,\Msun$) clusters along a dust lane. A comparison with new NICMOS observations reveals that only $16 \pm 6$\% of the IR-bright clusters in the Antennae are still heavily obscured, with values of $\Av > 3$ mag. 
\end{abstract}


\section{Introduction}

The Antennae (NGC 4038/39) are the youngest and nearest example of a
pair of merging disk galaxies in the Toomre (1977) sequence. As such, they have been observed in virtually every wavelength regime: radio (e.g., Wilson \etal\ 2000, 2003; Neff \& Ulvestad 2000; Hibbard \etal\ 2001; Zhu, Seaquist \& Kuno 2003; Schulz \etal\ 2007), infrared (e.g., Mirabel \etal\ 1998; Kassin \etal\ 2003; Wang \etal\ 2004; Brandl \etal\ 2005, 2009; Mengel \etal\ 2005; Clark \etal\ 2007; Rossa \etal\ 2007; Gilbert \& Graham 2007; Mengel \etal\ 2008), optical (Rubin, Ford, \& D'Odorico 1970; Whitmore \& Schweizer 1995; Whitmore \etal\ 1999; Kassin \etal\ 2003; Bastian \etal\ 2006), ultraviolet
(e.g., Hibbard \etal\ 2005), and X-ray (e.g., Fabbiano \etal\ 2003, 2004; Metz \etal\ 2004; Zezas \etal\ 2006, 2007).  One important result has been the discovery that much of the star formation occurs in the form of massive compact star clusters (Whitmore \& Schweizer 1995; Whitmore \etal\ 1999), often referred to as ``super star clusters.'' Some of these clusters have properties expected of young globular clusters (i.e., mass, effective radius, etc.).  Hence, it may be possible to study the formation (and disruption, see Fall, Chandar, \& Whitmore 2005, hereafter FCW05; Whitmore, Chandar, \& Fall 2007, hereafter WCF07; and Fall, Chandar, \& Whitmore 2009, hereafter FCW09) of globular clusters in the local universe rather than trying to determine how they formed some 13 Gyr ago.

One of the primary challenges for previous observations of the Antennae was the difficulty of distinguishing faint young clusters from individual young stars, which can be as luminous as $\Mv \approx -9$. The Advanced Camera for Surveys, with its superior spatial resolution over a wider field of view than the Wide Field and Planetary Camera 2 (WFPC2), provides an opportunity to improve this situation. The ability to better distinguish faint clusters from stars provides several new science opportunities.  The first is an extension of the cluster luminosity and mass functions to lower values.
A large number of studies have found that the young cluster luminosity function (LF) in a variety of star-forming galaxies is a power law of the form $\phi(L)dL \propto L^{\alpha} dL$ with a value of $\alpha\approx -2$ (e.g., see compilations in Whitmore 2003; Larsen 2005, 2006; de Grijs \& Parmentier 2007). However, recently there have been claims (e.g., Fritze-von Alvensleben 2004; de Grijs \& Parmentier 2007; Anders \etal\ 2007) that the initial cluster luminosity function has a turnover at faint magnitudes, and is closer to the Gaussian form found for old globular clusters. The ability to push the LF about 2--3 mag fainter provides a more stringent test of the power-law nature of the initial cluster LF.  Areas in the outskirts of the galaxy provide an especially sensitive test, since their low background levels allow cluster detections to fainter magnitudes. Most of the clusters in these tails, loops, and outer areas have intermediate ages (approximately 100--300 Myr) and have, therefore, survived the initial stages of cluster disruption or ``infant mortality'' associated with the removal of interstellar material (see Hills 1980; Lada, Margulis \& Dearborn 1984; Boily \& Kroupa 2003;  Bastian \& Goodwin 2006; and Fall \etal\ 2010 for theoretical discussions and Lada \& Lada 2003, FCW05, and WCF07 for observational evidence). 

Several studies have also found that the young cluster mass function (MF)
in a variety of galaxies is a power law of the form $\phi(M)dM \propto
M^{\beta} dM$ with a value of $\beta\approx -2$ (e.g., Zhang \& Fall 1999 in the Antennae; Hunter \etal\ 2003 in the Magellanic Clouds; Bik \etal\ 2003 in M51; and Chandar \etal\ 2010 in M83). While the LF and MF are sometimes considered to be interchangeable, this is not true  for populations of clusters that include a wide range of ages, and hence mass-to-light ratios. 
In fact, the observed similarity in the values of $\alpha$ and $\beta$ can be used as indirect evidence that the age and mass distributions are independent, at least up to ages of 300 Myr (e.g., Fall 2006).  This independence has been established more directly in FCW05, and can be expressed explicitly by the fact that the bivariate mass-age distribution $g(M,\tau)$ can be approximated by the product of the univariate mass and age distribution:  $g(M,\tau)\approx M^{\beta} \tau^{\gamma}$  (FCW05, WCF07, FCW09).   

There have also been claims for several galaxies, including the Antennae, that there is a cutoff to the upper mass with which clusters can form (e.g., Gieles \etal\ 2006; Bastian 2008; Larsen 2009), and that there are bends in the age distributions of some galaxies (e.g., Boutloukos \& Lamers 2003, and 
Lamers \etal\ 2005). However, based on our WFPC2 data, we found that both the mass function and the age distribution are simple power laws (see FCW09 for discussion). We suspect that these apparent discrepancies are due to selection effects in other studies, in particular to the exclusion of crowded regions  which contain most of the young clusters.  In this paper we examine whether or not there is evidence for such bends or cutoffs from our new ACS data set.

A second new science opportunity enabled by the high-resolution ACS images is to dissect the contents, both clusters and individual stars, in some of the spectacular star forming knots found in the Antennae. We will be able to assess whether some of the universal trends found for star-forming galaxies, for example the correlation between the absolute visual magnitude of the brightest cluster and the total number of clusters (e.g., Whitmore 2003), holds or not in individual sub-kpc regions of the Antennae. Age and mass estimates for the clusters in these regions will allow us to assess how star formation proceeds in these extreme environments, and whether there is evidence for sequential star formation. 

In this paper, we adopt a distance modulus of 31.71 mag for the Antennae,
as found by Schweizer \etal\ (2008) based on the Type Ia supernova 2007sr.
This modulus corresponds to a distance of 22~Mpc for a Hubble constant
of $H_0 = 72$ km s$^{-1}$ Mpc$^{-1}$. Note that this distance is slightly further than that adopted in our earlier papers (19.2 Mpc), and quite different from the shorter distance of 13.3 Mpc found by Saviane \etal\ (2008), based on an apparent misidentification of the tip of the red giant branch (see Schweizer \etal\ 2008 for details). At a distance of 22 Mpc, 1\arcsec\ is equivalent to 107 pc, and one $0\farcs05$ ACS/WFC pixel covers 5.33 pc.

The remainder of this paper is organized as follows: \S2 describes the observations and photometric reductions; \S3 gives details of our procedure to estimate the ages and masses of the clusters; and \S4  summarizes our method for separating stars and clusters (with more details included in Appendix~A). We compare the 50 most luminous, massive, and IR-bright clusters in \S5; examine the LF of star clusters in the Antennae in \S6, and then the mass and age distributions in \S7. In \S8 we perform a detailed analysis of the cluster populations in star-forming knots and other regions, and \S9 discusses patterns of star formation, including triggered star formation. A summary is provided in \S10. Future papers will include a more detailed study of the age, mass, and size distributions of the clusters
(Chandar \etal, in prep.) and will incorporate our NICMOS observations of the Antennae more fully (Rothberg \etal, in prep).  

\section{Data and Analysis}

\subsection{Optical ACS Observations}

Observations of the main bodies of NGC 4038/39 were made with the {\it Hubble Space Telescope\/} ({\it HST\/}), using the Advanced Camera for Surveys (ACS), as part of Program GO-10188. Multi-band photometry was obtained in the following optical broad-band filters: F435W ($\approx B$), F550M ($\approx V$), and F814W ($\approx I$). Observations in $V$ were made using the medium-width filter F550M rather than the more widely used broad-band filter F555W to minimize possible contamination from [\ion{O}{3}] and other nebular emission lines. Additional observations were made using the narrow-band filter F658N to study \Ha\ emission at the redshift of the Antennae. All optical observations were made with the Wide Field Camera (WFC) of the ACS, with a scale of $0\farcs049$ pixel$^{-1}$ yielding a field of view of $\sim\,${}$202\arcsec \times 202\arcsec$.  Four separate exposures were taken for each of the filters F435W, F658N, and F814W, yielding total exposure times of 2192~s, 2300~s, and 1680~s,
respectively. Six separate exposures were taken through the F550M filter, with a total exposure time of 2544~s.  The longer integration and larger number of dither pointings maximized the spatial resolution of the F550M observations. Slightly shorter ACS/WFC exposures were also taken at two positions in the Southern tail, using the filters F435W, F550M, F658N, and  F814W, and will be presented in a future paper.

Archival F336W photometry of the Antennae (Program GO-5962) was used 
to supplement our optical ACS/WFC observations. Two sub-pixel dither positions separated by $0\farcs025$ were observed through the F336W filter, yielding a total exposure time of 4500~s.

The flatfielded ACS/WFC images were co-added using the task MULTIDRIZZLE (Koekemoer \etal\ 2002) in PyRAF/STSDAS. A PIXFRAC value of 0.9 and a final scale of $0\farcs045$ pixel$^{-1}$ were used for each of the
filters. A second, higher resolution F550M image was constructed to take advantage of the additional pointings.  This ``super-resolution'' image was constructed using a PIXFRAC value of 0.7, and has a scale of $0\farcs03$ pixel$^{-1}$.  From here on we will use the term ``super-res'' to denote the $0\farcs03$-pixel F550M image.  The archival WFPC2/F336W images were co-added in a manner similar to that used by Whitmore \etal\ (1999), but using MULTIDRIZZLE rather than the older DRIZZLE task.  Each chip was treated separately, creating four final output images (PC, WF2, WF3, and WF4).  A PIXFRAC value of 0.8 was used for all four chips, resulting in final scales of $0\farcs035$ pixel$^{-1}$ for the PC chip and $0\farcs075$ pixel$^{-1}$ for each of the WF chips.
 
Figure~1 shows a color image of the Antennae produced by Lars Christensen (ESA).  The F435W image is shown in blue, F550M is shown in green, and a combination of the F814W and F658N (\Ha) images is shown in red. One intriguing aspect of this color image is that it makes it possible to tell the difference between regions that are red because they have strong \Ha\ emission (these have a pinkish tint) and regions that are red because they suffer strong reddening due to dust (these have an orange tint).  

Catalogs of cluster candidates were constructed from the F550M ``super-res'' image using the task DAOFIND in IRAF. The higher resolution of the ``super-res'' image is better suited to separate close objects which might otherwise remain blended in the lower resolution images. Removal of artifacts (e.g., hot pixel residuals, remaining cosmic rays) and background galaxies was aided by the calculation of a ``Concentration Index'' \ci\ for each object, which is taken to be the difference between magnitudes measured within a one and a three pixel radius (see Whitmore \etal\ 1999).  
We note that more concentrated objects, such as stars, have lower values of \ci\ (i.e., $\approx$1.45), while extended objects, such as clusters, have larger values of \ci\ (i.e., $>$1.52). Hence, this parameter might more properly be called a ``diffuseness'' index. However, for historical uniformity we prefer to retain the term ``Concentration'' Index. We also used the FWHM measured by the IRAF task RADPROF, and a parameter which we refer to as the ``Flux Ratio,'' which is defined to be the ratio of the flux in a 3-pixel radius aperture to the flux measured in the background annulus.  
This parameter was found to be useful for removing false detections in areas of dust. The final object catalog was assembled using parameter values of $\ci > 0.0,  1.5 $ pixels $ < {\rm FWHM} < 5.5$ pixels, and a Flux Ratio $>$\,1, and contained 60,790 candidate objects. For reference, a typical FWHM for stars in the Antennae is 1.7 pixels and for clusters is 2.0 pixels. 

It has been suggested that the \ci\ technique, which uses the difference
between magnitudes measured within two different apertures, may be sensitive to the precise center of an object at the sub-pixel level (Anders \etal\ 2006). If true, this would mean that objects of the same intrinsic size, but falling at the center of a pixel versus near the edge of a pixel, might have values of \ci\ that differ by as much as a magnitude or more according to Figure~29 in Anders \etal\ We tested for this effect by comparing our measured values of \ci\ for a number of isolated point sources believed to be foreground stars (i.e., Training Set 1 below). Figure 2a  shows little, if any, trend. That is, if we break the sample in half, objects with more central pixel values have a mean $\bar{\ci} = 1.437 \pm  0.014$, while objects centered more on the pixel edges have a value of $\bar{\ci} = 1.459 \pm 0.013$.  Hence, the difference in $\bar{\ci}$ is only $0.022 \pm 0.019$. We can increase the sample by including all objects with $\ci < 1.52$ and $V < 23$ mag. While this sample may include some unresolved clusters in addition to stars (as discussed in \S4), we again see no discernible trend, with more centrally positioned objects having a mean $\bar{\ci} = 1.459 \pm 0.005$, while objects with centers closer to the edge of a pixel have $\bar{\ci}=1.460 \pm 0.005$.  Hence, the difference is $0.001 \pm 0.006$.  These comparisons show that centering has a negligible effect on our measurements of \ci. We also note that, while using as large a radius as advocated by Anders~\etal\ (i.e., $\sim\,$5~pixels) might be optimal when objects are well separated, in crowded fields similar to the Antennae smaller apertures are clearly to be preferred.

The IRAF tasks XY2RD and RD2XY were used to convert object coordinates
in the ``super-res'' F550M image to the F435W, F550M, F658N, and F814W ACS/WFC image coordinate system, as well as to the F336W WFPC2 coordinate system. Any remaining small shifts in object positions were determined for each filter from the task GEOMAP.  Aperture photometry for the ACS/WFC data was performed using the IRAF task PHOT, with a 2-pixel radius and inner and outer background annuli of 4 and 7 pixels, respectively, corresponding to angular sizes of $0\farcs09$, $0\farcs18$, and $0\farcs315$ in radius. The VEGAMAG zeropoint values for ACS/WFC of 25.779, 24.867, and 25.501 mag were used for the F435W, F550M, and F814W images (Sirianni \etal\ 2005). Aperture photometry for the F336W WFPC2 data on the PC chip was conducted using an aperture with a 2-pixel radius and inner and outer background annuli of 4 and 7 pixels, respectively, corresponding to angular sizes of $0\farcs07$, $0\farcs14$, and $0\farcs245$, respectively. An aperture radius of 1.5 pixels, and inner and outer background annuli of 3 and 6 pixel radii, corresponding to $0\farcs112$, $0\farcs225$, and $0\farcs45$, respectively, were used for the WF2, WF3, and WF4 chips.

Aperture corrections, defined as the amount of light in magnitudes required to extrapolate from an aperture magnitude to a  total magnitude, were derived by making photometric measurements of isolated, high signal-to-noise (S/N) clusters in each filter (for the F550M filter this was done on both the normal and ``super-res'' ACS/WFC images) and for each individual CCD in the WFPC2 data. Aperture corrections for clusters were found to be significantly larger than those for isolated stars, for all instruments and filters. A median value was used for the aperture corrections for each filter. The aperture corrections used for the 2-pixel photometry of the normal ACS/WFC images correspond to 0.905, 0.890, and 0.915 mag for the F435W, F550M, and F814W images, respectively. The aperture corrections for the 3-pixel photometry of the ``super-res'' F550M image was 0.886 mag. These aperture corrections are for the 2- or 3-pixel apertures to a $0\farcs5$-radius aperture.  Aperture corrections from
$0\farcs5$ to infinity were taken from Sirianni \etal\ (2005).

Taking the different sizes of clusters into account properly would require different aperture corrections for each cluster. We experimented with making such corrections by deriving a formula relating aperture corrections and \ci\ indices for isolated, high S/N clusters. While this works well for many clusters, in crowded regions where values of \ci\ can be very uncertain the use of the formula adds considerable noise.  Hence, corrections for variable cluster sizes have not been made in the current paper, but will be revisited in a future paper (Chandar et~al., in preparation). We note that while this use of constant corrections typically makes individual absolute magnitudes of clusters uncertain by a few tenths of a magnitude (and more than a full magnitude in a few exceptional cases), the effects on colors are negligible since the latter are determined from fixed apertures.

Aperture corrections for the F336W images and the 2-pixel PC and 1.5-pixel
WF2, WF3, and WF4 apertures correspond to 1.041, 0.989, 0.937, and 1.055 mag, respectively.  These corrections take the photometry in 2- or 1.5-pixel radius apertures to $0\farcs5$.  The standard 0.1 mag correction from $0\farcs5$ to infinity was then added to the above corrections (Holtzmann \etal\ 1995). We also applied Charge Transfer Efficiency (CTE) corrections to the WFPC2/F336W photometry based on the prescription given by Dolphin
(2000).\footnote{We used the updated version from the webpage http://purcell.as.arizona.edu/wfpc2\_calib/.} We computed the X and Y coordinates for each object on the original images for each dither position.
For objects appearing in both dither positions we adopted weighted
averages of the CTE correction.

In addition to the data described above, long-slit \Ha\ observations using the Space Telescope Imaging Spectrograph (STIS) were planned as part of this proposal, similar to those reported by Whitmore \etal\ (2005).  Unfortunately, STIS stopped working before these observations could be obtained.

Completeness corrections were determined by adding artificial clusters in regions of different background levels to the F550M ``super-res'' image,
using the IRAF task MKOBJECT. These artificial clusters had the same median values of \ci\ and FWHM as real cluster candidates over a magnitude range of 18 to 28. The artificial clusters were then run through the same detection algorithms as used for real cluster candidates.  Recovery of the artificial clusters was based on two criteria: (1) whether the recovered
luminosity was within $\pm\,$0.1 mag of the input magnitude, and (2) whether the recovered artificial cluster was detected within $\pm\,$0.5 pixel of its input coordinates. The numbers of recovered artificial clusters were then sorted as a function of the background level (we used seven different background levels), binned in 0.25~mag bins, and compared with the initial numbers of artificial clusters. Completeness curves were constructed by smoothing the binned data using a boxcar smoothing algorithm and interpolating a curve through these points.  Figure~3 shows the completeness curves for the different background levels.  The dashed horizontal line marks the 50\% completeness level.  The completeness limits in $V$ at 50\% for the seven background levels are: 26.5, 26.4, 26.0, 25.3, 24.7, 24.2, and 23.3 mag, respectively.

In the rest of the present paper we will sometimes use the loose terminology ``$U$'' for the F336W filter for brevity, and similarly ``$B$'' for F435W, ``$V$'' for F550M, and ``$I$'' for F814W.

\subsection{Near-Infrared NICMOS Observations}

Infrared observations were made using the Near Infrared Camera and Multi-Object Spectrometer  (NICMOS) camera on {\it HST\/} as part of Program GO-10188.  Observations were made using the NIC2 camera with the F160W, F187N, and F237M filters, and the NIC3 camera with the F110W, F160W, F164W, F187N, and F222M filters. Only the NIC3 results are briefly discussed here. See Rothberg \etal\ (in prep.) for a more detailed presentation of the NICMOS data.

Two sets of NIC3 observations of the ``overlap region'' were carried out, one including the northern nucleus (NGC 4038) and the other the southern nucleus (NGC 4039).  The NIC3 data were not reduced using the standard pipeline products. Instead, raw files were obtained from the archives and were first processed with the routine CALNICA, which flattens the data and performs instrumental calibration of the data, including dark current subtraction, non-linearity correction, conversion to count rates, and cosmic-ray identification. The output data were then processed with the task PEDSUB, which subtracts quadrant-dependent residual biases. Next, hot pixels were identified manually and masked. Finally, foreground stars were identified and masked. The F222M observations included an additional set of observations of a blank field for the purpose of background subtraction, since thermal contributions from the telescope are important in this bandpass.  

The NICMOS observations were combined using the MULTIDRIZZLE task in STSDAS with the default values. The linear scale of NIC3 images is $0\farcs20$ pixel$^{-1}$. However, the NIC3 observations were dithered, and the increased sampling made it possible to produce higher resolution data.  The data in all three filters were rescaled to a final linear scale of $0\farcs15$ pixel$^{-1}$ using a PIXFRAC value of 0.8. The data were rotated to a position angle of 0$^{\circ}$, and all bad pixels were ignored in the final combination. No interpolation was used to correct for saturated or bad pixels. 
Testing of interpolation showed that leaving this turned on in MULTIDRIZZLE
produces spurious results in cluster detection and photometry.  

Photometry was conducted on the final data once they were divided by an
ADCGAIN value of 6.5. The photometric zeropoints used for the photometry correspond to 22.610, 21.818, and 20.134 mag in the F110W, F160W, and F222M filters, respectively, and are on the VEGAMAG system. The aperture sizes used for the candidate-cluster photometry were 1.5 pixels in each filter, which corresponds to an angular size of $0\farcs225$ and a spatial size of 24.1 pc. Aperture corrections were measured to compensate for missing flux. Unfortunately, due to the limited FOV no high S/N isolated clusters could be identified for measuring aperture corrections. Hence, archival data for Program 9997 (P.I.\ Dickinson, Photometric Recalibration
of NICMOS) were retrieved and processed in the exact same manner as the target observations for the purpose of obtaining aperture corrections.  The FWHM of cluster candidates on the NIC3 images of the Antennae were found to be very similar to point sources from Program 9997, justifying the use of
point-source aperture corrections for the slightly resolved (at least by ACS
standards) clusters in the Antennae. Two sets of aperture corrections were applied. The first corrected the  $0\farcs225$ aperture to a $1\farcs1$ aperture used for the photometry zeropoint determination.\footnote{See http://www.stsci.edu/hst/nicmos/performance/photometry/postncs\_keywords.html/.} These aperture corrections correspond to 0.548, 0.628 and 0.723 mag for the F110W, F160W, and F222M filters, respectively. Additional aperture corrections of 0.059, 0.090, and 0.121 mag for the F110W, F160W, and F222M filters, respectively, were used to extrapolate to infinity.

\section{Age Dating the Clusters}

We age date the clusters using our ACS observations in the $U\!BV\!IH_{\alpha}$ filters, following procedures very similar to those used for our earlier WFPC2 data (e.g., FCW05). We estimate the age, extinction, and mass of each cluster by comparing the integrated photometry described above with predictions from Charlot \& Bruzual (2007; hereafter CB07, private communication; also see Bruzual \& Charlot 2003) population synthesis models. Figure~4 shows a $U\!-\!B$ vs.\ $V\!-\!I$ two-color diagram for the 160 brightest observed sources in the $V$ filter ($m_{V} \leq 21.0$; $M_{V} \leq -10.7$). At these high luminosities, all of the objects should be clusters.  The measured colors are compared with the CB07 population synthesis models, which are provided on the {\em HST} VEGAMAG system, with open star symbols marking the predicted locations of $10^6$, $10^7$, $10^8$, $10^9$, and $10^{10}$\,yr old model clusters, starting from the upper left. The arrow in the diagram shows the reddening vector for $\Av = 1.0$ mag. Overall, the model predictions match our cluster photometry relatively well. At the top end there are a fair number of clusters that fall above the models, due to either photometric uncertainties, inappropriate models, or a combination of the two.  The open circles represent our measurements of Bastian \etal\ (2009) clusters and show that these clusters span a relatively large range in color--color space and, hence, in age. A few of the Bastian \etal\ clusters are not in our sample of the brightest 160 clusters, whence they do not feature a data point at the center of the circle. Two clusters (T54 and T270) are highlighted and will be discussed in more detail below. 

While two-color diagrams such as shown in Figure~4 are useful for visualizing
the evolution of clusters, and hence are used extensively in this paper, we note that we actually use a $\chi^2$ minimization technique for age-dating. This means that points outside the models (e.g., the ``high'' points in Figure~4) can be reasonably well fit. We estimate the age $\tau$ and extinction \Av\ for each cluster by performing a least $\chi^2$ fit that compares observed magnitudes with magnitudes predicted from CB07 stellar population models with solar ($Z=0.02$) metallicity, appropriate for young stars and clusters in the Antennae (Christopher \etal\ 2008; Bastian \etal\ 2009). The best-fit combination of $\tau$ and \Av\ returns the minimum $\chi^2$: $\chi^2(\tau,A_V) = \sum_{\lambda} W_{\lambda}~(m_{\lambda}^{\mathrm{obs}} - m_{\lambda}^{\mathrm{mod}})^2$, where the sum runs over all available broad- and medium-band ($UBVI$) filters and the F658N narrow-band filter, but requires a minimum of three measurements (including the $V$ band) to estimate age and extinction. The weights are related to the photometric uncertainty $\sigma_{\lambda}$ through $W_{\lambda} = [\sigma_{\lambda}^2 + (0.05)^2]^{-1}$. The F658N filter includes both continuum and (nebular) line emission, and is dominated by line emission for the youngest objects ($\tau \lea \mbox{several}\times 10^6$~Myr) and by continuum emission for older objects with $\tau \gea 10^7$~yr. This enables us to use the narrow-band filter as a fifth data point in many cases, regardless of the age of a cluster. We find that including the narrow-band filter helps to break the age--extinction degeneracy, leading to more accurate age estimates in a number of cases. In a forthcoming paper
(R.\ Chandar \etal, in prep.) we will provide a more comprehensive
description of our age-dating procedure. 

Figure~5 (left panel) compares our photometric age estimates for 15 clusters
with the spectroscopic ages determined by Bastian \etal\ (2009) from
absorption- and emission-line features in ground-based integrated cluster spectra. This comparison shows that the photometric and spectroscopic age estimates agree within the uncertainties for all but two of the clusters, T54 and T270. Our age estimates are older in both cases. For T270 the discrepancy is probably due to a spatial resolution effect. Whereas we do not see any H$\alpha$ emission in our small-aperture measurements, consistent with our age estimate of $\log\tau = 7.86$, the ground-based spectrum obtained through a much larger aperture shows strong emission lines (see Figure~3 by Bastian \etal). This interpretation is consistent with a blowup of the region shown in the lower right panel of Figure~5 (originally from our Figure 1), which reveals H$\alpha$ emission close to, but not coincident with the cluster. Although Bastian \etal\ exclude emission-lines from their fits, it appears that light from younger stars surrounding the cluster may contribute to the stellar continuum within their larger aperture, thereby affecting their results, since they find an age of $\log\tau < 6.8$ for T270. For T54 the color image shows no surrounding H$\alpha$ emission (see Figure 5). This object could be either an older cluster with low extinction or a younger cluster with higher extinction. The location of this cluster in the two-color diagram (Figure~4) is consistent with both age estimates, our $\log\tau = 8.16$ and Bastian \etal's $\log\tau = 6.9$. However, the lack of other nearby clusters and of any \Halpha\ emission makes the younger age estimate somewhat suspect.

Other authors (e.g., Anders \etal\ 2004) have found that including near-IR
filters in the age-dating process can be useful. We have chosen to not include the NICMOS observations in our age determinations in the current paper for a variety of reasons, including the following two:  Our NICMOS images cover only $\sim\,$50\% of the area covered by the ACS images, and the difference in spatial resolution between the NICMOS and ACS images would lead to serious difficulties in crowded areas. We plan to revisit this issue in future papers.

\section{Differentiating Stars from Clusters}

In Whitmore \etal\ (1999), one of our primary difficulties was differentiating individual stars from star clusters. The new ACS data presented here, with their higher spatial resolution, provide an opportunity to better separate the two. The concentration index \ci\/, the difference between magnitudes measured within a one and a three pixel radius, is one of the primary tools we use to help differentiate stars from clusters based on their image 
size.\footnote{We experimented with other size estimates (e.g., the FWHM 
measured by the IRAF task RADPROF, and the effective or half-light radius $r_e$ estimated from the ISHAPE software package; Larsen 1999), but found that---while these methods provide superior performance in certain circumstances (e.g., isolated objects with high S/N)---they were less robust in many cases (e.g., very crowded regions).  We also tried estimating the residual light after a ``mean cluster PSF'' had been subtracted from every object, and found that the performance was comparable, but not superior, to the simple \ci\ index.} 

We begin by identifying three ``training sets'' of objects: 
(1) a set in a relatively isolated area primarily containing foreground stars and ancient globular clusters;
(2) one in an area of the Northwest Extension (Outer~9 in Figure~17;
see also Figure~5 in Whitmore \etal\ 1999) containing primarily
intermediate-age clusters ($\sim$100--300~Myr); and
(3) one in a crowded area (Knot~S) containing both individual young stars
and clusters. The lessons learned from these three training sets will then be applied to objects in the entire galaxy. Many additional details and information on the training sets are given in Appendix~A.

Based on Training Set 1 (Figure~6) we find that overall the concentration
index \ci\ separates well most stars from old globular cluster candidates, 
with $1.37 < \ci < 1.52$  ($\bar{\ci}=1.448$, sample rms = 0.049) for stars and $1.62 < \ci < 1.78$ ($\bar{\ci}=1.676$, rms = 0.052) for old globular clusters.  We therefore adopt a value of $\ci=1.52$ as our boundary between unresolved and resolved objects. Stars and ancient-cluster candidates also separate fairly nicely in $U\!-\!B$ vs.\ $V\!-\!I$ two-color diagrams (Figure 6b). This relatively good separation is important for cases where it is difficult to differentiate stars from clusters based on size alone, such as in very crowded regions. See Tables~1 \& 2 and Appendix~A.1 for more details concerning Training Set 1.

We use Training Set 2,  which is dominated by intermediate age 
($\sim$100--300~Myr) clusters in the outer portions of the Antennae, to assess how our measurements of \ci\ and age behave at fainter magnitudes.
In this age range, the brightest stars should have $\Mv \approx -3$, well below our detection threshold. We find that essentially all objects brighter than $\Mv = -7$ have \ci\ values expected for extended objects, while at fainter absolute magnitudes the scatter increases and a few objects have $\ci < 1.52$ despite the fact that they are probably star clusters (Figure~7).
One interesting result is that we find a possible correlation between \Mv\
and \ci\ for these intermediate-age clusters, in the sense that brighter clusters are typically larger, as might be expected. The cluster sizes in the Antennae will be addressed in more detail in a future paper (R.\ Chandar \etal, in prep.). Figure~7 also shows that age estimates begin to fail for $\Mv \geq -7$. See Table~3  and Appendix~A.2 for more details concerning Training Set 2.  

Training Set 3 provides a more difficult, but also more typical case of a crowded region including both individual young stars and clusters. Figure~8  shows all objects in and around Knot~S, plotted in four two-color diagrams, each for a different range of luminosity. The figure starts with the brightest objects ($\Mv < -10$) in the top left panel and ends with relatively faint objects ($-8 < \Mv < -7$) in the bottom right panel. The objects with $\ci < 1.52$ (i.e., profiles indistinguishable from stars) are shown as open squares, while those with $\ci > 1.52$ (i.e., resolved) are shown as solid circles.  
Evolutionary tracks of CB07 solar-metallicity model clusters are shown as solid lines. Padova models for stars brighter than $\Mv = -6$ are shown by a string of dots in the top left panel. The arrow shows the reddening vector for $\Av = 1.0$ mag. Based on these models and comparisons with our observations discussed below, we divide the two-color diagram into four distinct regions: \\
(1) ``cluster space'' ($U\!-\!B < 0.6$ and $V\!-\!I > 0.1$), \\
(2) ``foreground-star space'' ($U\!-\!B > 0.6$ and $V\!-\!I > 0.9$), \\
(3) ``yellow-star space'' ($U\!-\!B > 0.6$ and $0.1 < V\!-\!I < 0.9$), and    \\
(4) ``blue-star/cluster space'' ($V\!-\!I < 0.1$).

Several conclusions can be drawn from Figure~8. The first is that the brightest objects (i.e., $\Mv < -10$) are all consistent with being young ($<$\,10 Myr) clusters with relatively little extinction (i.e., they lie close to the models).  This is reassuring, since we have used $\Mv = -9$ [i.e., the Humphreys \& Davidson (1979) limit] to differentiate clusters from stars in the past (e.g., Whitmore \etal\ 1999). The upper-right panel (showing objects with $-10 < \Mv < -9$) has three point-like objects that fall in or near what we call yellow-star space. These appear to be luminous yellow supergiant stars. We also find several point-like objects in cluster space, indicating that---based on their position in the two-color diagram---some clusters at the distance of the Antennae appear sufficiently concentrated that they cannot be distinguished from stars based on their size alone. Our final, and perhaps most important, conclusion is that in all cases down to $\Mv = -7$ more than 60\% of the objects are resolved, and hence are clusters. 
This is actually a  conservative lower limit since, as we just noted, many of
the unresolved objects are also clusters based on their position in the
two-color diagram.

These results have a number of important ramifications, the most important
being that the number of faint clusters continues to rise down to at least
$\Mv \approx -7$. This provides a counter example to published claims that the initial cluster LF in some galaxies has a Gaussian-like distribution (e.g.,
de Grijs, Parmentier, \& Lamers 2005; Anders \etal\ 2007), since this would require that essentially all the faint objects be individual stars. We will address this issue more fully in \S6 below, when we examine the LF in more detail.

See Tables~4 \& 5 and Appendix~A.3 for more details concerning Training Set 3, and Appendix~A.4 for an extension of this methodology to all of NGC 4038/39 covered by the ACS images.

\section{Fifty Most Luminous, Massive, and IR-bright Clusters}

In this section we compare the 50 most luminous (in $V$: Table~6), most massive (Table~7), and most IR-bright (in $F160W$: Table~8) clusters.  Part of our motivation for doing so is to facilitate comparisons with other studies, since we suspect that at least some of the differing conclusions reached by various authors (see Introduction) result from excluding young clusters in crowded environments. In general these three sets are relatively disjoint, with the most luminous clusters tending to be the youngest (median $\log\tau = 6.78$), the IR-bright clusters being older (median $\log\tau = 7.57$; after removing those with questionable ages as indicated in Table 8), and the most massive clusters tending to be the oldest (median $\log\tau = 7.84$).

Figure~9 shows the logarithmic age vs.\ mass diagram with circles marking the 50 most massive clusters.  Note that there are some objects in the diagram with apparent masses that would qualify them for the top 50, but which have been excluded (i.e., are not circled in Figure~9 or included in
Table~7) based on a visual review of the images and the two-color diagrams.
These are caused by a variety of interlopers and artifacts (e.g., stars, galactic
nuclei, incorrect extinction corrections). The fraction of artifacts is quite low up to values of $\log\tau < 8.6$, which will be used in the following sections as an upper limit to reliable age dating (e.g., when constructing the mass function). The most important artifact beyond this age is due to clusters that are probably quite young (e.g., are near regions with strong \Halpha\/, although they themselves do not have \Halpha\ emission) and are heavily
extincted by dust (e.g., are in the overlap region). The age-dating algorithm often classifies these as intermediate-age or old clusters due to their red colors and lack of \Halpha\ emission, hence giving them larger masses than they actually possess. This is the well-known age vs.\ extinction degeneracy.
As long as a cluster shows \Halpha\ emission ($\log\tau < 6.8$) it is possible
to break this degeneracy, but once there is no \Halpha\ emission, it is
difficult to distinguish an intermediate-age cluster from  a young cluster
with strong extinction.  Luckily, most young clusters expel most of the dust around them within about 10 Myr, hence it is primarily the clusters with extensive {\it foreground\/} extinction where this is a problem (i.e., in the overlap region).
 
Figure~10 shows all three sets of clusters (most luminous in blue, most massive in red, and most IR-bright in white).  Note that the IR observations with NICMOS only cover about half of the field of view (see Figures~11 and 12). For this reason we only include the 25 most IR-bright clusters, so that at least in the regions which are covered in all three sets the density of points are similar and hence can be more easily compared.

As expected, the western loop (i.e., the right 1/3 of Figure~10a) has a higher number of luminous clusters (19 of 50) than of massive clusters (9 of 50) because of the predominance of young clusters in this region.
Conversely, the area around Regions 4, 5, 6, and 7 and Knots~E and F (i.e., in the vicinity below the red circle 34 in Figure~10a) has the highest number
of massive clusters (19  of 50) compared to luminous clusters (11 of 50) because of the preponderance of intermediate-age clusters in this region. 
This implies that  there has been more star formation in this region than anywhere else in the galaxy during the past few hundred Myr. 

Figures~11 and 12 show the NICMOS F160W observations on the top (with and without positions marked by circles and rank order numbers), and the
$V$ observations on the bottom for comparison, for the two NIC-3 pointings.
While most of the IR-bright objects have clear optical counterparts, there are a few IR-bright objects with relatively faint counterparts (i.e., 3, 16, 24, 28 in Figure~11 and 12, 18, 26 in Figure~12). There is only one object (12) out of the 50 with a marginal detection (i.e., $\Mv = -5.58 \pm 0.30$; see Table~8). More quantitatively, only 8 of the 50 brightest IR-sources ($16 \pm 6$\%) have values of $\Av > 3$, with the highest value being $\Av \approx 7$ (assuming an intrinsic color of \vi\ = 0.5 for all the clusters). This is in good agreement with the result by Whitmore \& Zhang (2002) that 85\% of the bright thermal radio sources have optical counterparts. {\it Hence, our optical sample is not missing a large fraction of clusters due to dust obscuration\/}.

\section{Luminosity Function}

In this section we utilize our new technique for differentiating clusters from stars based on a combination of size and color information, and we revisit the shape of the cluster LF in the main bodies of the Antennae. In Figure~13 we use a magnitude limit of $\Mv < -7$, since the completeness
correction is $\leq\,$50\% for this range. It also becomes difficult to distinguish the stars from clusters for fainter magnitudes (e.g., see Figure~7).
We note that corrections have {\em not} been made for extinction when
constructing the LFs, for reasons further discussed below. We also manually removed six bright  ``interlopers'' (five foreground stars and the nucleus of NGC 4039) from the sample to insure that they do not artificially bias the bright end of the cluster LF.

Figure~13a shows the LF for all detected objects, i.e., for a {\em mix} of individual stars and clusters. Open symbols show the uncorrected data while filled symbols show completeness corrected data. To determine the slope of the LF via a least-squares fit, we restrict the range of the fit to $\Mv = -15$ to $-$7 (or $-$15.2 to $-$7.2 if one accounts for a foreground extinction of $\Av = 0.15$). This yields a good fit with a single power law, $dN/dL \propto L^{\alpha}$, where $\alpha = -2.26 \pm 0.03$. We next use what we have learned in \S4 to produce subsamples designed to differentiate the clusters from the stars, to the degree possible. Figure~13b shows the result of using only the concentration index as a size criterion, $\ci > 1.52$. Figure~13c shows the result of using only a color criterion, $U\!-\!B < 0.6$, to select objects in cluster space or in blue-star/cluster space as shown in Figure~8. Figure~13d shows the result of using both the size and color criteria for clusters. This is our preferred cluster catalog and represents our ``best guess'' LF.  As the figure demonstrates, the results of the various selections and fits are fairly similar, with values of the power-law index $\alpha$ ranging from $-$2.26 when using all objects to $-$2.13 for our best LF. This reflects the fact that most of the objects, even at $\Mv = -7$, are
still clusters rather than stars.

The fits in Fig.~13 use variable bin sizes, where there are an equal number of clusters in each bin, since Ma\'{\i}z Apell\'aniz (2008) finds that using the standard fixed-bin-size technique will artificially steepen a LF.  We empirically confirm this bias, with the variable-bin method systematically giving slightly higher values of $\alpha$ (i.e., flatter slopes), with a mean
difference of $\sim\,$0.1. In the cases below where comparisons are made with past values of $\alpha$ derived from fixed binning, a correction of 0.1 is made. We use the variable-binning technique as our preferred method throughout the rest of this paper, although some later figures (e.g., Figure~15) show both techniques for comparison.  

Figure~14 attempts to extract the {\em stellar} LF by restricting the sample to objects with $\ci < 1.52$ and $\Mv < -10$ (the latter criterion to avoid foreground stars).  As expected, for stars the slope is steeper, with $\alpha = -2.47 \pm 0.04$.  

The possibility of a bend (at the few-$\sigma$ level) in the cluster LF obtained from WFPC2 data was discussed by Whitmore \etal\ (1999). The new ACS data do not appear to show any similar bend. More specifically, if we break the total luminosity range for clusters into two segments ($-15 < \Mv < -10.4$ and $-10.4 < \Mv < -7$), and use  the $U\!-\!B < 0.6$ criterion for distinguishing clusters, we find power-law exponents of $\alpha = -2.22 \pm 0.17$ and $\alpha = -2.23 \pm 0.02$, respectively. Hence, we do not find supporting evidence for any bend or break in the LF of star clusters in the Antennae when using the new, higher-quality ACS observations.

These results contradict recent claims for a turnover in the Antennae cluster
LF, based on a re-analysis of our WFPC2 data (Anders \etal\ 2007). Clearly, the much deeper and higher-resolution ACS data show no evidence for a turnover, even if we restrict our sample to only include resolved objects.
The peak found near $\Mv \approx -8.5$ mag by Anders \etal\ appears to be caused by the introduction of a selection criterion that requires objects to have a $U$-band photometric uncertainty of $\sigma_U\leq 0.2$~mag.  
While completeness corrections could compensate for this effect in principle
(as Anders \etal\ 2007 attempt to do), in practice it is difficult to accurately model the completeness of a cluster system as complicated as the Antennae's.
Using the Anders \etal\ technique, even a very small mismatch in the true
vs.\ estimated completeness corrections can result in an apparent turnover
near the completeness threshold, which appears to have happened in their
analysis.

We search for any hint of a flattening of the cluster LF at the faintest
magnitudes by attempting to extend the fits to $\Mv = -6$, at the risk of
pushing the completeness corrections to where they become large. At the bright end we use a restrictive cutoff of $\Mv = -11$ to increase the sensitivity of the least-squares fit to faint magnitudes. Using the $U\!-\!B < 0.6$ criterion for clusters, since the $C > 1.52 $ criterion becomes unreliable beyond $\Mv = -7$,  we find values for $\alpha$ that run from $-2.24 \pm 0.04 $  (for $\Mv = -11$ to $-$8) to $-2.22 \pm 0.03$ ($\Mv = -11$ to $-$7) to $-2.13 \pm 0.02$ ($\Mv = -11$ to $-$6). Hence, we find weak evidence for a potential small amount of flattening of the cluster LF beyond $\Mv = -7$. However, due to the greater dependence on the completeness correction at these faint light levels, we consider this result very uncertain.  

A note on error estimates for the power-law index $\alpha$ is in order. While the formal internal statistical errors typically are quite small (e.g., 0.03 for the entire sample when a wide range in magnitudes is used, as shown in
Figure~13), they do not include a number of possible errors, both internal and external (e.g., deficiencies in cluster-finding algorithms and uncertainties in background and completeness corrections). A more realistic approach to determining actual errors is to compare the scatter in a variety of determinations. For example, using the four methods shown in Figure 13 yields  a mean $\bar{\alpha} = -2.20$ (rms = 0.05). If we divide the Antennae into three pieces: NGC~4039 ($\alpha \approx -2.17$), the overlap region between the two galaxies ($\alpha \approx -2.15$), and the western half of NGC~4038 ($\alpha \approx -2.18$), $\bar{\alpha} = -2.17$ (rms =0.02).  A comparison between an extinction-corrected LF with $\alpha = -1.99$ (WFC07) and an uncorrected LF with $\alpha = -2.07$  (Whitmore \etal\ 1999; down to $\Mv = -9$), both using WFPC2 data, yields a value of $\bar{\alpha} = -2.03$ (rms = 0.06). Note that including a correction for extinction has the problem that the faint end of the LF is automatically underpopulated due to the lack of objects below the completeness threshold, 
hence artificially flattening the LF.  For the purposes of the current exercise, we make the comparison between extinction-corrected and uncorrected values
only down to a value of $\Mv = -9$ in order to minimize the effect.
Throughout the rest of this paper we use uncorrected LFs.

Taking an average of all the different methods discussed in this section, our estimate for the slope of the cluster LF,  $dN/dL \propto L^{\alpha}$, is  $\bar{\alpha} = -2.17$ (rms = 0.07), based on an ensemble with 14 separate trials.  This is quite similar to our single best ACS value of $\alpha = -2.13 \pm 0.03$ from Figure 13d. We adopt the latter value as our best guess value, but use the rms value of 0.07  from the ensemble for our error estimate. 

These results are in reasonably good agreement with our earlier WFPC2 results of $-2.07 \pm 0.06$ (Whitmore \etal\ 1999; down to $\Mv = -9$)  and extend the luminosity range by $\sim\,$2 magnitudes.

\section{Mass and Age Distributions for Clusters in the Antennae}

We previously studied the mass and age distributions for star clusters in the Antennae using the WFPC2 observations (Zhang \& Fall 1999; FCW05; WCF07; FCW09). Here, we use the higher-resolution ACS images plus our new ability to separate clusters from stars to assess the accuracy and extend the range of our earlier results.

\subsection{Mass Distributions}

The mass function for a population of star clusters provides important information on the formation and dynamical evolution of the clusters, and in principle is more straightforward to interpret than the luminosity function,
where the dimming of older clusters can affect the slope.  A comparison between the mass and luminosity functions can yield insights into the age
distribution of the clusters as well as the processes responsible for the
formation and dissolution of clusters. However, it is important to keep in mind
that mass is a derived property, introducing additional biases and uncertainties. Hence we must be careful to only use it in age and mass ranges that are reliable.

The top three panels of Figure~15 show the mass functions for cluster
candidates (i.e., $\Mv < -9$) in the Antennae in three intervals of age, $\log\tau<7.0$, $7.0 \leq \log \tau < 8.0$, and $8.0 \leq \log \tau \leq 8.6$.
The corresponding lower mass limits range from $10^4$ to $10^5\,\msun$, respectively, as shown by the dashed red lines in Figure~9.

The bottom panels show the mass functions in the same intervals of age,
but with only resolved objects selected (i.e., $\ci>1.52$). Each panel shows the mass function resulting from two different bin selections: our preferred version with variable bin sizes, where there are an equal number of clusters in each bin (filled circles), and a version with approximately fixed bin sizes (open circles).

We find that over the plotted mass-age range, the mass function declines monotonically, with no evidence for a flattening or a peak. Similarly, we do not see any obvious steepening of the mass function at the high mass end as suggested by Gieles et~al.\ (2006), which would be expected if there was a physical cutoff to the masses with which clusters can form. Hence, the mass functions for clusters in the Antennae can be described by a single power-law of the form $dN/dM \propto M^{\beta}$ (or $dN/d\log M \propto M^{\beta+1}$ plotted logarithmically).  The solid lines are best fits to the variable-binning data (filled circles), while dashed lines are fits to the fixed-bin data (open circles). The values of $\beta$ given in Figure~15  are for the variable-bin data, and the mean value for the six panels is $\bar{\beta} = -2.10 \pm 0.13$. A more realistic value for the uncertainty in $\beta$ is 0.2, based on the discussion of uncertainties in $\alpha$ (\S6) and the fact that the mass is a derived property, thus introducing additional artifacts and uncertainties (e.g., uncertainty in the metallicity). 

The results presented here confirm and extend those based on previous WFPC2 data (Zhang \& Fall 1999; FCW09). We find that the mass function continues as a power-law with $\beta\approx -2$ to lower masses (i.e., $\sim 10^4\,$\msun\ for ages $<10^7$ yrs) and to older ages ($\sim 3\times10^8$~yr for $>10^5\,$\msun) than reported previously. This has a number of physical implications, which are discussed in detail in FCW09. In particular, the shape of the mass function does not change for the first $\sim 3\times10^8$~yr over the observed mass-age range, at least within the uncertainties,
which constrains various models for cluster disruption.

Figure 16 shows the MFs (top row) and LFs (bottom row) for four of the $\sim\,$0.5~kpc-size areas that will be discussed in \S8.  In general, the indices $\alpha$ and $\beta$ are essentially the same within the errors, as already found for the entire distribution of objects in the Antennae (i.e., $\alpha = -2.13 \pm 0.07$ and $\beta = -2.10 \pm 0.20$. Fall (2006) shows  that this similarity between the cluster LFs and MFs is only expected if the MF is a power law, and if the cluster age and mass distributions are approximately independent of one another. 

\subsection{Age Distributions}

The shape of the age distribution for star clusters provides clues to the formation and survival of the clusters. We previously found, based on WFPC2 data, that the age distribution of star clusters in the Antennae declines more or less continuously and in approximately power-law fashion, 
$dN/d\tau \propto \tau^{\gamma}$, with $\gamma\approx-1.0$ for the first few $10^8$~yr. We also found that the shape of $dN/d\tau$ was similar in
two different intervals of mass (FCW05, WCF07, FCW09), suggesting that the age distribution does not depend strongly on the mass of the clusters.
Here we confirm and extend these two important conclusions, using the
higher quality ACS data and the results from our detailed investigation of the 50 most massive clusters in the Antennae. A more detailed treatment will be presented by Chandar \etal\ (in preparation).

The cluster age distribution is the result of a combination of formation and disruption processes. A value of $\gamma=-1$ for the shape of the age distribution implies that there should be equal numbers of clusters in equal bins of $\log\tau$ above a given mass, which means that there are approximately ten times fewer clusters each subsequent decade in $\log\tau$.
If we assume that the star formation rate (SFR) is roughly constant, this 
implies a 90\% cluster disruption rate each decade in $\log\tau$. While the assumption of a constant SFR is clearly an oversimplification, especially for merging galaxies where bursts of star formation are expected, the rapid drop in $dN/d\tau$ by two orders of magnitude from 5 to 500 Myr, when compared to the predicted increase in the SFR from a model simulation of the Antennae (a factor of a few, see Mihos \etal\ 1993), suggests that the dominant influence for understanding the demographics of clusters are disruption processes rather than fluctuations in the formation rate. Other arguments supporting this conclusion are presented in FCW05 and WFC07.
Specifically, the $dN/d\tau$ diagram declines in a similar fashion throughout
the Antennae (see Figures 8 and 9 in WCF07), even though there are no hydrodynamical processes that could synchronize the formation rate of the clusters this precisely over such large separations. In addition, the $dN/d\tau$ diagram looks similar in many star forming galaxy (see WCF07, Mora \etal\ 2009, Chandar \etal\ 2010a, 2010b).

In WCF07 our best-fit model for the Antennae clusters suggested an infant
mortality rate of $\sim\,$80\% per $\log\tau$ decade, corresponding to a
value of $\gamma \approx -0.7$, which is only slightly flatter than the
canonical value of $-1$ discussed above.

Recently, Bastian \etal\ (2009) have attempted to take into account variations in the star formation rate in the Antennae when interpreting the cluster age distribution, by adopting the star formation rate found in the simulation by 
Mihos \etal\ (1993). This is an interesting approach. Yet, it is limited by the large uncertainties imposed by our lack of understanding of the star-formation process (e.g., shock-induced vs.\ density-induced star formation, see Barnes 2004), the fraction of gas in the progenitors, etc.
Perhaps most worrisome in using the Mihos \etal\ (1993) model for the
Antennae to predict the varying SFR is the fact that according to this model most of the star formation should occur in the nuclear regions, while in reality most of it is currently distributed throughout the disks of the two galaxies. A new N-body/SPH simulation by Karl \etal\ (2010) predicts more distributed star formation, similar to the observations,  and has a nearly constant SFR in the age range $6.7 < \log\tau < 8.7 $, in contrast to the Mihos \etal\ (1993) model where the SFR increases a factor of a few. This demonstrates the large uncertainties involved with making these simulations, and hence in using the Bastian \etal\ approach for estimating the cluster dissolution rate. 

Probably the best approach for determining the disruption rate of clusters will be to combine the data from a much larger sample of galaxies, especially non-interacting spiral galaxies, where the assumption of a roughly constant SFR is better justified statistically. The recent study by Mora \etal\ (2009) provides a promising start in this direction.  Using \textit{HST} imaging, they found power-law age  distributions for star clusters in five nearby spiral galaxies. Although they used a somewhat different technique, counting clusters brighter than a given $V$ band luminosity rather than above a given mass, they found that the age distributions are all consistent with $\approx\,$80\% of the clusters being disrupted every decade in age for $\tau \lea 10^9$~yr, nearly identical to our results for the Antennae (see their Figure 15).

Table~7 lists the 50 most massive star clusters found in the present study of the Antennae. We can use its data to derive a mass-limited ($\log M > 5.8$)
age-distribution up to $\log\tau = 8.6$ (i.e., limited by artifacts that manifest themselves at higher ages; see discussion in \S5). Making the first-order assumption of a roughly constant SFR during this period, a value of $\gamma=-1$ would imply that there should be approximately equal numbers of clusters in the three intervals $\log\tau = 6$--7, 7--8, and 8--9, 
while a value of $\gamma=-0.7$ (i.e., the preferred value from WCF07)
would imply a factor of 2 increase for successive intervals. Counting the clusters in Table~7, we find that the above three age intervals contain 10, 19, and 45 clusters (extrapolating from a logarithmic range 8--8.6 for the last bin), respectively, in good agreement with the prediction from an 80\% disruption model with roughly constant SFR.

Because Table~7 contains only 50 clusters, thus leading to low number
statistics, we also checked the number of clusters in a second range of masses, $5.3 \leq \log(M/M_{\odot}) \leq 5.7$.  Here we find that the three age intervals $\log\tau=6$--7, 7--8, and 8--9 contain 28, 41, and 126 clusters, respectively.  Hence the lower mass clusters have a similar age distribution as the higher mass clusters, increasing in number by approximately a factor of 2 in successive unit intervals of $\log\tau$, as predicted from an 80\% disruption model with roughly constant SFR.

To conclude, our new results confirm that the number of massive star clusters in the Antennae increases only slightly in bins of $\log\tau$ and, therefore, must decline fairly steeply as a function of linear~$\tau$, with an index of $\gamma\approx-0.7$. We have interpreted this steep decline as being primarily due to the disruption, rather than the formation, of clusters
(see also FCW05; WCF07; FCW09). In addition, the similarity of the relative numbers of clusters in different age intervals, but at different masses, also supports the idea that any disruption that affects these clusters does so in a manner nearly independent of the cluster mass, at least for the first few $10^8$~yr.

\section{Area-by-Area Analysis}

In this section we examine selected areas in the Antennae. Three different sets of areas have been selected for study, as shown in Figure~17. The first set consists of the bright star forming ``knots'' first identified by Rubin, Ford, \& D'Odorico (1970). These are labeled by letters A, B, C, through T.
The second set consists of less prominent, more extended star forming regions
similar to the ``other regions of interest'' first identified by Whitmore \&
Schweizer (1995). These ``regions'' are labeled by numbers 1--16. The third set consists of areas in the outer parts of the galaxy (e.g., tails and loops).
These are labeled as Outer~1, Outer~2, \dots, Outer~10. Table~9 collects information for each of the knots, regions, and outer areas.

In the following, we include all objects in our analysis rather than trying to select only clusters. This choice is based on the similarity we found in the results for the luminosity functions (Figure~13) between the various methods of selecting clusters and the total sample, and is---ultimately---due to the fact that most of the objects in our list are clusters down to at least $\Mv = -7$. Specifically, we do not wish to eliminate objects with $\ci < 1.52$ because many of them are unresolved young clusters, as seen from their position in cluster or blue-star/cluster space (\S4). Repeating the analysis using, for example, a $\ci > 1.52$ size criterion, results in only minor differences.

\subsection{General Properties}

We begin by examining general trends in the properties of the selected areas
and then discuss possible physical interpretations for these trends. Some of the most prominent correlations for clusters in the selected areas are shown in Figure~18. The top left panel shows the \Mv(brightest) vs.\ $\log N$ diagram, where \Mv(brightest) is the absolute magnitude of the brightest cluster in each area and $\log N$ is the number of clusters brighter than $\Mv = -9$ in that area. This is the strongest correlation ($8\sigma$-level of significance) of those shown in the figure. A version of this diagram for cluster populations in entire galaxies was first shown by Whitmore (2003) and has been subsequently studied by several other authors (Larsen 2002; WCF07; Bastian 2008; Larsen 2009). The diagram was one of the primary drivers in developing the ``universal'' cluster demographics model described in WCF07, since the observed correlation can be explained by a statistical ``size-of-sample'' effect, assuming all young cluster systems in galaxies have power-law LFs with a value of $\alpha \approx -2$. A comparison of the data from our area-by-area analysis with the best-fit line for cluster populations in entire galaxies (mergers, starbursts, dwarfs, spirals) from WC07 (dotted line) shows that the relationship is essentially identical. This again provides evidence for a universal relationship, at least down to the $\sim\,$0.5~kpc scales we have studied here.

The second best correlation is the $\log E(B\!-\!V)$ vs.\ $\log \tau$ relation
shown in Figure~18b ($7\sigma$-level of significance). This is similar to the result from Figure~10c in Whitmore \& Zhang (2002), but we now use the  median value of $\log E(B\!-\!V)$ for clusters within a defined area instead of plotting values of $\log E(B\!-\!V)$ for individual clusters. Similar results have also been found by Bastian \etal\ (2005) for M51 and Mengel \etal\ (2005) for the Antennae. Using areas instead of individual clusters, we find a typical $E(B\!-\!V)$ value of $\sim\,$0.3 mag (i.e., $\Av \approx 1$ mag) for areas dominated by the youngest clusters with logarithmic ages (in years) $\log\tau \approx 6.5$, declining to values of $E(B\!-\!V) \approx 0.1$ mag
($\Av \approx 0.3$ mag) for areas with logarithmic ages $\log\tau \approx 8.0$.

Figures~18c and 18d show values of $\alpha$ (the slope of the LF) and $\beta$ (the slope of the MF) as a function of the median age of the clusters in each region. We first note that for ages greater than $\log\tau = 7 $, the slope values are relatively constant, and the mean values of $\alpha$ and $\beta$ are roughly equal within the uncertainties (i.e., $\bar{\alpha} = -2.12 \pm 0.04$ (rms = 0.17) and $\bar{\beta} = -2.09 \pm 0.06$ (rms = 0.24)).  This is similar to what we found for the total distributions of clusters in the Antennae, as discussed in \S7.1, and provides further evidence that the mass and age distributions are independent of each other, as discussed in Fall (2006). 

Figures~18c and 18d also show tentative evidence for flatter slopes for  $\alpha$ and $\beta$ when $\log \tau < 7$. However, it is possible that this is due to observational biases, for example difficulties in making completeness corrections for the youngest clusters, which tend to lie in areas of higher background brightness. In principle, our completeness corrections should take care of this effect, but the combination of high background and severe crowding makes this task more difficult. Another potential problem is extinction,  since fainter clusters might suffer more extinction than highly luminous clusters due to their lack of massive stars needed to clear out the surrounding dust. We can test for this hypothetical effect fairly easily by comparing the colors of the bright and faint clusters.
This comparison shows that a very small effect does exist (i.e., $-$0.02 mag
in $U\!-\!B$ when comparing clusters in the knots brighter than $\Mv = -10$ with those in the range $\Mv = -10$ to $-$9), but this effect is too small to cause more than a very minor difference in the LFs.
 
\subsection{Area-by-Area Analysis of Outer Areas}

One of our primary goals for the outer areas, where the background is very low, is to push the LF to $\Mv\approx -6$.  This is particularly interesting because most of the clusters in these regions have intermediate ages, having probably been formed during the first encounter between the two galaxies (e.g., see Figure~17 of Whitmore \etal\ 1999).  We should therefore be able to search for a possible turnover in the LF that is expected to eventually appear if the power-law form of the LF for young clusters evolves
dynamically---assuming evaporation processes typical of the Milky Way---to the peaked form found for old globular clusters.  Based on Fall \& Zhang (2001) we might expect this turnover to occur around $\Mv \approx -4$, assuming that most of the outer clusters are 100--300 yr old and that two-body relaxation is the dominant process.

Figure~19 shows the data and values of $\alpha$ for the best fits for the three outer areas with the most clusters (Outer 2, 5, and 7). We find no clear evidence for a turnover in the LFs of the cluster population in the tails, consistent with an extrapolation from the Fall \& Zhang (2001) model.

\section{Patterns of Star Formation}

Star formation in merging galaxies is sometimes considered to be quite
chaotic, with gas clouds slamming into one another at hundreds of \kms.
The ``overlap region'' in the Antennae is often quoted as a prime example.
Yet, the periods and locations of violent interaction are relatively short-lived, being most pronounced during closest approaches. Hence, during much of a merger's evolution the distortions imposed by the interaction are less dramatic, resulting in a number of well defined patterns, both large- and small-scale (e.g., Renaud \etal\ 2009). These patterns can provide important clues to the processes of star and cluster formation.

\subsection{Large-Scale Patterns}

On the largest scales, a number of long ``linear'' (or gently curved) features
can be discerned in the Antennae. The most obvious examples are the two long tidal tails ($\sim$130 kpc in projection tip to tip) that give this system its nickname. Most of the clusters in the inner parts of these tails (the only parts covered by the central pointings; results from outer pointings along the tails will be reported in a future paper) have ages of roughly 200 Myr and were probably formed during the first-pass encounter between the two galaxies. However, an intriguing feature is the existence of ongoing star formation at a low level, as evidenced by the existence of about a dozen faint H$\alpha$ knots strung along the inner edge of the southern tail near its base, where it connects to the disk of NGC~4038 (i.e., Outer~5, see Figures~17 and 20a).

A possible continuation of this feature is the increasingly strong \Halpha\
``streak'' seen passing through knots~N and P and extending to the
``Western Loop'' that includes knots~G, R, S, T, M, and L (Figure~20b and 17). A faint but, again, increasingly strong dust lane is associated with this
H$\alpha$ emission. In the outer regions it is first aligned with the \Halpha\ knots themselves, but further into the galaxy it becomes offset to the inside of the \Halpha\ emission. A similar trend in alignment between \Halpha\ emission and dust is observed in M51. In fact, this whole aligned feature of \Halpha\ emission plus dust lane in the Antennae is reminiscent of, though perhaps less regular than, the spiral arms and outer tidal features in M51. 

Similar large-scale features are the ``spiral-looking tail'' in Region~1
(Figure~20e; $\sim$3.4 kpc)  and the long (Figure~20c; $\sim$8.2 kpc) dust lane running from the bottom of Outer~6 through Outer~3 and Outer~2, to near the end of Outer~1. A less prominent example is the nearly linear string of giant \ion{H}{2} regions following a line from Knot~F to E to Region~3 (Figure~20d). This nearly straight string was used to good advantage by Whitmore \etal\ (2005), who aligned the long-slit of the STIS spectrograph with it. This string also forms the eastern edge of the overlap region. Hence, its apparent regularity in both morphological linearity and a smooth velocity field, suggests that even the overlap region might be  less chaotic than commonly believed. 

Several other linear features are seen on slightly smaller scales. Examples are the two linear dust lanes that run from near the nucleus of NGC 4038 out to Regions 13 and 14 (Figure~20f), respectively, with regions of prominent star formation along one edge of each dust lane (e.g., Knot L).  
The side-by side arrangement of star formation and dust lanes is reminiscent of normal spiral arms (although the linear rather than spiral arrangement is not typical). 

Note that while clusters form currently at the highest rate in the chaotic
overlap region, other clusters are also forming throughout most of the two
galaxies. In particular, the velocity pattern of the ionized gas in most of the northern half of NGC 4038 is quite regular (Amram \etal\ 1992), yet a large fraction of the clusters are forming in this region at the present time.  Hence, high collision velocities are not required for making clusters, a result also found by Zhang \etal\ (2001) and Whitmore \etal\ (2005), and supported by theory (Renaud \etal\ 2008).

\subsection{Small-Scale Patterns---Evidence for Sequential Star Formation}

While the specific details of how star formation is triggered may still be obscure, it is clear that compression and shocks play an important role (e.g., along spirals arms). One method for studying triggered star formation is to age-date stars or star clusters and then look for patterns that might indicate the presence of sequential star or cluster formation. A classic example of this is 30~Doradus, where Walborn \etal\ (1999) found evidence for several generations of star formation. Does our age-dating of clusters in the Antennae provide similar evidence of sequential cluster formation? 

Figure~21 shows the pattern of ages obtained for clusters in Knot~S, along with an \Halpha\ image of Knot~S. There is some evidence for triggering as shown by the preponderance of 5--10 Myr clusters (green circles; including the central dominant cluster) positioned in what appears to be a hole in the \Halpha\/. The younger clusters (blue circles) are primarily to the left, as the star formation apparently works it's way into the dust reservoir (see the ACS/HRC image in Figure~28). There is also a smaller region of triggered formation in the upper right, as evidenced by both the age dating and presence of \Halpha\ in that area. We also note a scattering of older clusters throughout much of the region with ages 10--100 Myr (red circles), which may have triggered earlier episodes of star formation in this region. 

Knot~B (Figure~22) presents a similar, but even more dramatic example of
triggered cluster formation. In this case we find a single cluster with an apparent age of $\sim$50 Myr (and a mass of $\sim$10$^6\,\msun$) located off to the right side, but also a population of younger clusters with a total mass of $\sim${}$8\times 10^6\,\msun$, about {\em eight times greater than the mass of the cluster that appears to have triggered their formation}.
Note that the \Halpha\ shell is centered on the older cluster and that the young clusters are located where the shell intersects a major dust lane at the edge of the overlap region (see Figure~1). Dividing the radius of the \Halpha\ bubble by the age of the central cluster yields an expansion velocity of $\sim$6 \kms. This is smaller than values of 20--30 \kms\ derived for the bubbles in knots S and J by Whitmore \etal\ (1999), and than the direct measurements of \Halpha\ velocities by Amram \etal\ (1992). One possibility is that the age of this cluster is closer to 10--20 Myr, which lies in the range that is hard to measure (see Figure 9 and discussion in Whitmore \etal\ 1999).  This slightly smaller age would give expansion velocities in the range 15--30 \kms\ and would also be more compatible with the age gradient shown in Figure 22.  Presumably, the dust lane in the overlap region provides a much greater reservoir of raw material needed to form massive clusters than  is the case for Knot~S. An analogy might be that a match dropped on a field of dry grass might only result in a few minor brushfires, but a match dropped next to a tinder box can result in a much more impressive, explosive fire (e.g., Knot~B). 

In general, early star formation tends to occur on the edges of gas reservoirs and then work its way into the dust lanes, creating a linear rather than circular gradient of sequential star formation. This is sometimes known as the ``blister model'' (Israel 1978). Several other knots and regions in the Antennae (e.g., knots A, E, F, K, L, and T; and regions 3, 4, and 13) have this linear morphology of sequential cluster formation with the gradient in the direction toward a major dust lane, as shown in Figure~23 (e.g., knots E \& F and region 13). We note, however, that similar morphologies might also result from large-scale processes such as a density wave or collision between two large gas clouds. In fact, it is likely that both large- and small-scale
triggering is happening simultaneously.

Another interesting case is a dark, relatively circular dust cloud with a small amount of star formation on only one side of the cloud that is found near Outer~3 (see Figures~20g and 20c) in the southern outskirts of NGC 4039.
The lack of an obvious trigger in this case suggests the possibility that the cloud is falling back into the galaxy and the star formation results from the ram pressure being exerted by the interstellar medium. Perhaps velocity measurements with ALMA will be able to test this hypothesis. This appears to be a relatively unique case in the Antennae, but we note that other examples may be harder to discern due to the geometry of their encounter.
One possible second case might be near the center of Knot~N, with a dust
cloud falling back into the galactic disk along our line of sight (see Figure 20b). 

\section{Conclusions}

The Advanced Camera for Surveys on {\em HST\/} was used to obtain
high-resolution images of NGC~4038/4039, which allow us to better
differentiate between individual stars and compact star clusters.  We have used this improved ability to extend the cluster LF by approximately 2--3 mag at its faint end, when compared to the previous WFPC2 photometry.
In addition, the Near Infrared Camera and Multi-Object Spectrometer (NICMOS) was used to help study the clusters still largely embedded in dust. Following are our main results:

1. The cluster luminosity function continues as a single power law, $dN/dL \propto L^{\alpha}$, with $\alpha=-2.13 \pm 0.07$ down to the current observational limit of $\Mv \approx -7$. This value is uncorrected for extinction and based on variable-binning fits. Similarly, the cluster mass function is a single power law $dN/dM \propto M^{\beta}$ with $\beta = -2.10 \pm 0.20 $ for clusters with ages  $<\!3\times10^8$~yr, corresponding to lower mass limits that range from $10^4$ to $10^5\,\msun$, depending on the age range of the subsample. Hence the power-law indices for the luminosity and mass functions are essentially the same.

2. The LF for intermediate-age clusters (e.g., $\sim$200 Myr old clusters found in the loops, tails, and outer areas) does not show any hint of a turnover down to $\Mv \approx -6$. This is consistent with relaxation-driven cluster disruption models, which predict the turnover should not be observed until $\Mv \approx -4$ for this age range. 

3. The brightest individual stars in the Antennae reach absolute visual   magnitudes of $\Mv \approx -9.5$, a limit compatible with the so-called
Humphreys--Davidson (1979) limit observed in the Milky Way and nearby galaxies (see Appendix A for details).

4. Our photometrically determined cluster ages are in good agreement with spectroscopically determined ages from Bastian \etal\ (2009) for 13 of the 15 clusters in common.  In the other two cases, spatial resolution effects are likely to be responsible for the age differences, since the ground-based spectroscopic observations contain light from a much larger area than our \textit{HST} measurements. 

5. An area-by-area analysis shows a strong correlation between the absolute magnitude of an area's brightest cluster, \Mv(brightest), and the (logarithmic) number of clusters brighter than $\Mv = -9$ in that area, $\log N$.  The \Mv(brightest) vs.\ $\log N$ diagram for all areas of the Antennae studied is essentially identical to the corresponding diagram for entire galaxies, supporting the universality of this observed trend on a scale of a few hundred pc and larger. In addition, the reddening vs.\ age correlation found for individual clusters by Whitmore \& Zhang (2002), Mengel \etal\ (2005), and Bastian \etal\ (2005) has been confirmed for the area-by-area analysis as well. 

6. The same area-by-area analysis also reveals a tentative discovery of a new relationship between the power-law exponent $\alpha$ of the LF and the median age of clusters in each area, in the sense that younger clusters have flatter LFs (i.e., values of $\alpha\/ \approx -1.8$ for the youngest star- and cluster-forming knots and $-$2.1 for the outer areas).  However, it is possible that this is caused by observational biases.  The power-law indices for the LFs and MFs are in good agreement, and are similar to the results for the total population of clusters in the galaxy.

7. Using our age and mass estimates of clusters in the Antennae, we have    found evidence of sequential cluster formation in several regions. Often there is a massive, somewhat older cluster which appears to have triggered the formation of younger, lower-mass satellite clusters. The area around Knot~B is a particularly dramatic example of triggered star cluster formation, with evidence for one massive ($\sim$10$^6\,\Msun$) cluster,  with an age in the range 10--50~Myr, having triggered the formation of several younger massive clusters (up to $5\times 10^6$\,\Msun) in the direction of a strong dust lane, with an amplification factor of $\sim$8$\times$ in the total mass of triggered clusters.

8. NICMOS observations show that $16 \pm 6$\% of the clusters in the Antennae are still largely embedded in their dust cocoons, with values of 
$3 \la \Av \la 7$ mag. However, the top 50 IR-bright clusters {\it all\/} have optical counterparts in the much deeper ACS observations, hence the effect on our optically-based catalog derived from deep ACS images is minimal. 

In conclusion, the results presented here confirm and extend those based on
previous WFPC2 data (e.g., Whitmore \etal\ 1999, Zhang \& Fall 1999, Whitmore \& Zhang 2002, FCW05, WCF07, FCW09). In addition, similarities between our results for the Antennae  and those for a growing number of other star-forming galaxies (e.g., Mora \etal\ 2009)  support the universality model for the formation and disruption of star clusters.  

\acknowledgments
We thank St\'ephane Charlot for making the CB07 population-synthesis models
for \textit{HST} passbands available ahead of publication, and Lars  Christensen for the use of his beautiful color image of the Antennae.  We thank Mike Fall, Nate Bastian, Mark Gieles, and Soeren Larsen for helpful 
discussions and comments on the manuscript.  We also thank an anonymous referee for several comments that helped improve the paper. Support for Program GO-10188 was provided by NASA through a grant from the Space Telescope Science Institute, which is operated by the Association of Universities for Research in Astronomy, Inc., under NASA contract NAS5-26555.

\bigskip\bigskip\bigskip\bigskip
\appendix
\centerline{\bf Appendix}

\section{Differentiating Stars and Clusters in the Antennae}

With their higher spatial resolution across the entire field of view when
compared to earlier WFPC2 data, our new ACS data provide a better
opportunity for separating individual stars from clusters. In this Appendix we describe in some detail three ``training sets'' designed to help teach us how to achieve this separation. These three training sets consist of objects in:
(1) an area dominated by foreground stars and old globular clusters;
(2) one dominated by intermediate-age clusters; and
(3) one containing a crowded mix of young stars and clusters.

\subsection{Training Set 1 -- Isolated Foreground Stars vs.\ Old Globular Clusters}

Training Set 1 includes reddish and yellow-red objects ($V\!-\!I > 0.5$) in relatively low-background areas around the edges of the visually luminous galaxy disks. Most of these objects were identified as candidate foreground stars of our Galaxy in Table 4 of Whitmore \etal\ (1999).  We contrast this set of 26 candidate stars with the 11 candidate old globular clusters from Table 3 of Whitmore \etal\ (1999). Information about the candidate foreground stars is given in Table 1. Information on the 11 candidate old globular clusters used as part of Training Set 1 is included in Table 2 (top group).

Figure~6a shows a plot of the absolute magnitude \Mv\ vs.\ the concentration
index \ci\ (i.e., the difference in magnitude using 1 and 3 pixel radius
apertures) for the candidate foreground stars and old globular clusters of
Training Set 1. We find that in general there is good separation between the two samples, with all of the candidate foreground stars lying in the range $1.37 < \ci < 1.52$ ($\bar{C} = 1.448$ mag, sample rms = 0.049 mag) and
all but two of the candidate old globular clusters lying in the range $1.62 < \ci < 1.78$ ($\bar{C} = 1.676$ mag, rms = 0.052 mag). Note that the discrepant concentration indices for two of the candidate old globular clusters were already known (\#1 and \#9 in Table~2, corresponding to \#8 and \#1, respectively, in Table~3 of Whitmore \etal\ 1999), but the objects were retained since they had colors appropriate for old globular clusters.

The measured \ub\ and \vi\ colors for our training-set objects are compared
with a CB07 evolutionary track for a model star cluster in Figure~6b. Note that only about half of the candidate stars (open circles) are detected in the $U$ filter and can, therefore, be placed on this diagram. Most of the candidate globular clusters (filled circles) fall very close to the CB07 model predictions. There is only one discrepant object (\#1) which falls well below the track, where most of the candidate stars rather than globular clusters are found. Hence object \#1 is likely to be a star, based on both its color and its small value of \ci.  The other globular cluster candidate (\#9) that has a value of \ci\ overlapping with those measured for individual stars has colors that are indistinguishable from those of other old globular clusters. Turning our attention to the candidate stars, we find that all but two lie in the bottom-right part of the two-color diagram, well separated from the globular clusters. While most of the candidate stars and globular clusters are separated
in the two-color diagram, it is clear that {\em some} stars can have colors similar to old globular clusters (typically early F-type stars).  Hence, we can use a combination of size and color information to more reliably differentiate between stars and clusters. The dotted line in Figure~6b shows the region of the \ub\ vs.\ \vi\ diagram which we call ``foreground-star space.''

Based on the results above, we include in Table~2 a new set of candidate
old globular clusters (middle and bottom groups). The criteria used for their selection are: age estimates in the range $8.8 < \log\tau < 10$ (note that an examination of objects with estimated $\log\tau > 10$ shows many of them to be either red foreground stars or distant background galaxies, hence our use of $\log\tau < 10$), concentration indices in the range  $1.57 < \ci < 1.80$, and locations away from regions with extensive dust (since a heavily reddened young cluster could masquerade as an old globular cluster).

\subsection{Training Set 2 -- Isolated Intermediate-Age Clusters}

The second training set consists primarily of intermediate-age clusters
($\sim$200 Myr) in what was defined as the Northwestern Extension in
Figure~6 of Whitmore \etal\ 1999. This area is also designated as 
``Outer~9'' in Figure~17. In a $\sim$200~Myr-old population, the brightest stars should have absolute magnitudes $\Mv \approx -3$, well below our detection threshold. Hence, the vast majority of objects in this sample should be clusters, with perhaps a foreground star or two also included.

This area contains a loop of material that is believed to have been extracted from NGC 4038 at roughly the same time as the long tidal tails (Whitmore \etal\ 1999). The loop features three of the best examples of intermediate-age clusters (\#\#~58549, 58909, and 59721 in Figure~24).
These clusters turn out to have nearly identical properties, as shown in
Table~3 (first three entries) and in Figure 25. The wide FOV of the ACS allows us to include many other clusters with similar properties. One likely bright foreground star (STAR-60076 in Figure~24) has been removed from the sample based on its small value of $\ci = 1.389$, although we note that it has colors similar to those of intermediate-age clusters. We recommend that a spectrum of this object be obtained to make a firm determination of its true nature.

Figure~7 shows the values \ci\ and $\log\tau$ for the objects of Training Set 2, after the objects have been sorted into three subsamples according to their absolute magnitudes \Mv. Note that all objects brighter than $\Mv = -8$ have values of \ci\ consistent with them being extended clusters, based on our results for Training Set 1  (i.e., $\ci > 1.52$). At fainter magnitudes we find that the increased observational scatter results in some estimates of $\ci < 1.52$ and $\log\tau < 8.0$. Since we do not expect any intermediate-age stars in this region to have magnitudes brighter than our detection threshold, we interpret such values of \ci\ and $\log\tau$ as being due to observational scatter, suggesting that the reliability of our \ci\ and age estimates is limited below $\Mv \approx -7$. A supporting piece of evidence is that there is no \Halpha\ emission in this area, which would be expected if there actually were stars present with $\Mv \approx -7$  and $\log\tau < 8$.
One other interesting result is the apparent correlation between \Mv\ and \ci\ seen in Figure~7a, in the sense that brighter clusters tend to be larger, as might be expected. 

Figure~25 shows the \ub\ vs.\ \vi\ diagrams for the objects of Training Set 2, again sorted by absolute magnitude. Note that the scatter for the three brightest clusters (crosses in top panel) is surprisingly small.  We again find that a threshold of $\Mv \approx -7$ marks the point at which a relatively large fraction ($\sim\,$20\%) of the objects would be misclassified (i.e., are in regions of color--color space expected to be occupied only by stars).
However, the midpoint of the color distributions remains about the same
over the entire magnitude range. This shows that the objects are all very similar and, therefore, likely intermediate-age clusters. For example, we are not suddenly seeing a large percentage of blue or yellow/red stars. Note, however, that there is a small offset between the observations and the model tracks in Figure~25, with the midpoint of the data lying roughly
0.15~mag blueward of the tracks in \vi.

\subsection{Training Set 3 -- Crowded Area of Knot S with both Stars and
Clusters}

The third training set consists of 23 objects around Knot~S that were 
examined independently by four of us (BCW, RC, BR, and FS) with a variety
of techniques that we have used to differentiate between stars and clusters in the past (e.g., visual examination using the IMEXAMINE tools in IRAF,
varying contrasts on the DS9 display, circular shape, lack of nebulosity, etc.).
This crowded area represents the opposite extreme from the areas containing
Training Sets 1 and 2, with both the high density of objects and the high
background level making size measurements more difficult. In general, the four independent assessments agreed at about the 85\% level. Our final list of objects is based on the majority agreement from the independent assessments and consists of 12 candidate stars and 11 candidate clusters. 
Figure~26 shows the area around Knot S with the 23 objects marked while
Table~4 includes information about the objects.

Next, Figure~27 shows the \Mv\ vs.\ \ci\ and \ub\ vs.\ \vi\ diagrams for the objects of Training Set 3. Nine of the 12 candidate stars have concentration indices in the range $1.37 < \ci < 1.52$, appropriate for stars (see Figure~6).
The one object (\#20) below this range is the faintest one in the sample
($\Mv = -7.3$) and also lies within a few pixels of a very bright object.  Hence the photometry for this object is suspect. This finding is consistent with our results from Figure~7, where we found that estimates of \ci\ and $\log\tau$ for objects with $\Mv \approx -7$ were suspect even when the objects were isolated.

Another object (\#18) has a much larger value of \ci\ than the other candidate stars and is almost certainly a cluster, based on its \ci\ value and the fact that it lies right on the CB07 evolutionary track in Figure~27, at an age of $\sim$4 Myr. This was one of the three objects in crowded subgroups added to the original sample of 20 training-set objects in order to determine whether stars and clusters can be differentiated visually in this environment.
While the correct assignments appear to have been made for the objects
in the other two crowded subgroups (\#\#~10 and 21, see Table 4), in the  case of \#18 the visual assignment of object type appears to have been incorrect.  

All of the objects in Training Set 3 that were visually classified as clusters have values of $\ci > 1.52$, with a mean $\bar{C} = 1.719$ ($\sigma_1 = 0.082$) only slightly larger than that for the old globular clusters of Training Set 1. Hence, in general our visual classification agrees with the \ci\ criterion
at the 80--90\% level. 

Figure~27b shows that only two of the 11 candidate stars (\#\#~4 and 22,
after having removed \#18) lie below the CB07 model tracks in or near the region where young yellowish stars lie, as indicated by the small dots marking Padova models for stars brighter than $\Mv = -6$. The remaining objects (e.g., \#\#~2, 5, 18, and 23) have colors expected for young clusters.
Hence, {\em size alone is not always sufficient to differentiate between stars and clusters.} Many clusters are so compact that they remain indistinguishable from stars even at the improved resolution provided by ACS images. It is also possible that in some clusters the light from a single supergiant star dominates the light profile sufficiently so as to result in a smaller concentration index.

It is important to understand how an object gets to a position like \#23's in the \ub\ vs.\ \vi\ diagram of Figure~27, with a blue color in \ub\ ($-$0.68 mag), but a yellow color in \vi\ ($+$0.62 mag). A single star cannot have such colors, as shown by the Padova stellar models. What is needed is a combination of red or yellow supergiants, which dominate the light in \vi, and blue supergiants, which dominate the light in \ub. This composite nature of cluster light is the reason why clusters with ages from 5 to 8 Myr---when the earliest, most massive red supergiants appear---evolve so rapidly to the right, as is seen in the long horizontal stretch from $\vi = -0.1$ to $+$0.8 mag in the \ub\ vs.\ \vi\ diagram. The position of some candidate stars in this region of the diagram is indisputable evidence that they are actually clusters, masquerading as stars based on their small \ci\ due to the likely dominance of a single red or yellow supergiant (see, e.g., Drout \etal\ 2009 for a discussion of yellow supergiants).

Figure~8 shows two-color diagrams for all measured objects in and around Knot~S, rather than just for the 23 objects of Training Set 3. The major conclusions based on these diagrams are discussed in \S4. We wish to make an additional point here, related to the discussion above. While most of the resolved objects in the $-10 < \Mv < -9$ diagram hug the CB07 evolutionary track very nicely, five objects lie substantially to its left. We believe most of these are cases where there is a mixture of light from a young cluster with one (or perhaps two) individual very bright star(s), hence resulting in colors that are intermediate between those of the cluster and stellar models. Ubeda \etal\ (2007) discuss this topic in their study of NGC 4214.  This color ``stochasticity'' does not happen for the most luminous clusters ($\Mv < -10$) because such clusters have enough stars so that a few bright stars cannot greatly affect the total color. This has also been noted by other authors (e.g., Cervino, Valls-Gabaud, \& Mass-Hesse 2002).

In Figure~8 we divided each two-color diagram into four regions: cluster space, foreground-star space, yellow-star space, and blue-star/cluster space.
The first three follow from the discussion above. However, the fourth region, blue-star/cluster space, is more problematic, since both individual blue supergiants and the youngest clusters have similar colors according to the models (i.e., the upper-left regions of Figures~8 and 27b).

In principle, it should be possible to distinguish stars from clusters in this region of the two-color diagram based on their concentration index.  However, as Figure~7 illustrates, the photometric uncertainties make this difficult for objects fainter than $\Mv \approx -7$. Nevertheless, we note a clear statistical enhancement of objects in blue-star/cluster space for objects in the range $-8 <$ \Mv\ $< -7$ in Figure~8. We suspect that this is indeed caused by the onset of a large number of young blue stars within this absolute-magnitude range (see \S A.4 below for further discussion).  A similar enhancement of yellow stars is not seen in yellow-star space, probably because the yellow supergiants are too faint to be observed in the $U$ filter.

Figure~28 shows an ACS/HRC image taken as part of Proposal GO-10187, a study of Supernova 2004gt that went off in Knot~S in 2004. The supernova is the very bright, red, and saturated object to the lower left of Knot~S marked with a cross. The improved spatial resolution of the ACS/HRC provides an even better opportunity to distinguish between stars and clusters than do our dithered ACS/WFC observations. In Figure 29, which is a graytone version of Figure 28, we show the locations of the most luminous candidate yellow and blue stars around Knot~S, based on the $\ci < 1.52$ criterion. Table 5 lists these stars and gives their positions and photometric
parameters.

Finally, Figure~30 shows a close up view of the second-brightest candidate yellow star. As in several other cases, the dominant yellow-red object appears
to be embedded in a faint  cluster. This is consistent with the object's intermediate position in the two-color diagram between the CB07 cluster tracks and the Padova stellar models. Note also the preponderance of very blue or very yellow-red objects at slightly fainter magnitudes in the ACS/HRC image (Figure~28).  Most of these objects are likely to be individual stars that have fallen out of our sample, generally due to the lack of a reliable WFPC2 measurement in the $U$ band (Filter F336W).

\subsection{Distinguishing Stars and Clusters Galaxy-Wide}

We now use what we have learned from Training Sets 1 through 3 to extend
our study of objects to the entire galaxies. Figures~31a--d plot all objects in the Antennae with $\Mv < -9$ that fall in cluster space, yellow-star space, foreground-star space, and blue-star/cluster space, respectively.  

Figure~31a shows that while the vast majority of objects in {\em cluster
space} are resolved (89\% with \ci\ $>$ 1.52), there are also objects present in this part of the two-color diagram that are so compact as to be undifferentiable from stars based on their size alone, as discussed in \S A.3 above.

Next, Figure~31b shows the dramatic difference in the fraction of unresolved objects in {\em yellow-star space} (79\%), demonstrating that---while apparent size alone is not a perfect discriminator between stars and clusters---it does provide a good first-order separation, especially for bright objects with $\Mv < -9$. However, one should keep in mind that there probably also are a few heavily reddened clusters in this region of the diagram, although they appear to be rare. Note that there are only four candidate yellow stars brighter than $\Mv = -9.5$, hence this absolute magnitude appears to be an effective upper limit to the luminosity of yellow supergiants. This upper limit is known as the Humphreys--Davidson (1979) limit.

The fraction of unresolved objects in {\em foreground-star space} is an even higher 92\%, as seen in Figure~31c. Note that the objects with apparent ``absolute magnitudes'' brighter than $\Mv = -10$ are almost certainly all foreground stars in our own Galaxy, whence their ``absolute magnitudes''---computed for the adopted distance to the Antennae---are meaningless. The objects fainter than $\Mv = -10$ are likely to be a mix of foreground stars, yellow stars in the Antennae, and a few heavily reddened clusters as, e.g., the two objects with $\ci > 1.52$.

The fraction of objects in {\em blue-star/cluster space} with $\ci > 1.52$ is essentially the same (87\%) as in cluster space (89\%), demonstrating that at these bright magnitudes few---if any---of the objects are individual stars.
However, when we change our lower limit for \Mv\ from $-9$ to $-8$ mag
and then on to $-7$ mag (not shown), we find that the fraction of objects with $\ci < 1.52$ increases from 13\% to 29\% and on to 32\%, respectively.
While a few percent of these increases may be due to observational
uncertainties (e.g., the fraction of objects in cluster space with $\ci < 1.52$ also increases from 11\% to 20\%), it is clear that we are beginning to see a larger fraction of individual blue stars at fainter absolute magnitudes in this part of the diagram.

\clearpage

%######   TABLES   ###################



\clearpage

%---------------
\begin{deluxetable}{rrccrcccc}
\tabletypesize{\small}
\tablewidth{0pc}
\renewcommand{\cola}{&}
\tablecaption{Training Set 1 -- Old Globular Clusters\label{table:globs}}
\tablehead{
\colhead{\#} &
\colhead{ID} &
\colhead{RA} & 
\colhead{DEC} &
\colhead{$M_V$} &
\colhead{$C$} &
\colhead{$U\!-\!B$} &
\colhead{$V\!-\!I$} & 
\colhead{Old ID}\\
 (1) &\colhead{(2)} & (3) & (4)  & \colhead{(5)}   & (6)  & (7)   & (8) & (9)\rlap{\tablenotemark{a}}
}
\startdata
\multicolumn{9}{c}{Previous Good Candidates}\\
\colzero     1\cola  4150\colh  12:01:51.4714\colb $-$18:53:47.899\colc    $-$9.27\cold     1.41
\cole     1.23\colg 0.91\colf      8\rlap{\tablenotemark{b}}\eol
\colzero     2\cola  4169\colh  12:01:51.5055\colb $-$18:53:47.739\colc    $-$9.28\cold     1.64
\cole     0.29\colg 0.86\colf       6\eol
\colzero     3\cola  5129\colh 12:01:51.8439\colb $-$18:53:42.270\colc    $-$9.42\cold     1.64
\cole     0.01\colg 0.80\colf       3\eol
\colzero     4\cola  5641\colh 12:01:53.1949\colb $-$18:53:39.703\colc    $-$8.36\cold     1.63
\cole     0.12\colg 1.10\colf      \llap{1}0\eol
\colzero     5\cola  9160 \colh  12:01:51.0299\colb $-$18:53:24.726\colc    $-$8.63\cold     1.69
\cole     0.09\colg  0.94\colf      9\eol
\colzero     6\cola  9443\colh  12:01:50.8450\colb $-$18:53:23.616\colc    $-$9.12\cold     1.64
\cole     0.24\colg 1.07\colf       7\eol
\colzero     7\cola  10881\colh  12:01:49.6483\colb $-$18:53:18.824\colc    $-$8.12\cold     1.75
\cole     0.21\colg  0.89\colf     \llap{1}1\eol
\colzero     8\cola  19740\colh  12:01:57.2806\colb $-$18:52:49.207\colc    $-$9.33\cold     1.68
\cole     0.24\colg  0.86\colf       5\eol
\colzero     9\cola  20383\colh  12:01:54.0195\colb $-$18:52:46.552\colc   
$-$10.89   \cold     1.48
\cole     0.31\colg  0.79\colf        1\rlap{\tablenotemark{b}}\eol
\colzero    10\cola  42197\colh  12:01:51.8296\colb $-$18:52:04.096\colc    
$-$9.45\cold     1.77
\cole     \llap{$-$}0.14\colg  0.77\colf       2\eol
\colzero    11\cola  60284\colh  12:01:53.0748\colb $-$18:51:15.764\colc    
$-$9.28\cold     1.65
\cole     0.10\colg  0.80 \colf    4\eol
\multicolumn{9}{c}{Additional Good Candidates}\\
\enddata
\tablenotetext{a}{See Table 3, Whitmore et al.\ (1999).}
\tablenotetext{b}{{L}ikely star, based on small value of $C$. See Appendix A.1 for discussion.}

\end{deluxetable}
\clearpage

%---------------
\begin{deluxetable}{lcccrcrccccccc}
\tabletypesize{\scriptsize}
\tablewidth{0pc}
\tablecaption{Training Set 2 (Outer~9) -- Intermediate-Age Clusters \label{train 2}}

\tablehead{
\colhead{\#} &
\colhead{ID} &
\colhead{RA} & 
\colhead{DEC} &
\colhead{$M_V$} &
\colhead{$C$} &
\colhead{$U\!-\!B$} &
\colhead{$V\!-\!I$} & 
\colhead{$E(B\!-\!V)$} & 
\colhead{Log $\tau$} & 
\colhead{Mass} \\
(1) & (2) & (3) & (4)  & \colhead{(5)}   & (6)  & \colhead{(7)}   & (8) & (9) & (10) & (11) 
}
\startdata
\cola     1\colb    58909\colc  12:01:50.9787\cold $-$18:51:30.197\cole   $-$10.22
\colf     1.81\colg    $-$0.01\colh     0.41\coli     0.00\colj     8.41\colk     3.22 E5\eol
\cola     2\colb    58549\colc  12:01:51.0594\cold $-$18:51:31.526\cole   $-$10.10
\colf     1.81\colg    $-$0.08\colh     0.38\coli     0.00\colj     8.36\colk     2.74 E5\eol
\cola     3\colb    59721\colc  12:01:50.6764\cold $-$18:51:26.030\cole    $-$9.96
\colf     1.84\colg    $-$0.03\colh     0.39\coli     0.00\colj     8.41\colk     2.56 E5\eol
\cola     4\colb    59474\colc  12:01:49.7051\cold $-$18:51:27.682\cole    $-$9.30
\colf     1.71\colg     0.30\colh     0.46\coli     0.00\colj     8.56\colk     1.78 E5\eol
\cola     5\colb    60197\colc  12:01:49.3203\cold $-$18:51:18.795\cole    $-$9.05
\colf     1.72\colg   ---~~\colh     0.57\coli     0.04\colj     8.26\colk     1.05 E5\eol
\cola     6\colb    59798\colc  12:01:49.9133\cold $-$18:51:25.384\cole    $-$8.89
\colf     1.89\colg     0.08\colh     0.52\coli     0.00\colj     8.51\colk     1.10 E5\eol
\cola     7\colb    59904\colc  12:01:50.0653\cold $-$18:51:24.245\cole    $-$8.72
\colf     1.68\colg     0.43\colh     0.43\coli     0.00\colj     8.56\colk     1.02 E5\eol
\cola     8\colb    59797\colc  12:01:51.3711\cold $-$18:51:25.379\cole    $-$8.18
\colf     1.74\colg    $-$0.11\colh     0.35\coli     0.00\colj     8.31\colk     4.50 E4\eol
\cola     9\colb    58828\colc  12:01:51.2279\cold $-$18:51:30.517\cole    $-$8.11
\colf     1.67\colg     0.00\colh     0.65\coli     0.10\colj     8.26\colk     5.29 E4\eol

\multicolumn{11}{c}{Likely Star}\\
\enddata

\tablenotetext{a}{This magnitude is meaningless if the object is a foreground star, but is included in case the object turns out to be a cluster. See text for discussion.}

\end{deluxetable}

%---------------

\begin{deluxetable}{rcccrccccccccc}
\tabletypesize{\scriptsize}
\tablewidth{0pc}
\tablecaption{Training Set 3 (Knot S) -- Mixed Stars and Clusters \label{train_3}}

\tablehead{
\colhead{\#} &
\colhead{ID} &
\colhead{RA} & 
\colhead{DEC} &
\colhead{$M_V$} &
\colhead{$C$} &
\colhead{$U\!-\!B$} &
\colhead{$V\!-\!I$} & 
\colhead{$E(B\!-\!V)$} & 
\colhead{Log $\tau$} & 
\colhead{Mass} \\
(1) & (2) & (3) & (4)  & \colhead{(5)}   & (6)  & (7)   & (8) & (9) & (10) & (11) 
}
\startdata
\multicolumn{11}{c}{Candidate Clusters}\\
\enddata
\tablenotetext{a}{These objects are likely to be clusters, based on \ci, hence
reddening and age values are included. }

\end{deluxetable}

%------------

\begin{deluxetable}{lcccccrr}
\tablewidth{0pc}
\renewcommand{\cola}{&}
\tablecaption{Training Set 3 (Knot S) -- Brightest Candidate Yellow and Blue Stars \label{table:globs}}
\tablehead{
\colhead{\#} &
\colhead{ID} &
\colhead{RA} & 
\colhead{DEC} &
\colhead{$M_V$} &
\colhead{$C$} &
\colhead{$U\!-\!B$} &
\colhead{$V\!-\!I$} \\
(1) & (2) & (3) & (4)  & (5)   & (6)  & \colhead{(7)}   & \colhead{(8)} 
}
\startdata
\multicolumn{8}{c}{Yellow Stars}\\

\colzero     1\cola  38311 \cola  12:01:50.4612\colb $-$18:52:10.702\colc    $-$9.36\cold     1.46
\cole     0.33\colg 0.51      \eol
\colzero     2\cola  33853  \cola  12:01:50.7798\colb $-$18:52:17.531\colc    $-$9.35\cold     1.44
\cole     1.23\colg 0.63         \eol
\colzero     3\cola  33032 \cola  12:01:50.3345\colb $-$18:52:18.862\colc    $-$9.32\cold     1.40
\cole     0.45\colg 0.63         \eol
\colzero     4\cola  34751 \cola  12:01:50.2708\colb $-$18:52:16.030\colc    $-$8.85\cold     1.35
\cole     0.48\colg 0.35         \eol
\colzero     5\cola  34526 \cola  12:01:50.4021\colb $-$18:52:16.375\colc    $-$8.54\cold     1.37
\cole     0.50\colg 0.12         \eol
\colzero     6\cola  34034 \cola  12:01:50.2675\colb $-$18:52:17.256\colc    $-$8.16\cold     1.44
\cole     0.66\colg 0.50         \eol

\multicolumn{8}{c}{Blue Stars}\\
\enddata

\end{deluxetable}

\begin{deluxetable}{rcrrccccccc}
\tabletypesize{\small}
\tablewidth{0pc}
\tablecaption{Properties of Knots, Regions, and Outer Areas \label{regions}}
\tablehead{
\colhead{ID} &
\colhead{Class\tablenotemark{a}} & 
\colhead{$N$} &
\colhead{$N_9$\tablenotemark{b}} &
\colhead{$M_V(Bright)$} &
\colhead{Log $\tau$\tablenotemark{c}} &
\colhead{$\sigma$\tablenotemark{d}} & 
\colhead{$E(B\!-\!V)$\tablenotemark{c}} & 
\colhead{$\alpha$} & 
\colhead{$\beta$} & 
\colhead{Range\tablenotemark{e}}\\ 
(1) & (2) & (3) & (4)  & (5)   & (6)  & (7)   & (8) & (9) & (10) & (11) 
}
\startdata

\multicolumn{11}{c}{Star Forming Knots}\\
\enddata

\tablenotetext{a}{1 = has very dominant cluster(s), 2 = has dominant cluster(s), 3 = has no dominant cluster(s), --- = not relevant since a general region rather than single cluster complex.}
\tablenotetext{b}{Number of clusters with $M_V < -9$.}
\tablenotetext{c}{Median}
\tablenotetext{d}{Dispersion in Log $\tau$. Note that low values signify a 
single- rather than multi-age population.}
\tablenotetext{e}{Range in Log $\tau$ used in determination of $\alpha$ and $\beta$.}

\end{deluxetable}

\clearpage

\begin{figure}%1
  \centering
  \includegraphics[scale=0.85]{new-f1.eps}                       % - BCW name
  \caption{
Color image of NGC 4038/4039 produced by Lars Christensen (ESA). The F435W image is shown in blue, the F550M image in green, and a combination
of the F814W and \Halpha\ images in red.
  \label{fig01}}
\end{figure}

\begin{figure}%2
  \centering
\includegraphics[scale=1.0]{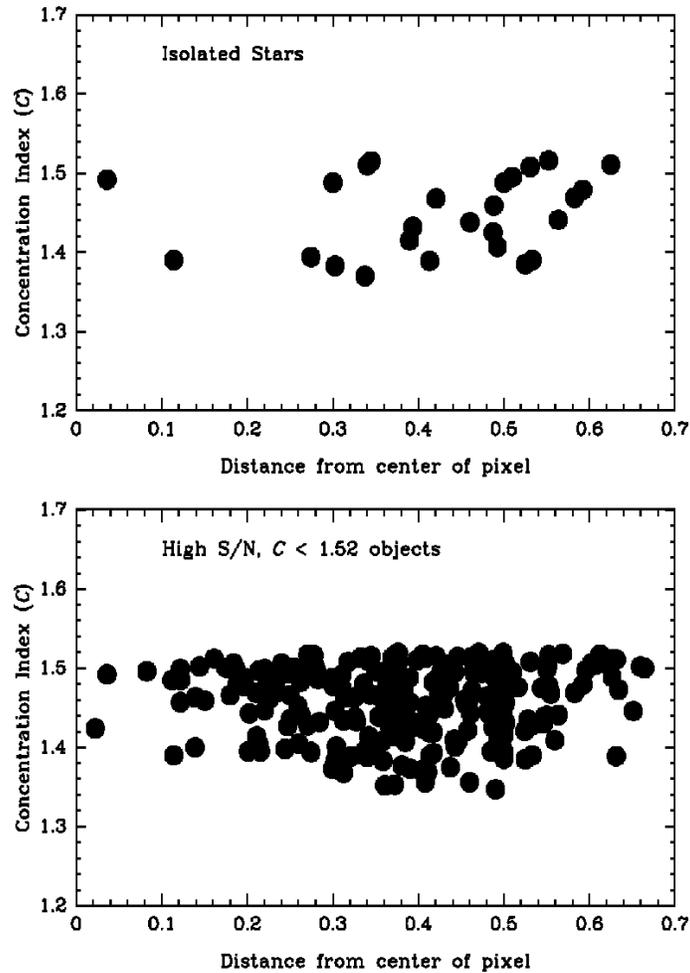}    % - BCW name
  \caption{
Values of the concentration index \ci\/, the difference between magnitudes measured within a one and a three pixel radius, are plotted versus the positions of objects relative to the pixel center for (a) a sample of 26 isolated foreground stars and (b) candidate stars selected for having values of $\ci < 1.52$ and $V < 23$ mag. Note that there are no clear trends, contrary to the claims made by Anders \etal\ (2002).
  \label{fig02}}
\end{figure}

\begin{figure}%3
  \centering
  \includegraphics[scale=1.0]{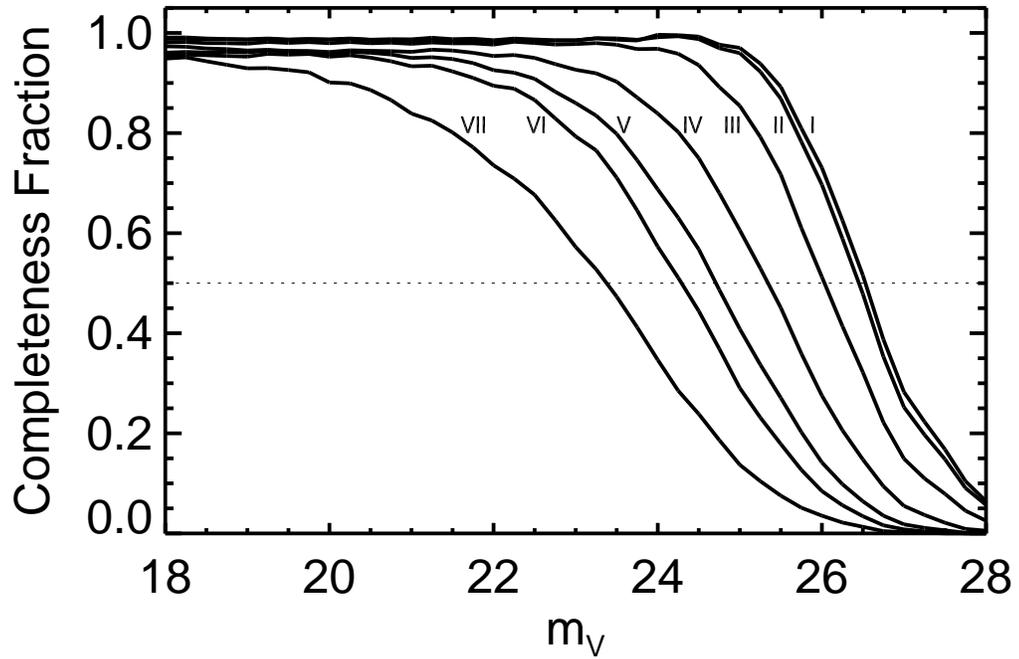}                       
  \caption{Completeness curves as determined from artificial object
experiments for different background levels.  The background levels (in units of electrons s$^{-1}$) are: $0   \le {\bf I}  \le 0.014$; $0.014 < {\bf II} \le 0.020$; $0.020 < {\bf III} \le 0.040$; $0.040 < {\bf IV} \le 0.060$; $0.060 < {\bf V}  \le 0.090$; $0.090 < {\bf VI} \le 0.120$; and ${\bf VII} > 0.120$.
  \label{fig03}}
\end{figure}

\begin{figure}%4
  \centering
\includegraphics[scale=0.7]{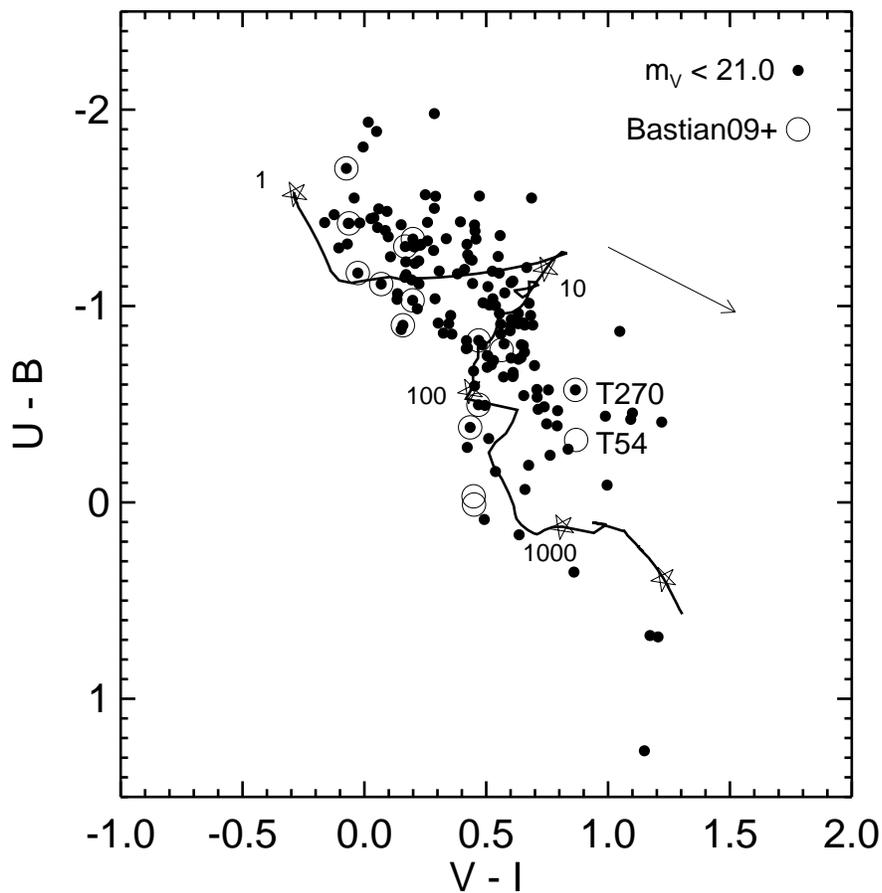}                  
  \caption{\ub\ vs.\ \vi\ diagram for the 160 brightest candidate clusters in the Antennae.  The solid line shows the evolutionary track of a CB07 model cluster of solar metallicity, with open stars denoting cluster ages of $10^6$, $10^7$, $10^8$, $10^9$, and $10^{10}$ yrs (labeled in Myr in the figure).  
Circled points also have spectroscopic observations by Bastian \etal\ (2009), providing independent age estimates. A few of the Bastian \etal\ clusters are not in our sample of the brightest 160 clusters, whence they do not feature a data point at the center of the circle.  The arrow shows the 
reddening vector for $\Av = 1.0$ mag. For details, see text.
  \label{fig04}}
\end{figure}

\begin{figure}%5
  \centering
\includegraphics[scale=0.7]{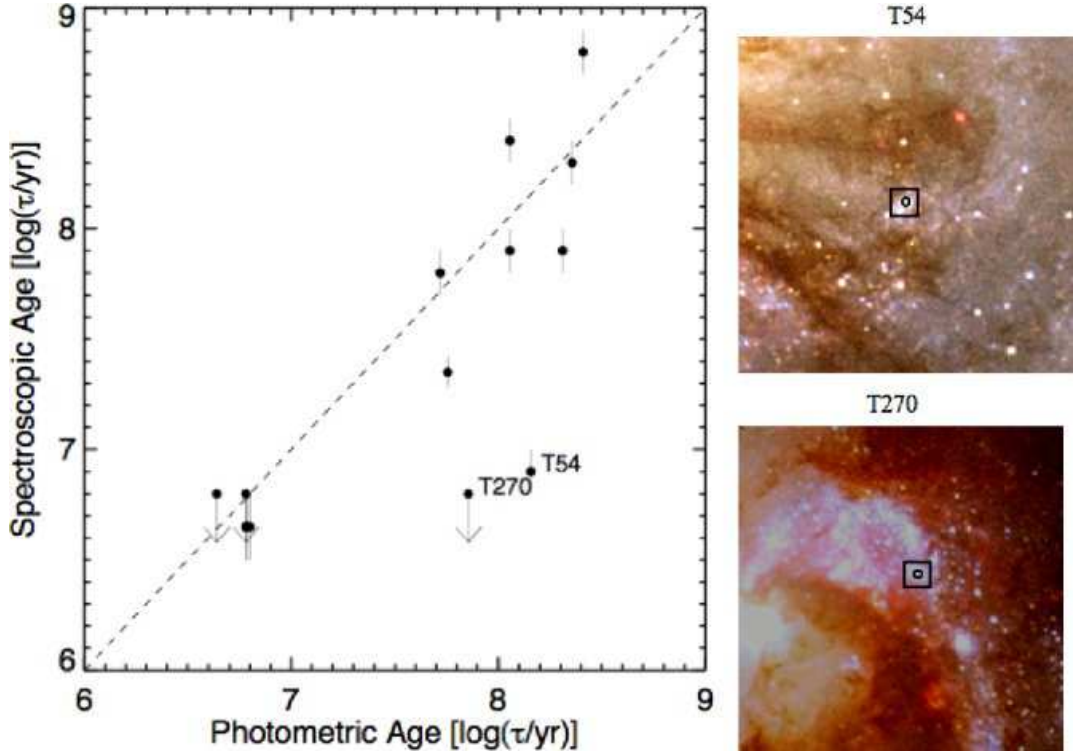}                  
  \caption{Comparison between spectroscopic estimates of cluster ages
by Bastian \etal\ (2009) and our photometric age estimates for 15 clusters (note that four data points overlay each other near [6.78, 6.65]). Images of the two outliers T54 and T270 are shown to the right, with our aperture size (radius = $0\farcs09$) marked by the small circles, and the approximate aperture (slit width = $0\farcs75$; no correction for ground-based seeing) used by Bastian \etal\ (2009) marked by the larger squares.  For details, see text.
  \label{fig05}}
\end{figure}

\begin{figure}%6
  \centering
\includegraphics[scale=0.7]{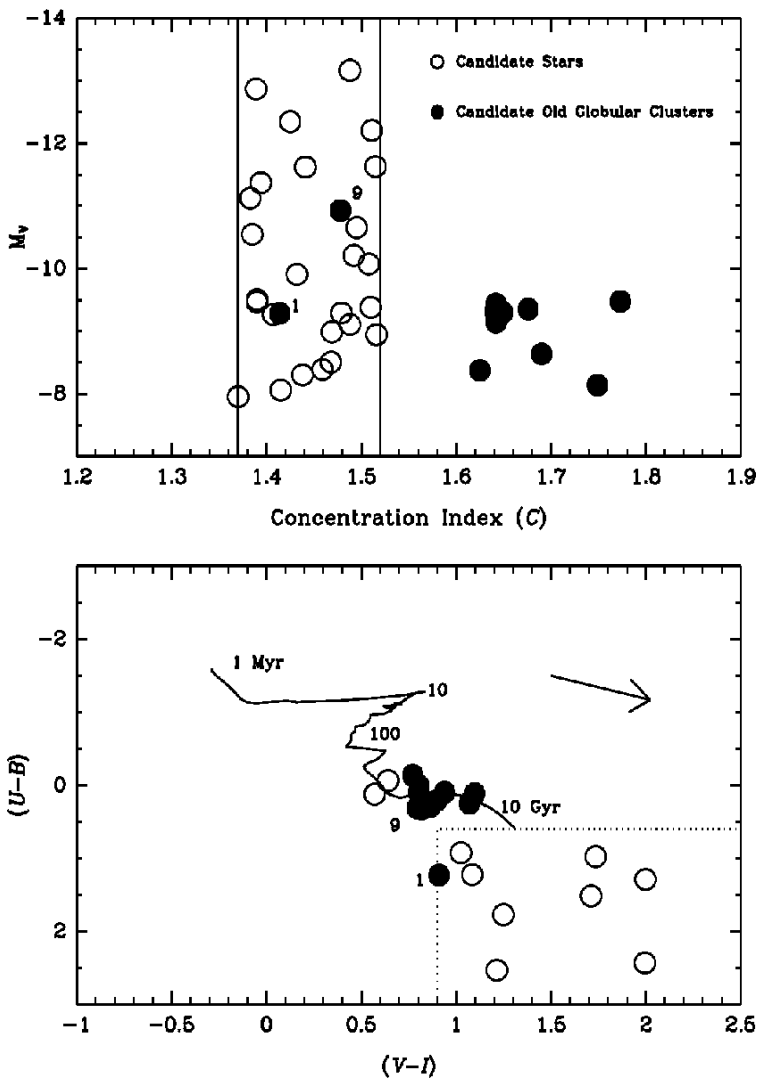}                       
  \caption{Properties for Training Set 1, which consists of hand-selected
stars (open circles) and candidate old globular clusters (filled circles) from Whitmore \etal\ (1999) in the field of the Antennae galaxies. The upper panel shows the computed absolute magnitude \Mv\ plotted vs.\ concentration index \ci, while the lower panel shows a two-color plot, \ub\ vs.\ \vi, for the same objects. The arrow marks a reddening corresponding to $\Av = 1$ mag,  assuming a standard Galactic extinction law, while the solid line shows the evolutionary track of a CB07 model star cluster of solar metallicity.
  \label{fig06}}
\end{figure}

\begin{figure}%7
  \centering
\includegraphics[scale=0.7]{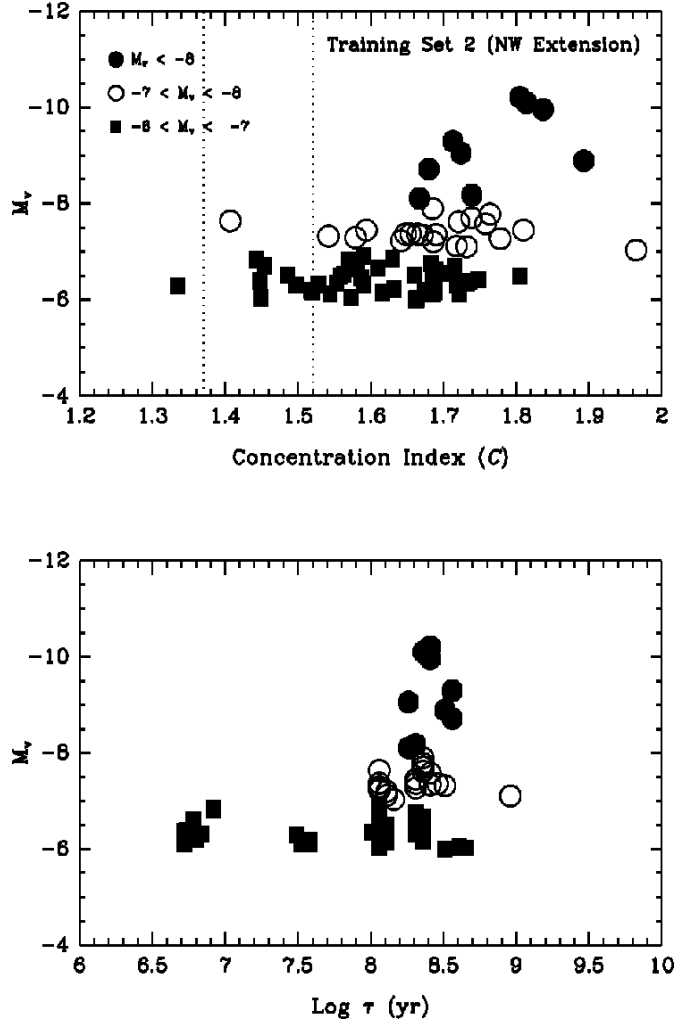}                       
  \caption{Properties for Training Set 2 (in NW Extension), which is
dominated by intermediate-age clusters. The upper panel shows that bright sources have concentration indices \ci\ expected of clusters. At fainter magnitudes increased observational scatter results in some objects falling into (or even below!) the star category. This demonstrates that using size alone to distinguish stars from clusters breaks down at $\Mv \approx -7$ for isolated objects. Similarly, the bottom panel shows that the age determination breaks down at $\Mv \approx -7$, since none of the plotted objects are believed to have ages of $\log\tau < 8.0$. 
  \label{fig07}}
\end{figure}

\begin{figure}%8
  \centering
\includegraphics[scale=0.6]{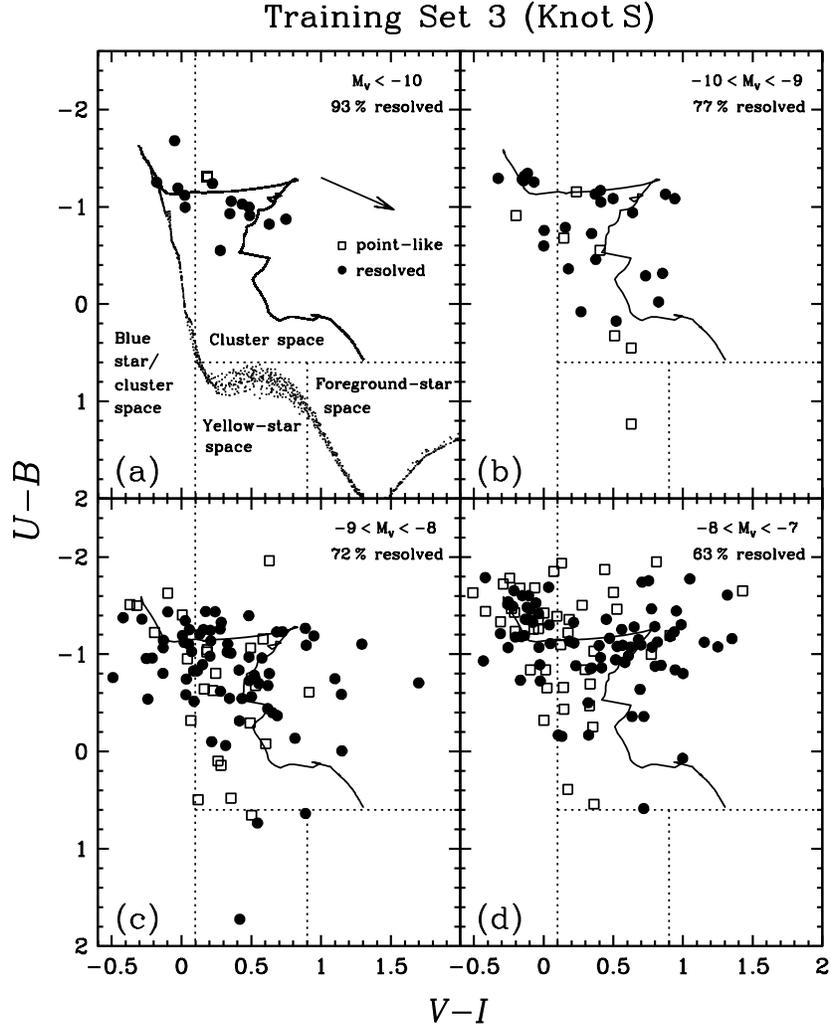}  
  \caption{\ub\ vs.\ \vi\ two-color diagrams for objects around Knot~S.
Resolved objects (i.e., $\ci > 1.52$) are shown as filled circles and unresolved objects as open squares. The four panels show objects in different luminosity ranges. The solid lines in each panel show CB07 evolutionary tracks for clusters of solar metallicity, while the dots in Panel (a) show Padova model colors for stars brighter than $\Mv = -7$.  The dotted lines show the regions of the diagram used to define cluster space, foreground-star space, yellow-star space, and blue-star/cluster space, as discussed in the text.
  \label{fig08}}
\end{figure}

\begin{figure}%09
  \centering
\includegraphics[scale=0.7]{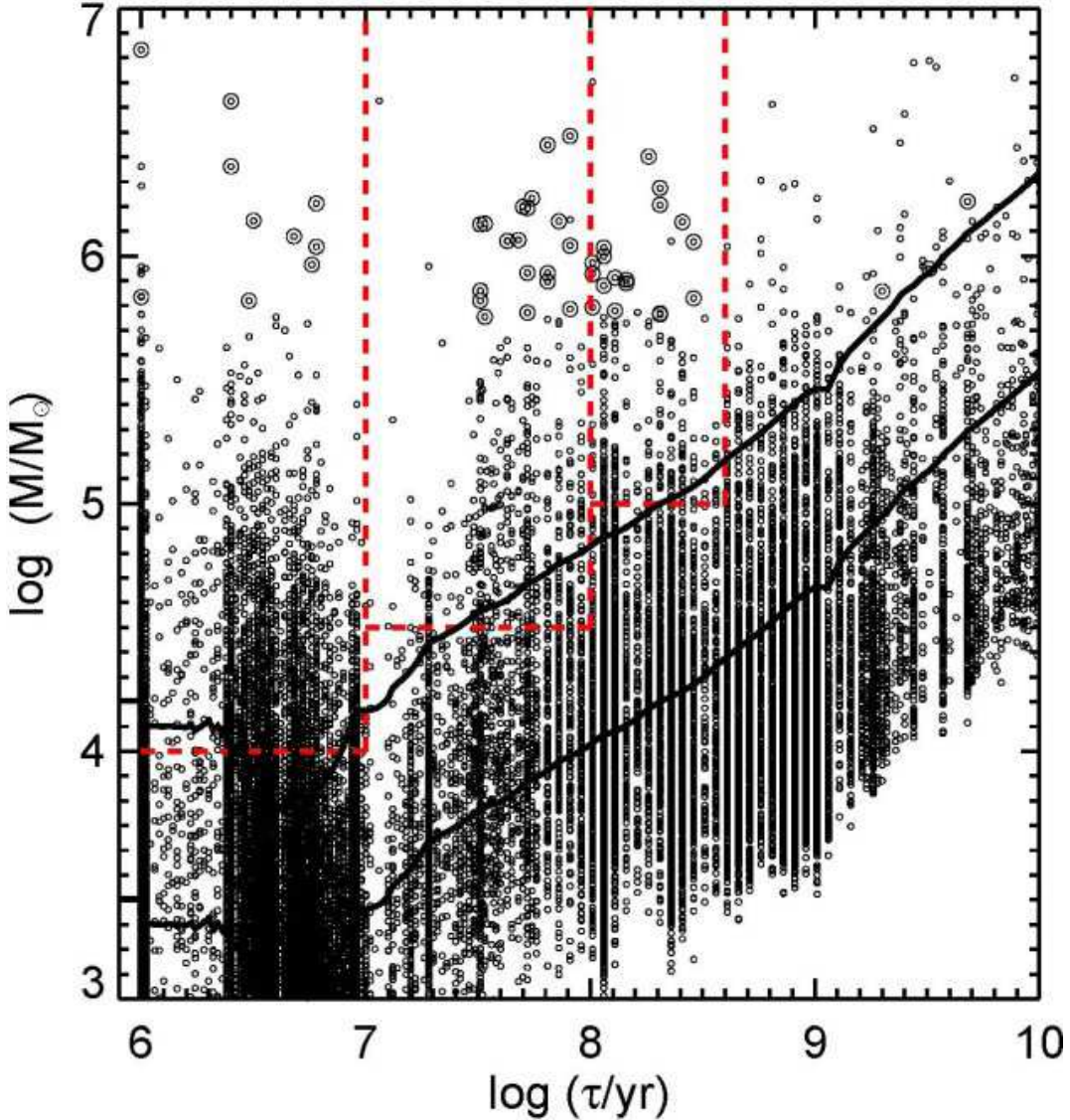}
  \caption{Log Mass vs.\ $\log\tau$ diagram for clusters in the Antennae.
The solid black lines show completeness curves for $\Mv = -9$ (to exclude
stars based on luminosity alone) and $\Mv = -7$ (where age and size estimates become difficult). The dashed red lines show the bins used for the MF determinations in Figure~15. The 50 most massive candidate clusters from Table 7 are circled. Some artifacts (e.g., stars, the nuclei of the two galaxies, etc.) have been removed from the sample based on visual inspection.
In addition, many of the apparently massive sources  with $\log\tau > 8.6$
are likely to be younger clusters with strong foreground extinction, as discussed in the text, and hence are not included in Table 8 or circled in this figure. 
  \label{fig09}}
\end{figure}

\begin{figure}%10
  \centering
\includegraphics[angle=-90,scale=0.57]{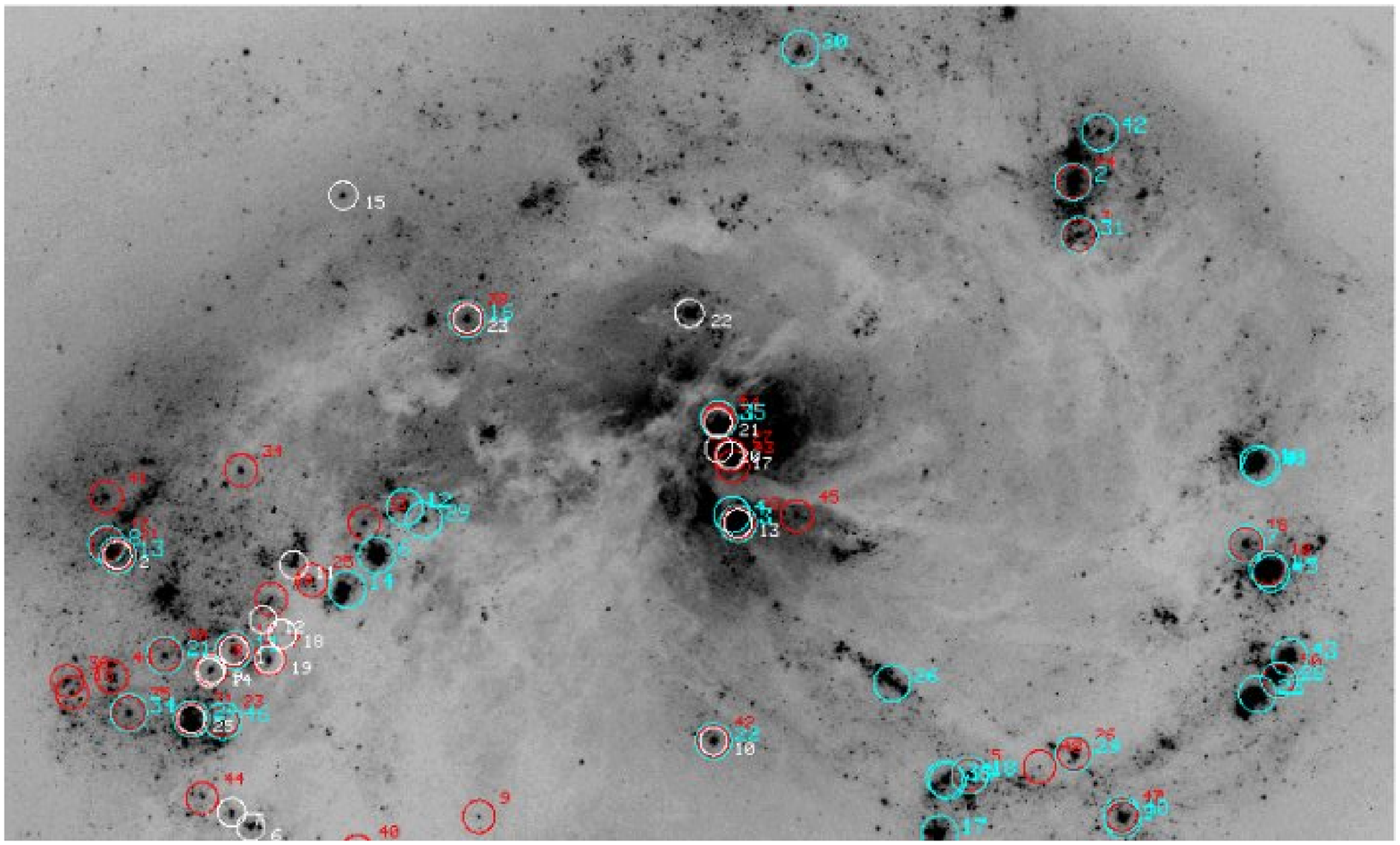}\\
\smallskip
\includegraphics[angle=-90,scale=0.57]{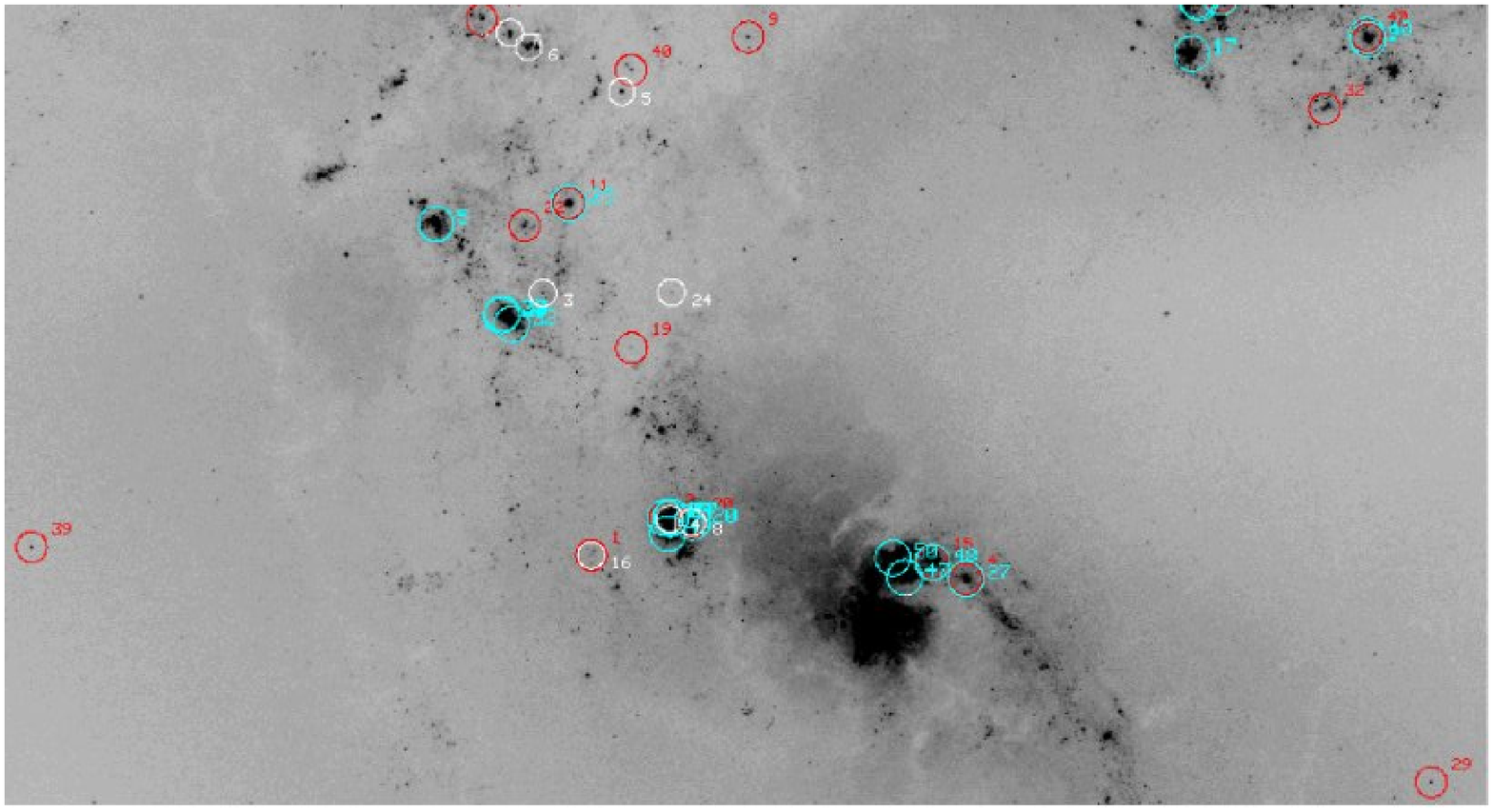}
  \caption{(a) Top half of the Antennae (NGC 4038) with the 50 most
luminous clusters (blue circles) and the 50 most massive clusters (red
circles) marked (see Tables 6 and 7). (b) Same information as in (a), but for the bottom half of the Antennae (NGC 4039). Also marked are the 25 most IR-bright clusters (white circles, see Table~8). Their number has been scaled down since the FOV for NICMOS observations was only about half that covered by the ACS observations (see also Figure~11). 
  \label{fig10}}
\end{figure}

%\setcounter{figure}{9}
%\begin{figure}
%  \centering
%\caption{
%  \label{fig10b}}
%\end{figure}

\begin{figure}%11
  \centering
\includegraphics[scale=0.8]{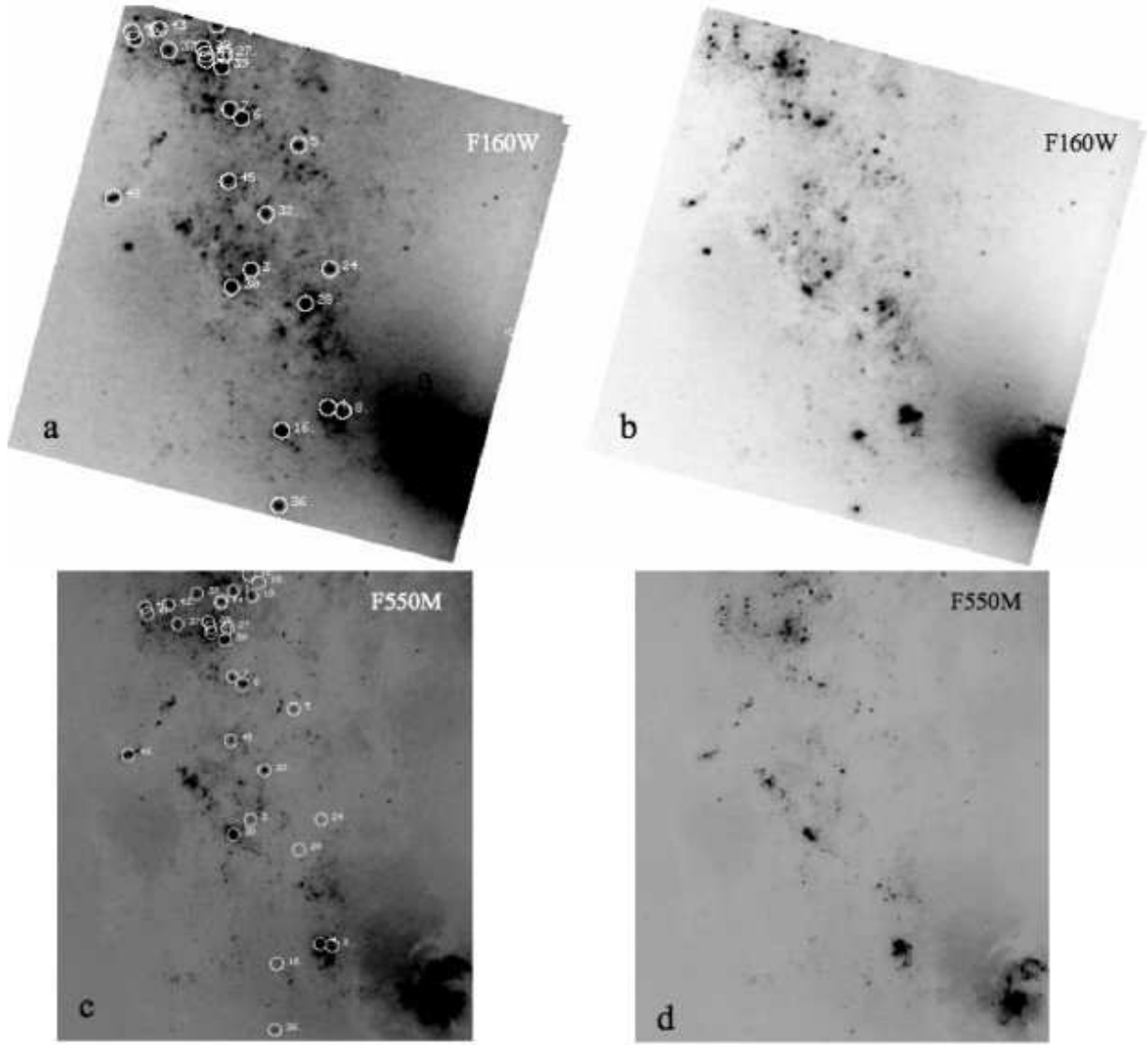}
  \caption{NICMOS field No.~1, which includes the nucleus of NGC~4039.
Panels a and b show different-contrast renditions of the F160W image, with and without the brightest IR clusters from Table 8 identified by circles.
Panels c and d show similar FOVs, taken from the F550M image obtained
with ACS.
  \label{fig11}}
\end{figure}

\begin{figure}%12
  \centering
\includegraphics[scale=0.8]{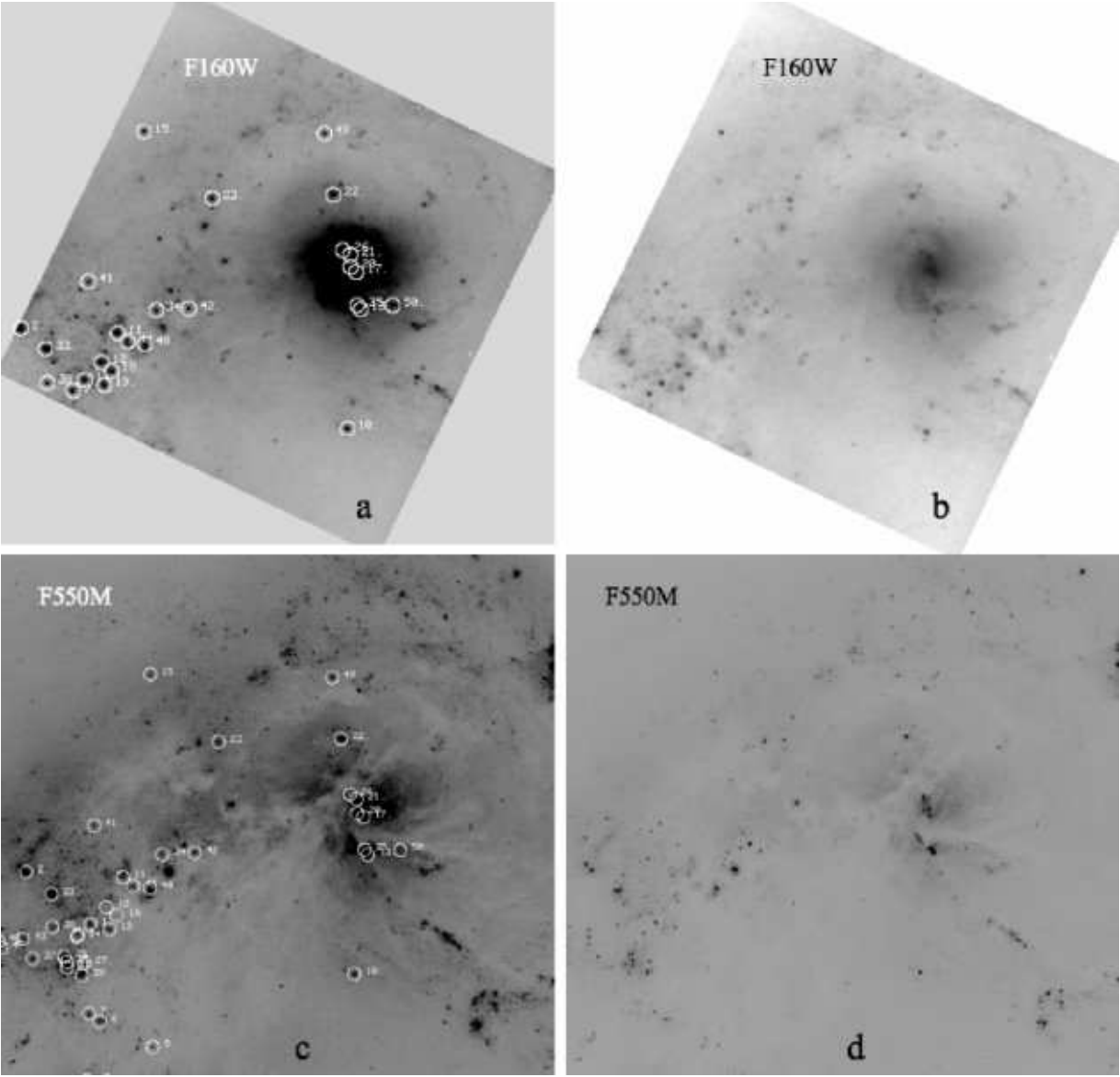}
  \caption{Similar to Figure~11, but for NICMOS field No.~2, which includes the nucleus of NGC 4038.
  \label{fig12}}
\end{figure}

\begin{figure}%13
  \centering
\includegraphics[scale=0.75]{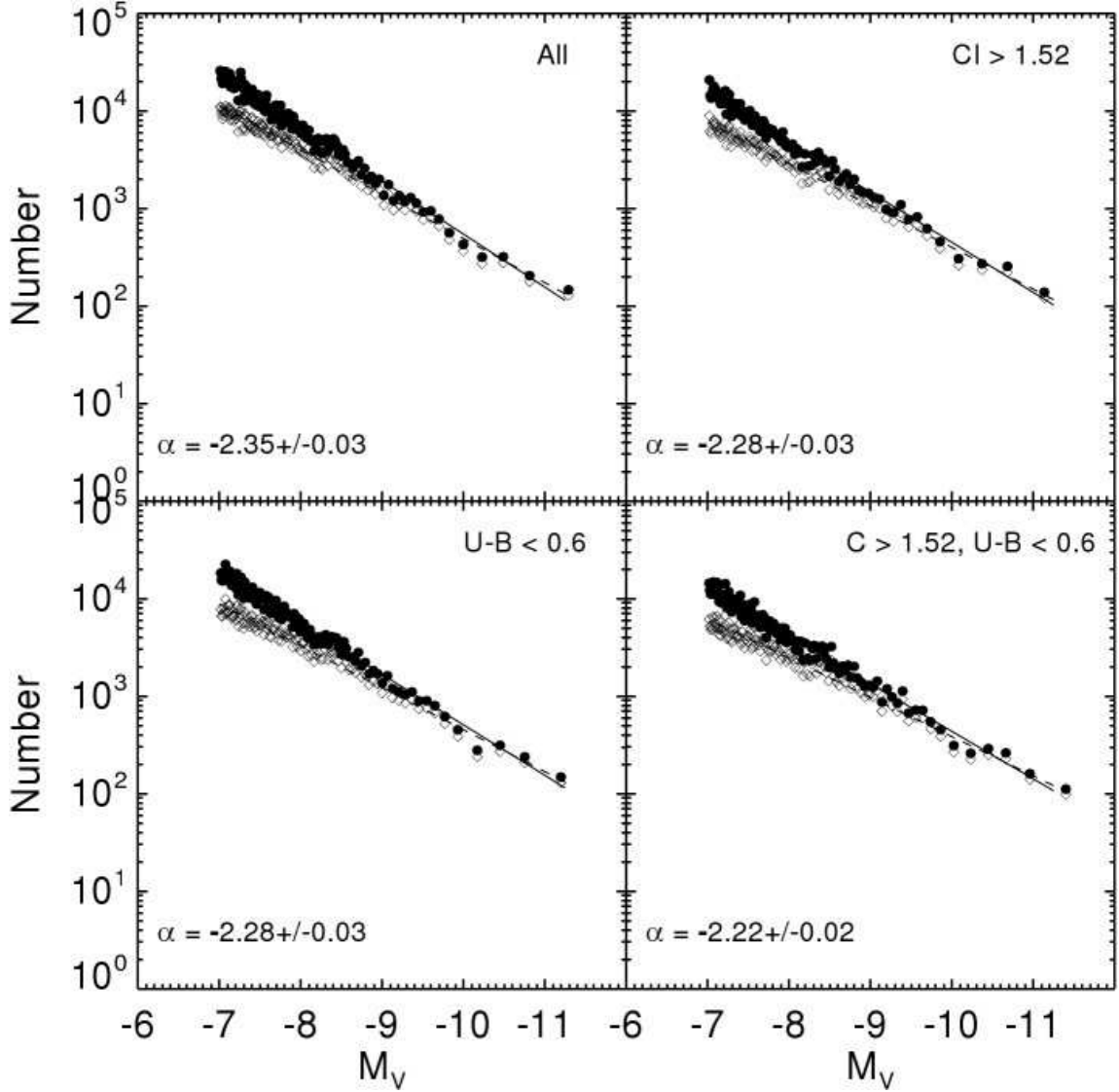}
  \caption{Luminosity functions for objects across the entire galaxy: 
(a) LF for all detected objects (i.e., a mix of individual stars and star clusters);
(b) LF only for resolved objects ($\ci > 1.52$);
(c) LF for objects with \ub\ colors that place them in ``cluster space'' or ``blue-star/cluster space''; and
(d) LF for objects selected by both size and \ub\ color to be likely young clusters.
Variable-binning is used with open circles showing observed data and filled circles showing completeness corrected data. Values of $\alpha$ are for the completeness corrected fits to the data. 
  \label{fig13}}
\end{figure}

\begin{figure}%14
  \centering
\includegraphics[angle=0,scale=0.4]{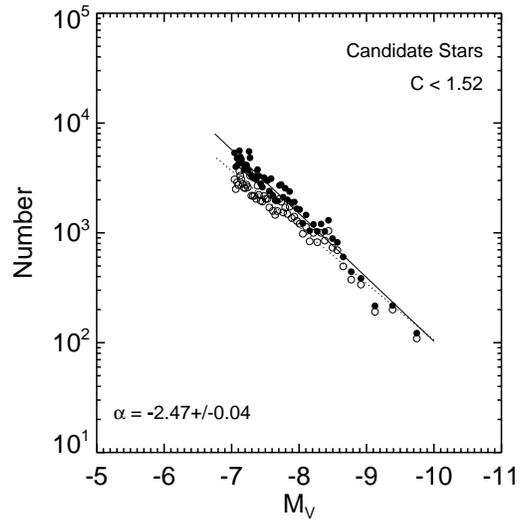}
  \caption{Luminosity function for candidate stars ($\ci < 1.52$) across the entire galaxies.
  \label{fig14}}
\end{figure}

\begin{figure}%15
  \centering
\includegraphics[angle=0,scale=0.8]{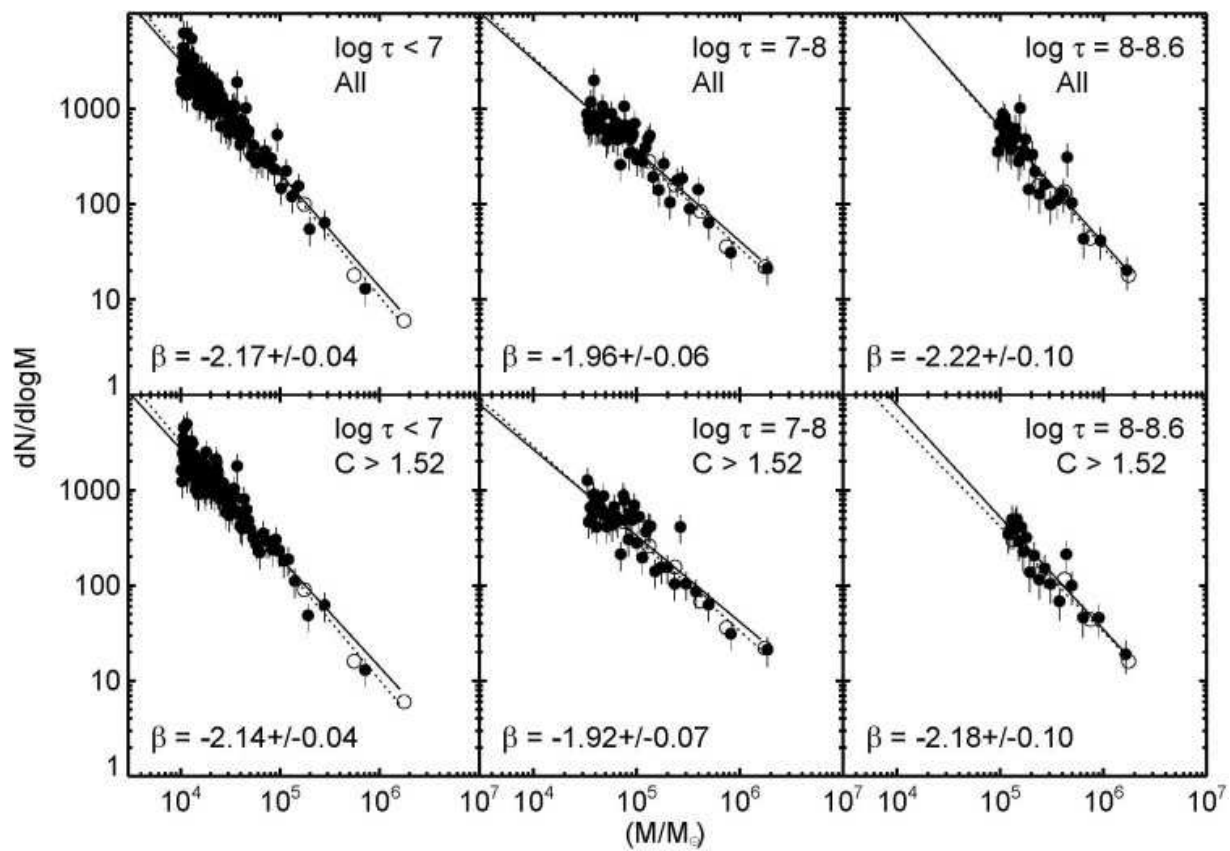}
  \caption{The top panels show the mass functions for clusters in the Antennae for three different age intervals, using the bins shown by the dashed red lines in Figure~9. The bottom panels show the corresponding mass functions for only the resolved clusters ($\ci > 1.52$).  The solid lines mark best fits to the variable-binning data (filled circles), while the dashed lines mark best fits to the fixed-binning data (open circles). The values of $\beta$ given are for the variable-binning data. 
  \label{fig15}}
\end{figure}

\begin{figure}%16
  \centering
\includegraphics[angle=-90,scale=0.6]{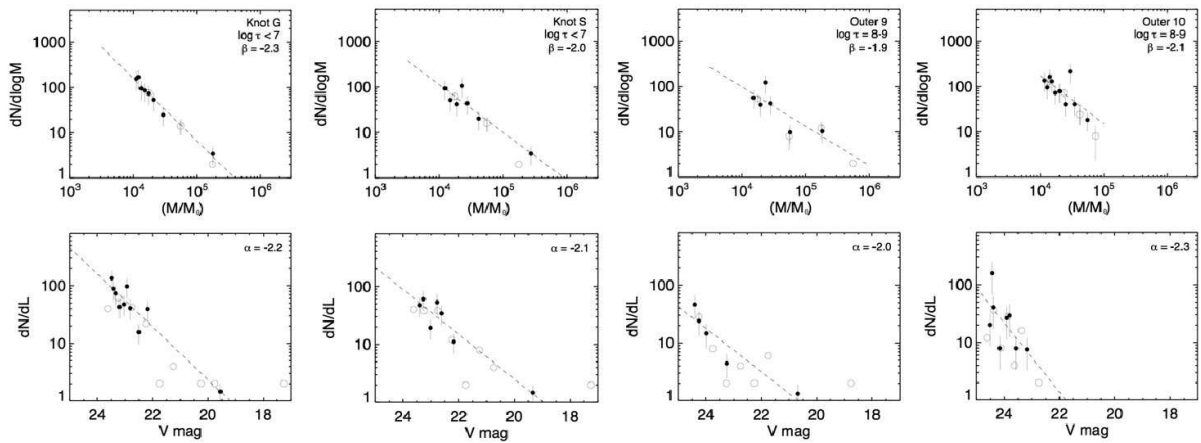}
  \caption{(top row) Cluster mass functions for age ranges corresponding to the dominant-age population in each of four representative areas, and (bottom row) the corresponding cluster luminosity functions. The dashed lines mark best fits to the variable-binning data (filled circles).
  \label{fig16}}
\end{figure}

\begin{figure}%17
  \centering
\includegraphics[scale=0.8]{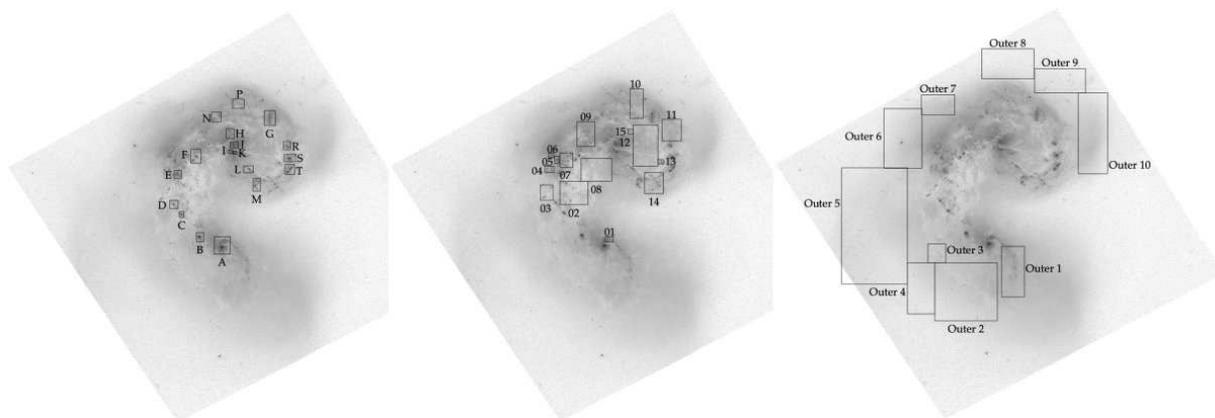}
  \caption{Locations of the star forming knots (letters), regions (numbers),
and outer areas in the Antennae.
  \label{fig17}}
\end{figure}

\clearpage   % Inserted July 1, 2009 at suggestion of Sharon Toolan to fix
             % the problem with including more than 16 figures

\begin{figure}%18
  \centering
\includegraphics[scale=0.7]{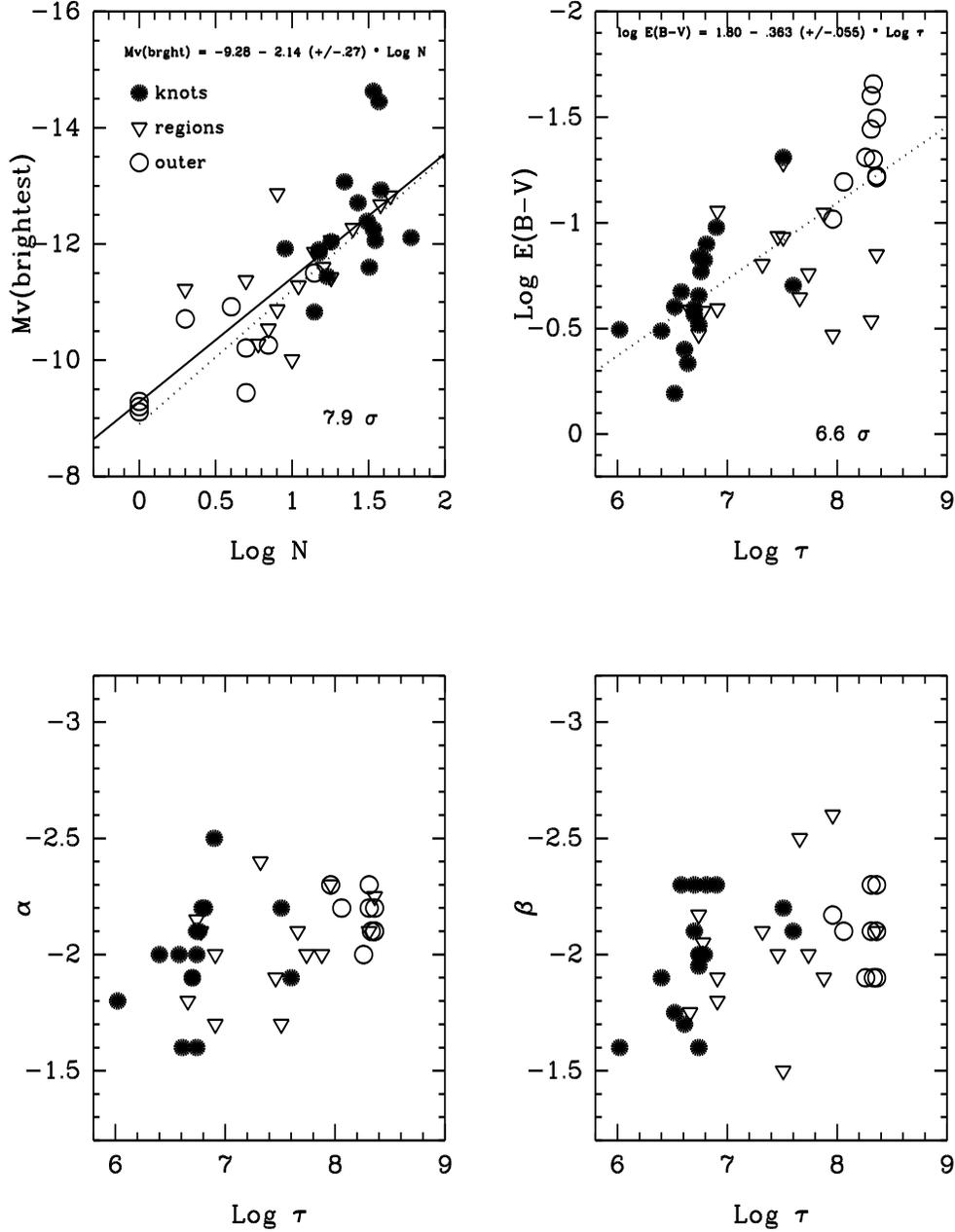}
  \caption{Various correlations for clusters in the knots, regions, and
outer areas shown in Figure~17. (a, upper left) Absolute magnitude \Mv\ of the brightest cluster in each area plotted vs.\ the logarithm of the number of clusters with $\Mv < -9$ in same area (the dotted line is the best-fit line for cluster populations in entire galaxies from WC07); (b, upper right) $\log E(B\!-\!V)$ vs.\ $\log\tau$ diagram for same clusters; (c, lower left) slopes $\alpha$ of the regional luminosity functions plotted vs.\ $\log\tau$; (d, lower right) slopes $\beta$ of the regional mass functions plotted vs.\  $\log\tau$.
The slopes of the best-fit lines for the top two diagrams are given and the significance ($\sigma$) of the correlation is given to the lower right in each diagram. For a detailed discussion, see text.
  \label{fig18}}
\end{figure}

\begin{figure}%19
  \centering
\includegraphics[scale=0.8]{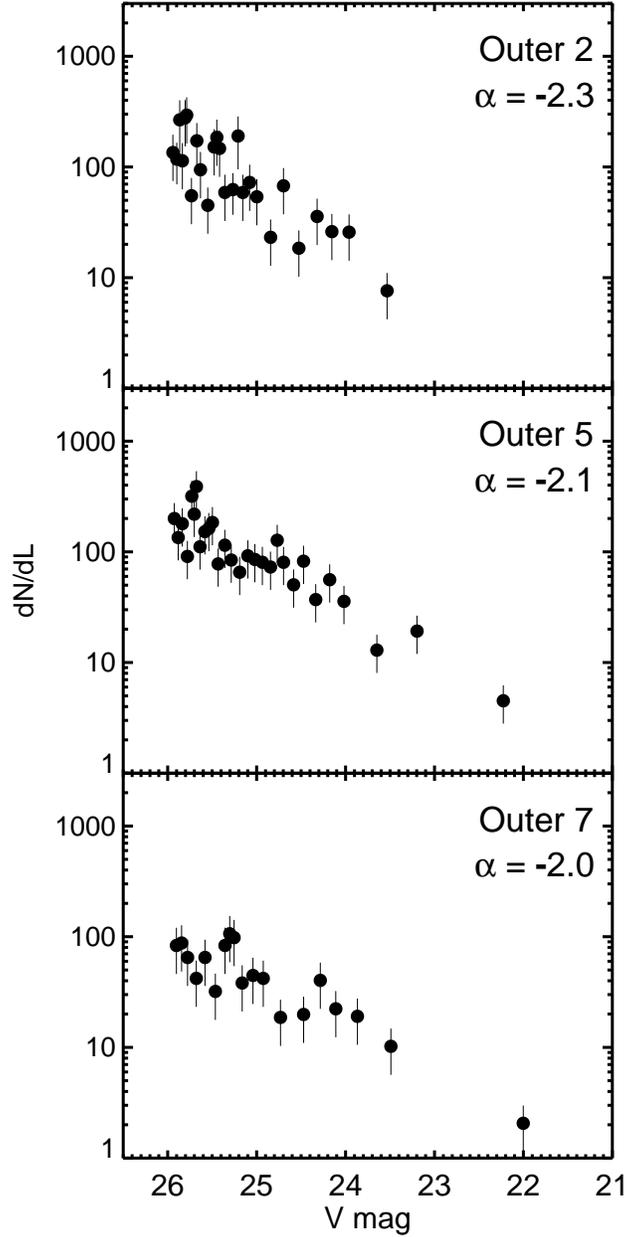}
  \caption{Luminosity functions for clusters in three outer areas. Note that there is no evidence of any turnover in the luminosity functions at fainter magnitudes. This is consistent with relaxation-driven cluster disruption models, which predict the turnover should not be observed until $\Mv \approx -4$ for clusters with ages of $\sim\,$200 Myr. This would correspond to a value of $V = 27.7$ in the diagram.
  \label{fig19}}
\end{figure}

\begin{figure}%20
  \centering
\includegraphics[scale=0.75]{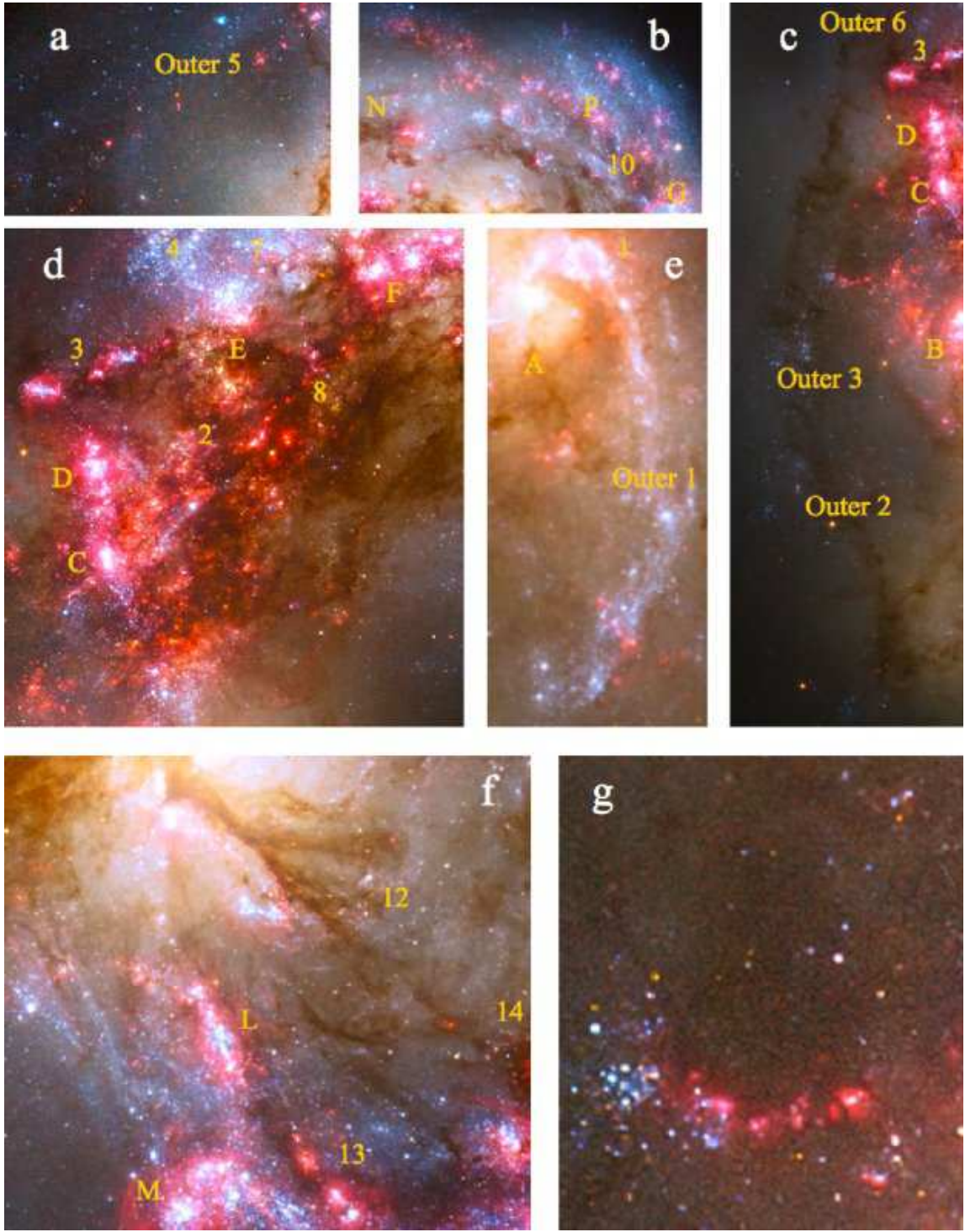}
  \caption{Mosaic of subareas in the Antennae (from Figure~1) that show interesting patterns of star formation. The capital letters and numbers in yellow correspond to the definitions of star formation knots, regions, and outer areas defined in Figure~17. For a discussion, see \S9.
  \label{fig20}}
\end{figure}

% -----------------
% Figure 13b  - Knot S, Halpha
\begin{figure}%21
\centering
\includegraphics[scale=0.6]{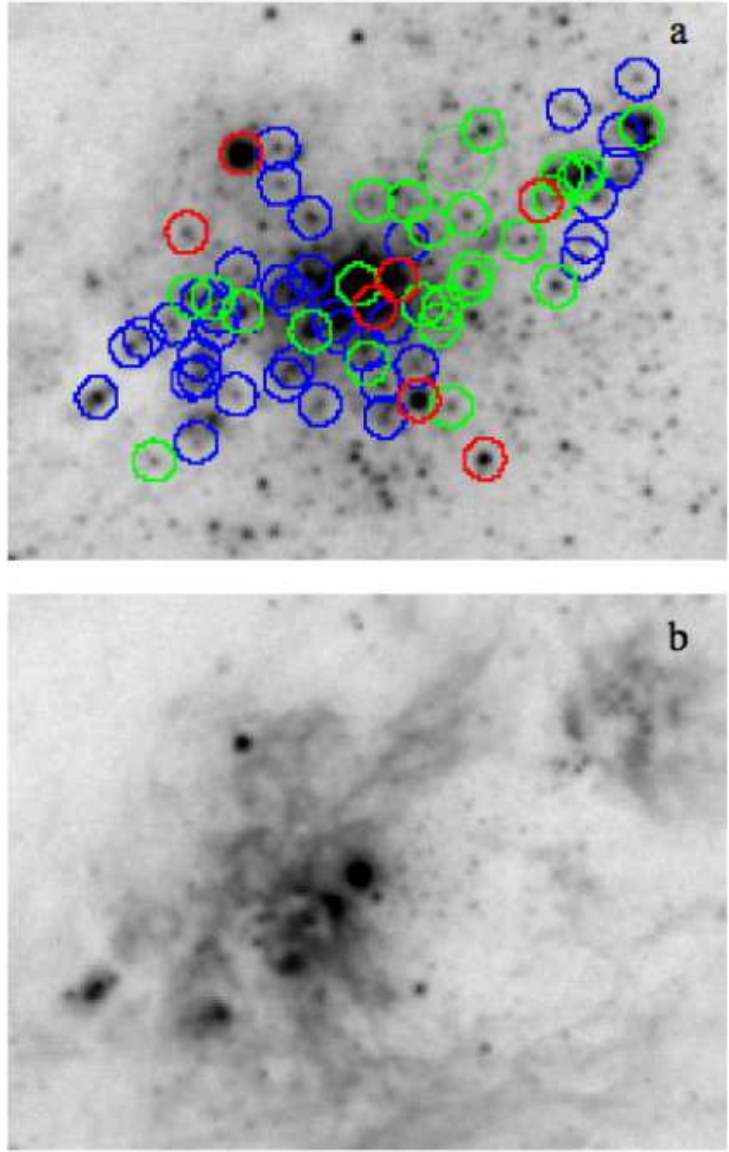}
  \caption{Top panel shows the $V$ image of Knot~S, with clusters color-marked by age (blue: $\log\tau < 6.7$; green: $6.7 < \log\tau < 7.0$; and red:   $7.0 < \log\tau < 7.7$). The bottom panel shows the corresponding \Halpha\ image of Knot~S.
}
\label{fig21}
\end{figure}

%--------
% Figure 19a - Region G image
\begin{figure}%22
\centering
\includegraphics[scale=0.6]{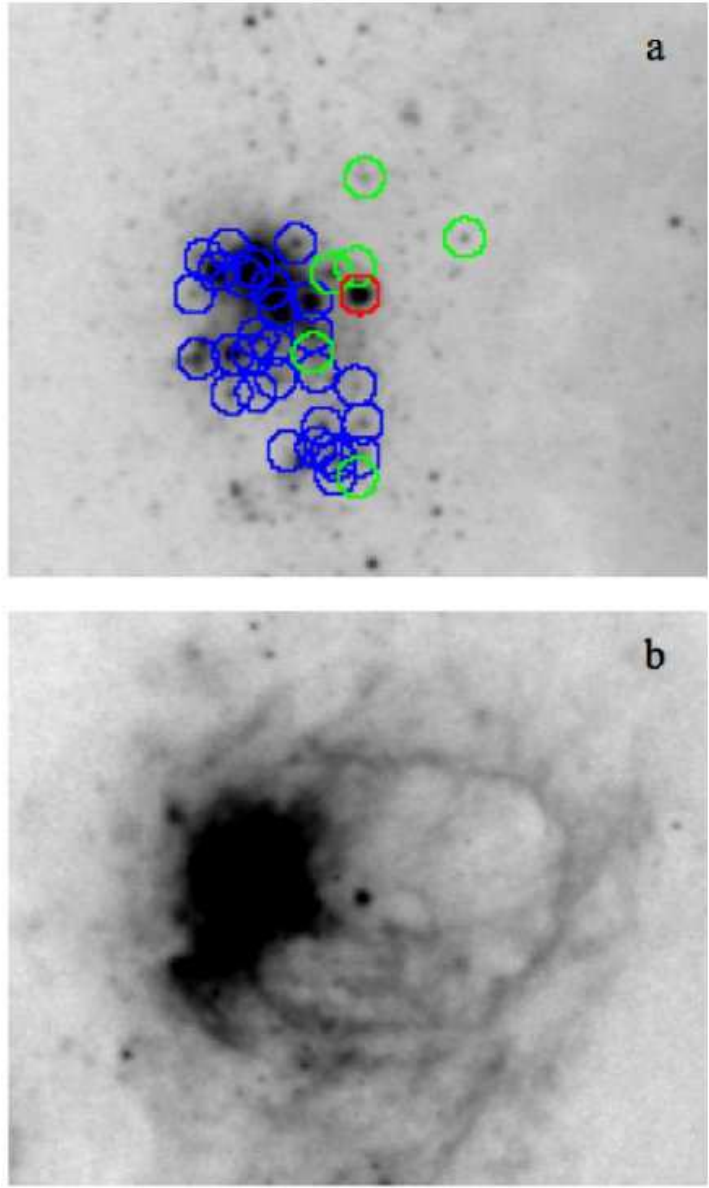}
  \caption{Similar to Figure 21, but for Knot~B.
The one intermediate-age cluster is circled in red and is the 20th most massive cluster from Table 7, with a mass of $\approx 10^6\,M_{\odot}$
and an age in the range 10--50 Myr. It appears to lie at the center of the prominent \Halpha\ shell (lower panel) and is likely to have triggered the formation of the younger clusters around it, including the second most massive cluster known in the Antennae, with a mass of $4.2\times 10^6\,M_{\odot}$ and age of $\sim\,$2.5~Myr.
}
\label{fig22}
\end{figure}

%--------
% Figure 19a - Triggered 4 plot
\begin{figure}%23
\centering
\includegraphics[scale=0.8]{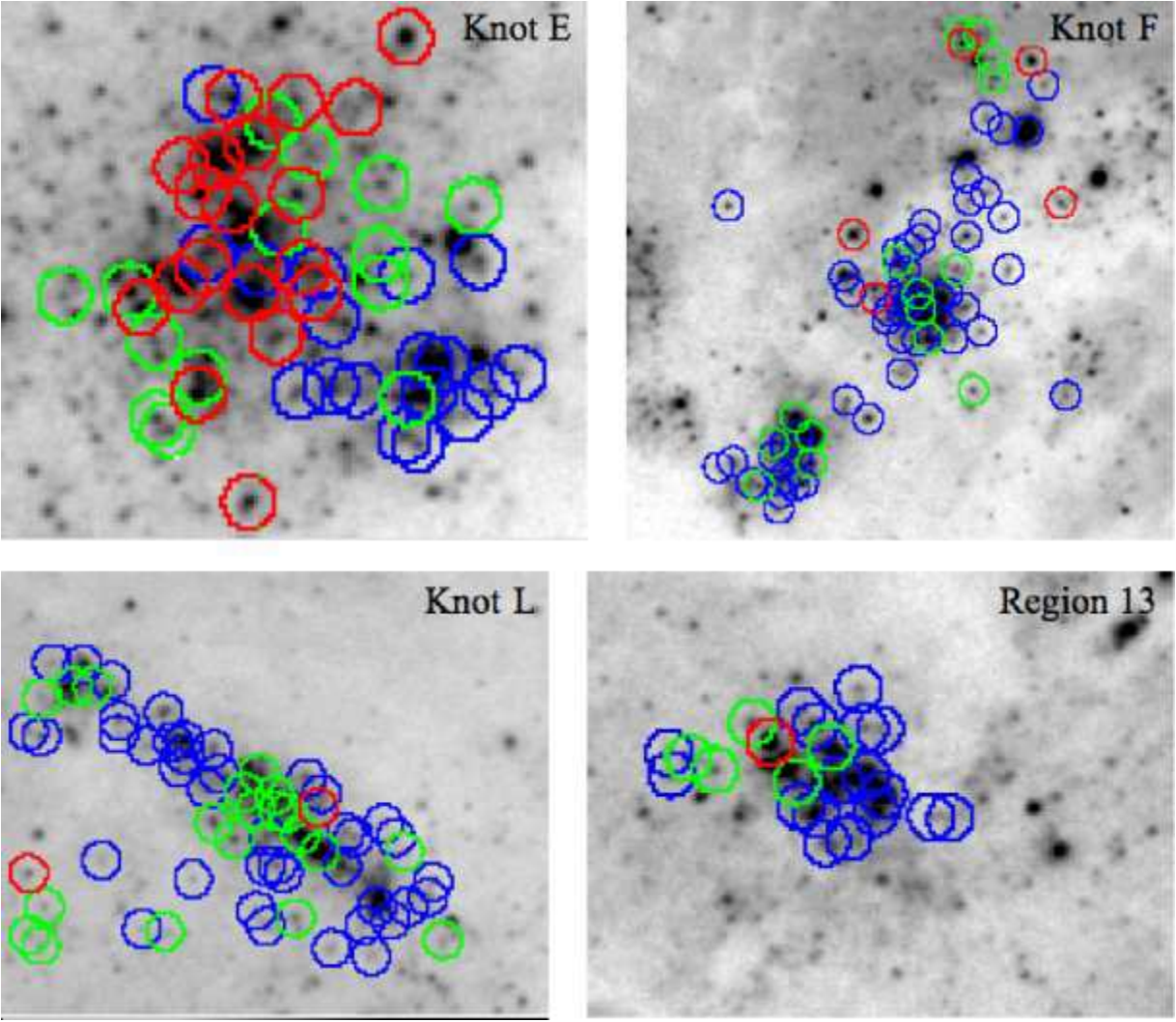}
\caption{Evidence for triggered cluster formation in knots E, F, L, and
Region 13. Clusters are color-marked by age (blue: $\log\tau < 6.7$; green: $6.7 < \log\tau < 7.0$; red: $7.0 < \log\tau < 7.7$).
}
\label{fig23}
\end{figure}

%--------
% Figure 6 - Training set 2 - images
\begin{figure}%24
\centering
\includegraphics[scale=0.7]{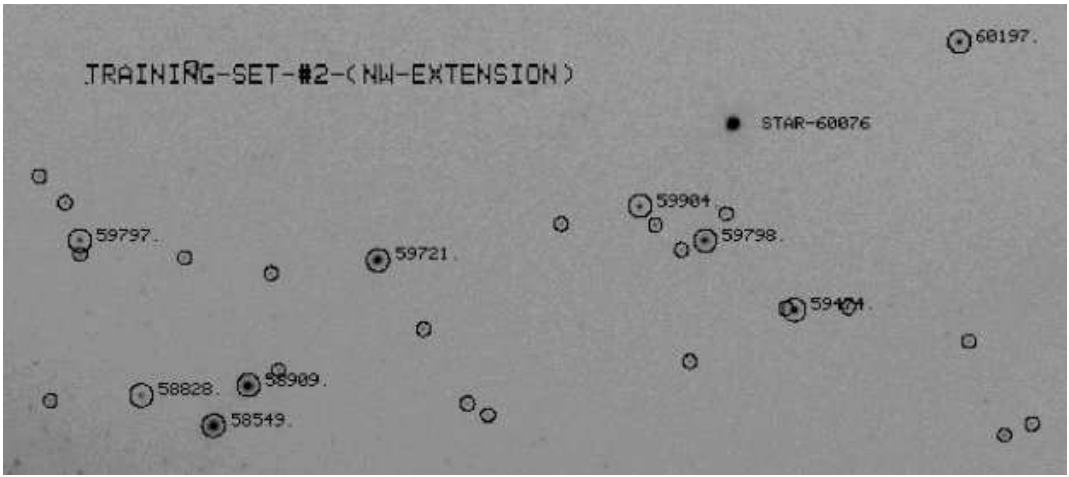}
\caption{Image of area containing Training Set 2. This area is marked ``Outer 9'' in Fig\-ure~17 and is called the ``NW Extension'' in Whitmore \etal\ 1999. The most luminous clusters ($\Mv < -8$) are marked by large circles with the ID number included (see Table 3), while the less luminous clusters in the range $-8 < \Mv < -7$ are marked by small circles.
Most of the fainter clusters are not visible here because of the chosen
contrast stretch. Note that the brightest object (60076) is probably a star since it has a concentration index of $\ci = 1.39$, although it has colors similar to the other clusters in the region. 
}
\label{fig24}
\end{figure}

%--------
% Figure 8 - Training set 2 - color-color
\begin{figure}%25
\centering
\includegraphics[angle=0,scale=0.7]{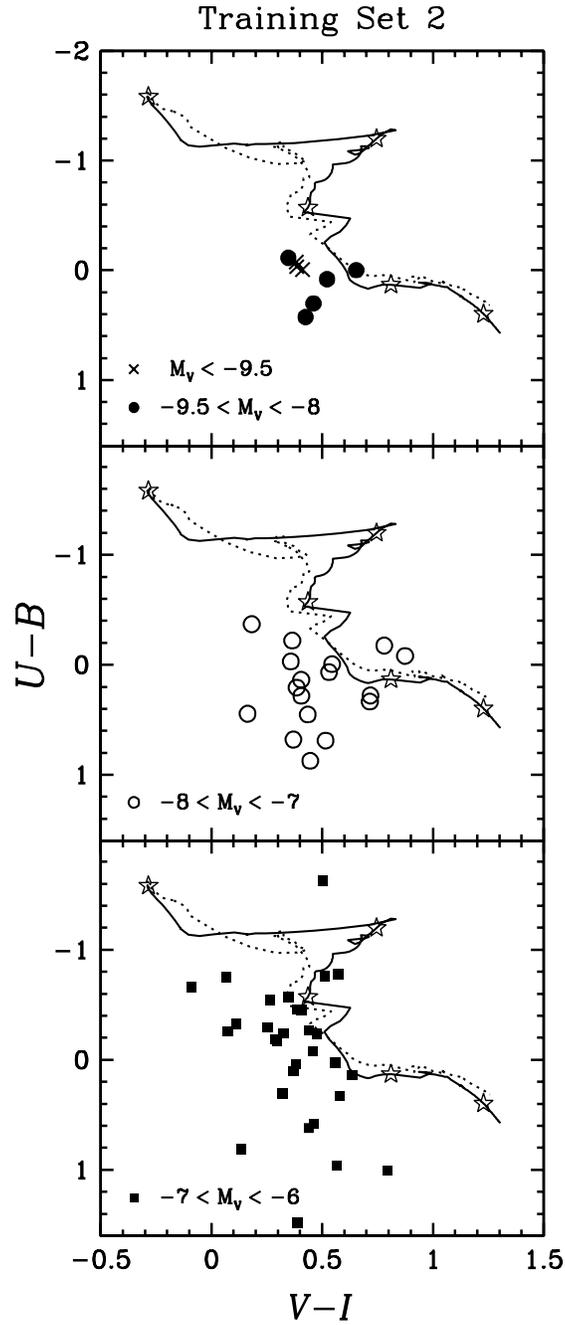}
\caption{\ub\ vs.\ \vi\ diagrams for Training Set 2, with objects sorted
by absolute magnitude \Mv. Note how small the scatter is for the three brightest clusters (crosses in top diagram) and how it increases for fainter clusters. The solid lines show evolutionary tracks for CB07 model clusters of solar metallicity, while the dotted lines show similar tracks for model clusters of $Z = 0.4\,Z_{\odot}$. 
}
\label{fig25}
\end{figure}

%--------
% Figure 9 - Training set 3 - images
\begin{figure}%26
\centering
\includegraphics[scale=0.8]{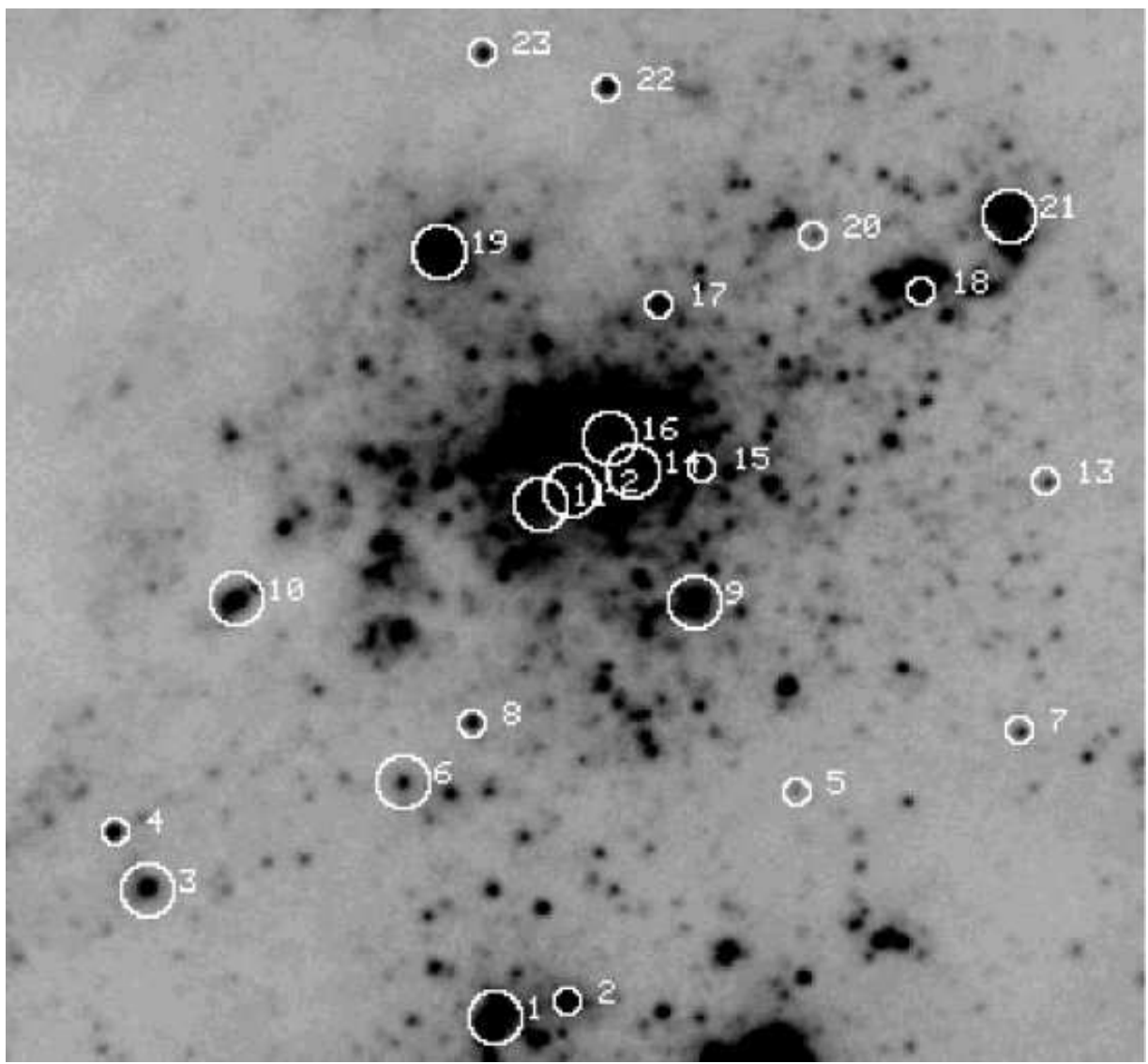}
\caption{Image of the area around Knot~S, showing the Training Set 3
objects discussed in the text and listed in Table~4. Large circles mark candidate clusters, while small circles mark candidate stars. 
}
\label{fig26}
\end{figure}

%--------
% Figure 10 - Plot  that goes with training set 3
\begin{figure}%27
\centering
\includegraphics[scale=0.7]{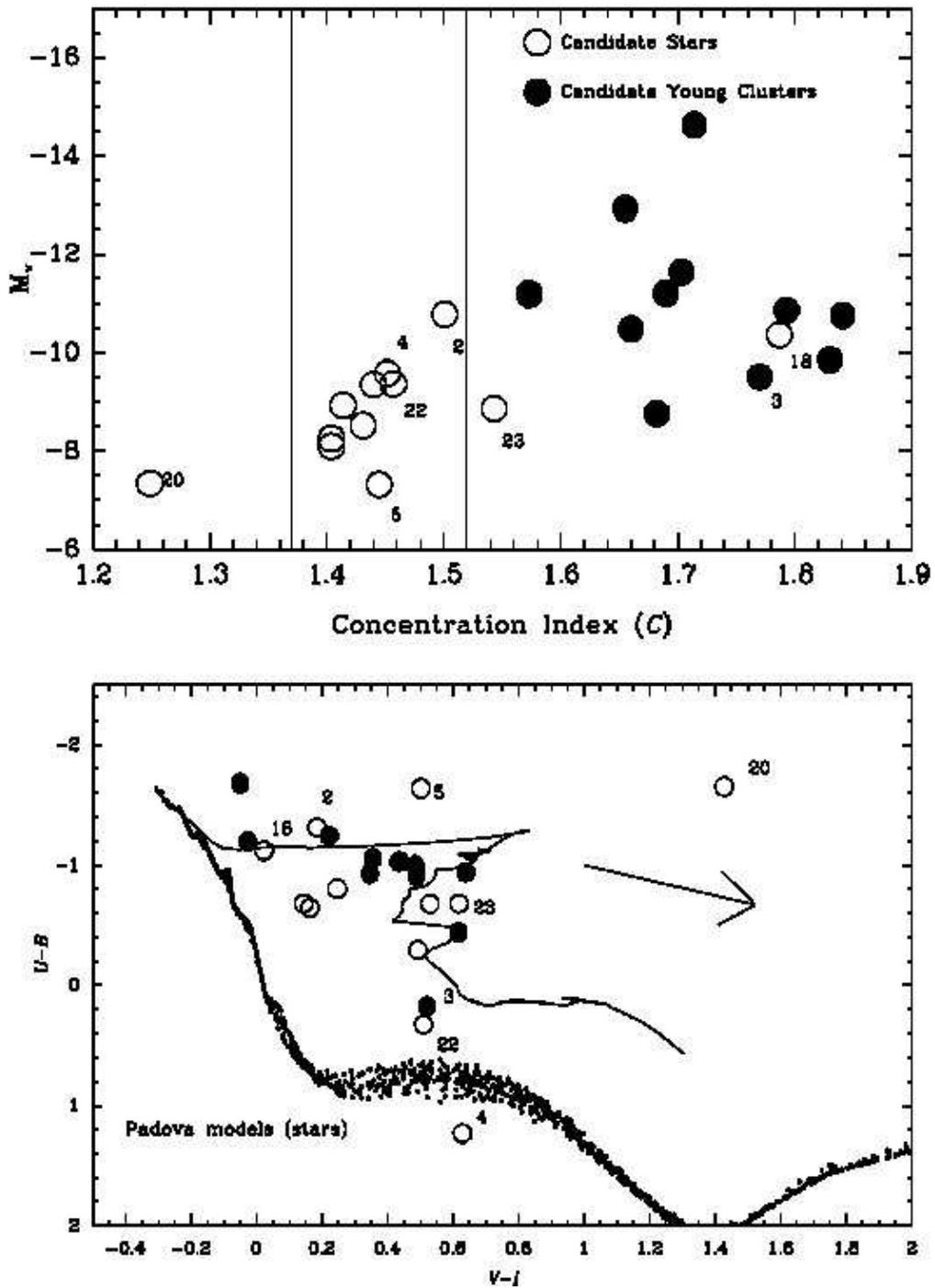}
\caption{Properties for Training Set 3.  The two diagrams are similar to
those shown in Figure 6 for Training Set 1. See text for a description 
of how the ``candidate'' stars and clusters were selected.
}
\label{fig27}
\end{figure}

%--------
% Figure tbd - HRC image
\begin{figure}%28
\centering
\includegraphics[scale=0.8]{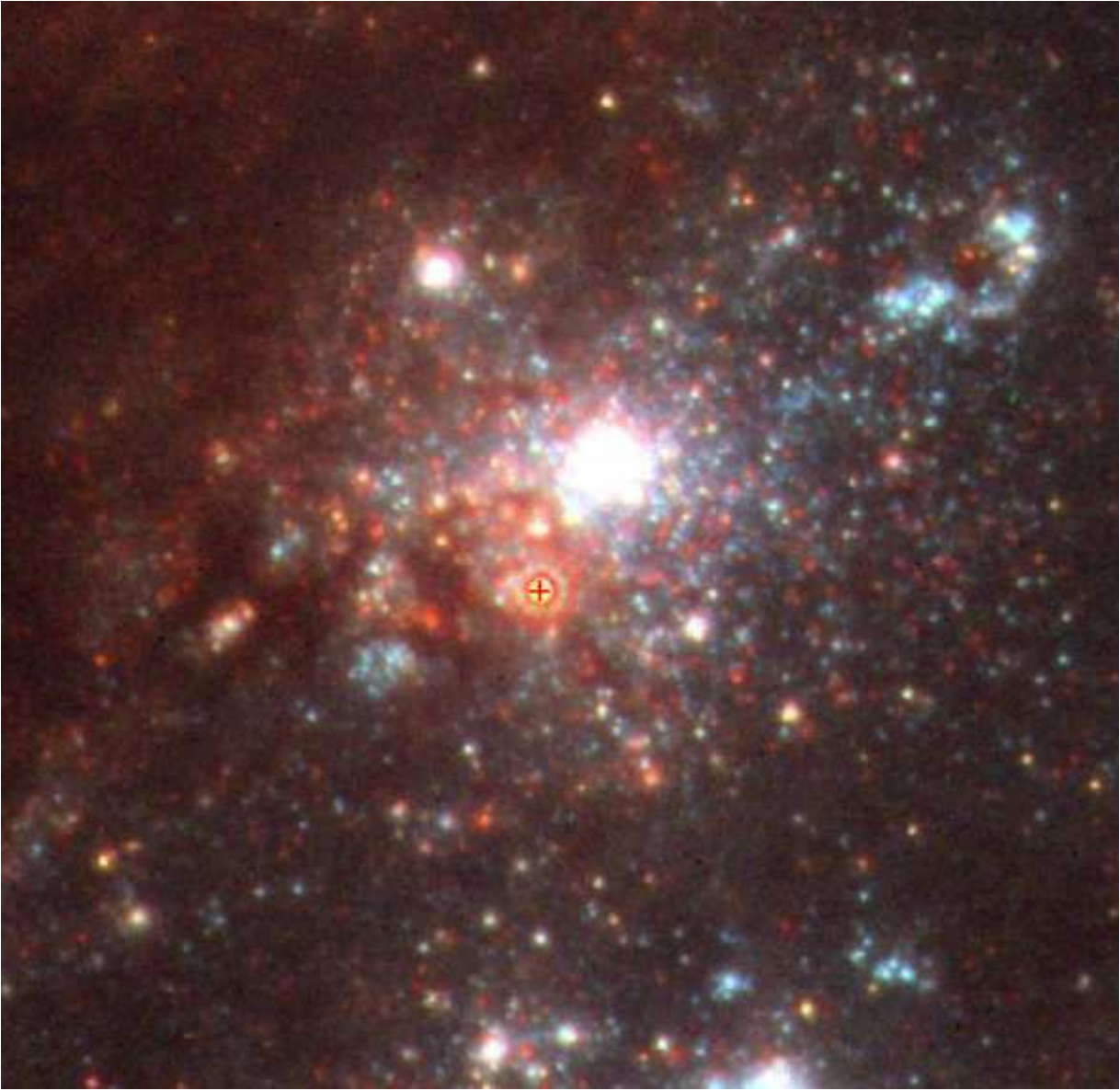}
\caption{ACS/HRC image of Knot S from the Hubble Legacy Archive (Whitmore \etal\ 2009).  The cross marks the position of Supernova 2004gt.}
\label{fig28}
\end{figure}

%--------
% Figure tbd - HRC image
\begin{figure}%29
\centering
\includegraphics[scale=0.8]{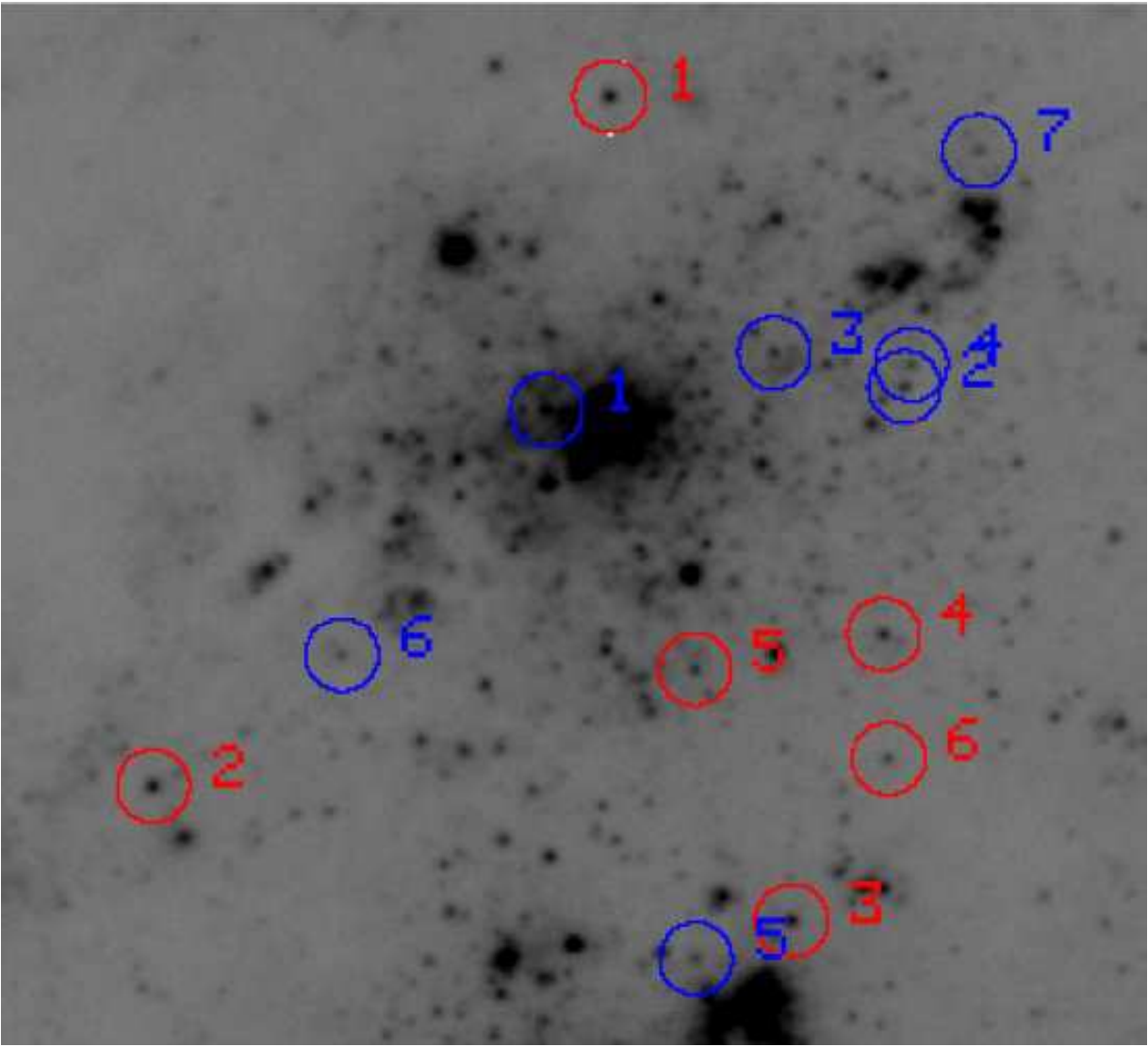}
\caption{Brightest candidate blue and yellow stars in Knot~S, taken from Table~5.}
\label{fig29}
\end{figure}

%--------
% Figure tbd - HRC image
\begin{figure}%30
\centering
\includegraphics[scale=0.8]{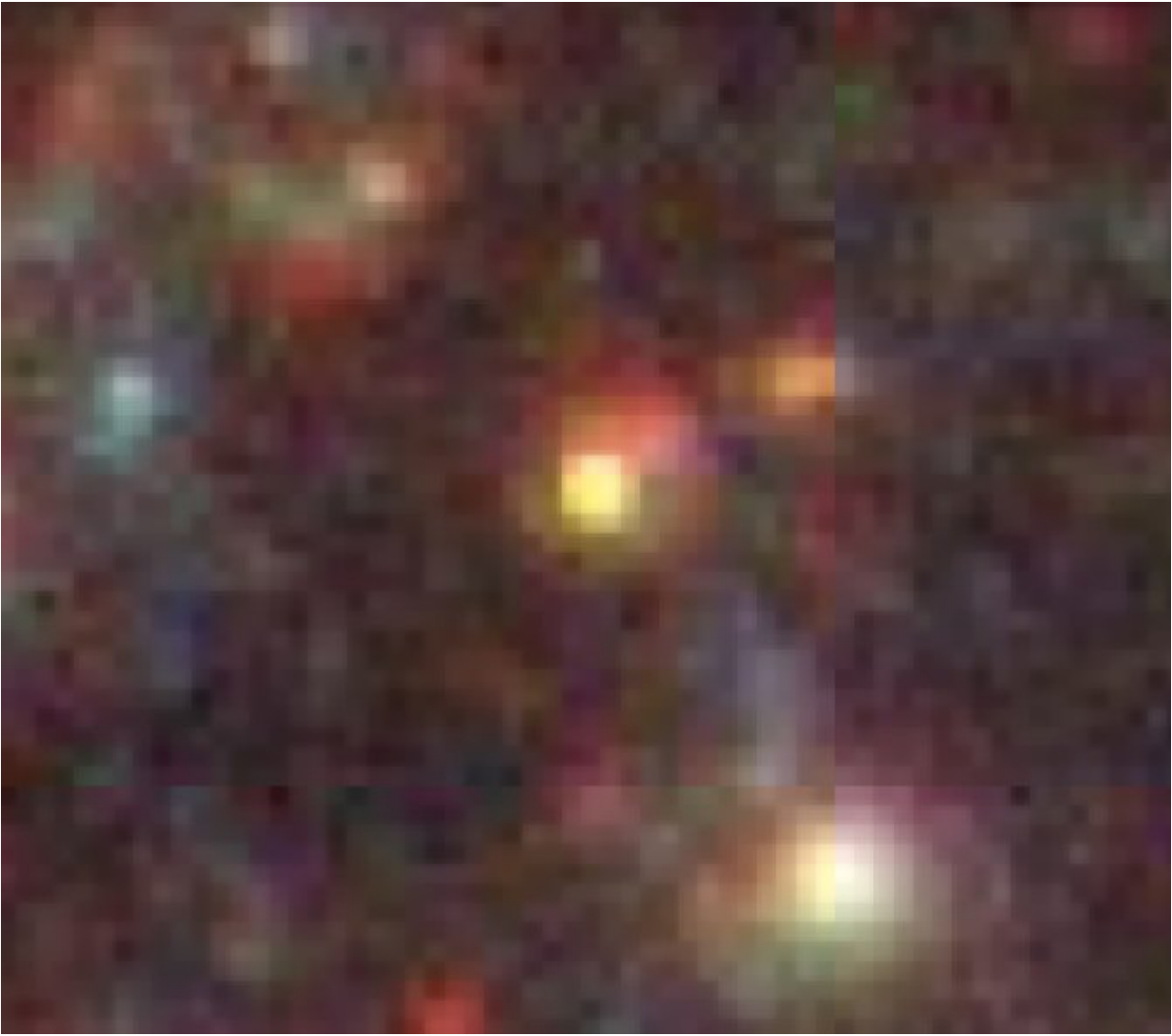}
\caption{Enlargement from Figure 28 showing the second-brightest candidate yellow star of Table 5. Note the concentrated light profile of the yellow star (near the center of the image), when compared to a normal blue cluster near the bottom right, and the associated reddish cluster that the star probably resides in. Most of the other yellow and red stars in Figure 28 show similar morphologies, although not as pronounced since the bright stars generally appear more centered on the associated faint clusters. Use the interactive display of the Hubble Legacy Archive (hla.stsci.edu, image name is {\em HST\_10187\_03\_ACS\_HRC\_F814W\_F555W\_F435W}) to
examine this image in more detail.}
\label{fig30}
\end{figure}

%--------
% Figure 11 - clusterspace
\begin{figure}%31ab
 \centering
\includegraphics[scale=0.5]{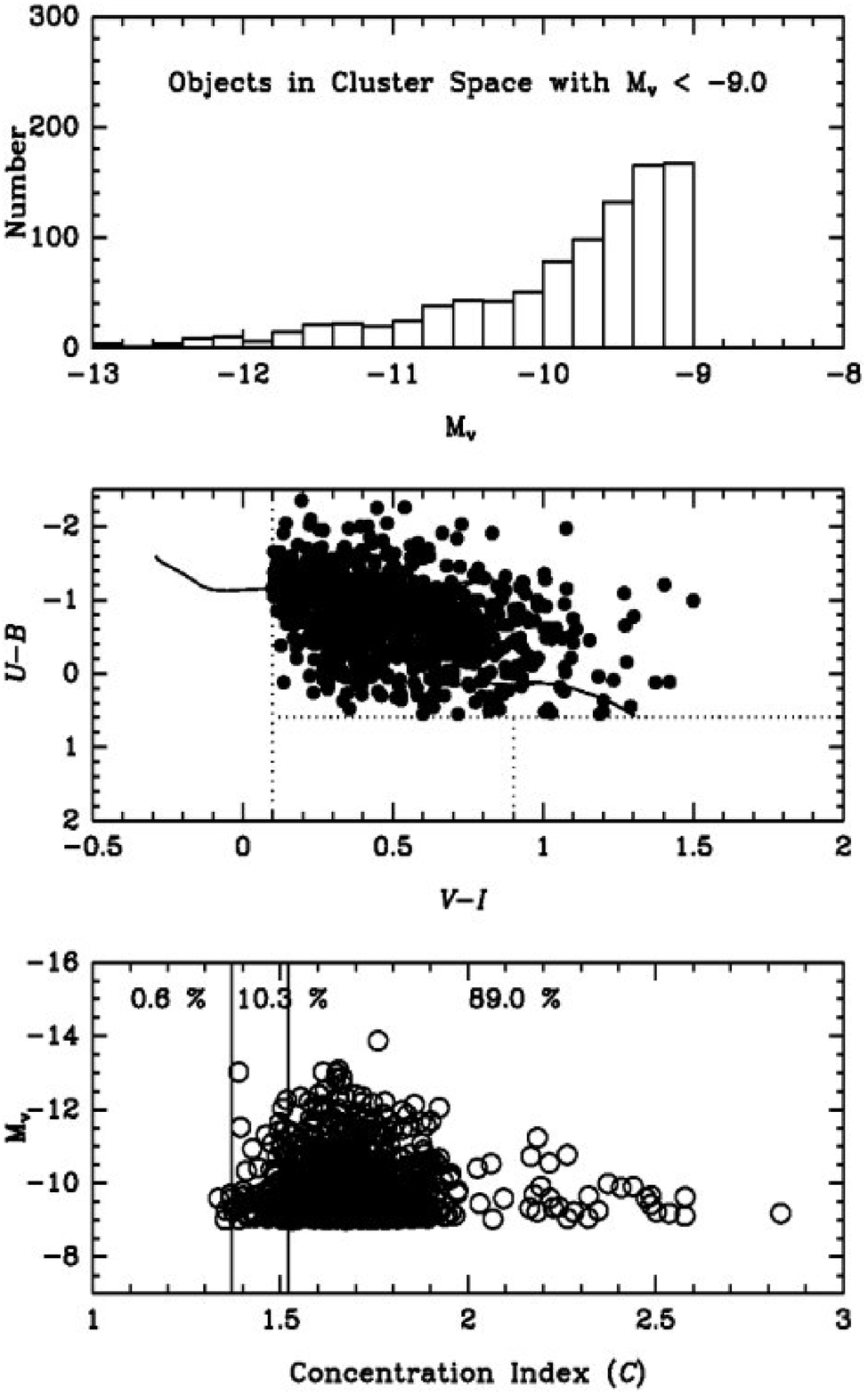}~\includegraphics[scale=0.52]{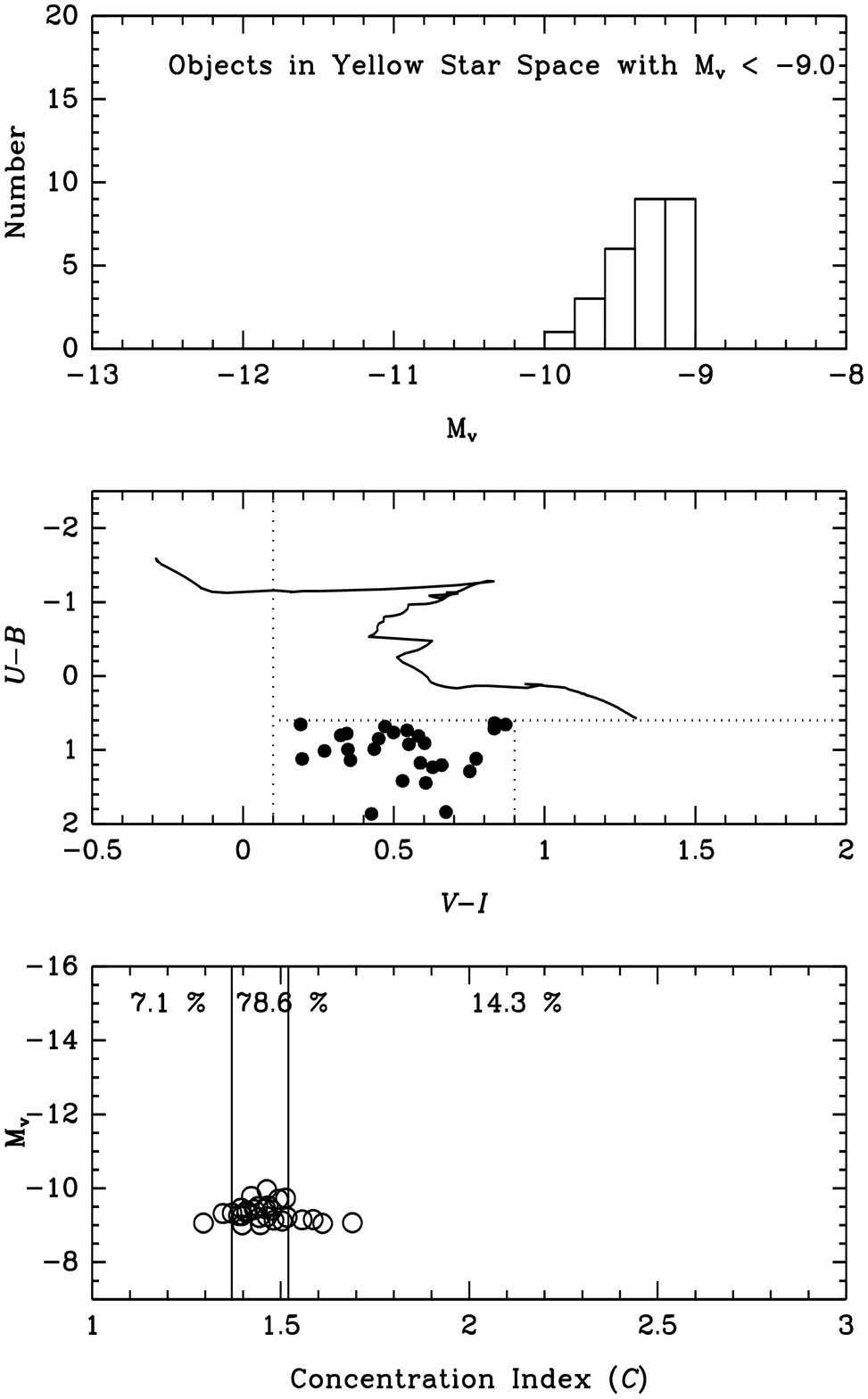}
\caption{(a-left) All objects in the Antennae with $\Mv < -9$ that fall in the ``cluster space'' portion of the two-color diagram (as described in the text).  The top panel shows the number of these objects as a function of luminosity, the middle panel shows their two-color plot, and the bottom panel shows their distribution in \Mv\ vs.\ \ci\ and the percentage of objects in each of three bins. (b-right) Same as (a), but for objects in ``yellow-star space.''
}
\label{fig31a}
\end{figure}

%--------
% Figure - red star space 
%\setcounter{figure}{30}
%\begin{figure}
%\centering
%\includegraphics[angle=0,scale=0.6]{f31b.eps}
%\caption{(b) Same as Figure~31(a), but for objects in ``yellow-star space''.
%}
%\label{fig31b}
%\end{figure}

% Figure - foreground star space
\setcounter{figure}{30}
\begin{figure}%31cd
\centering
\includegraphics[scale=0.5]{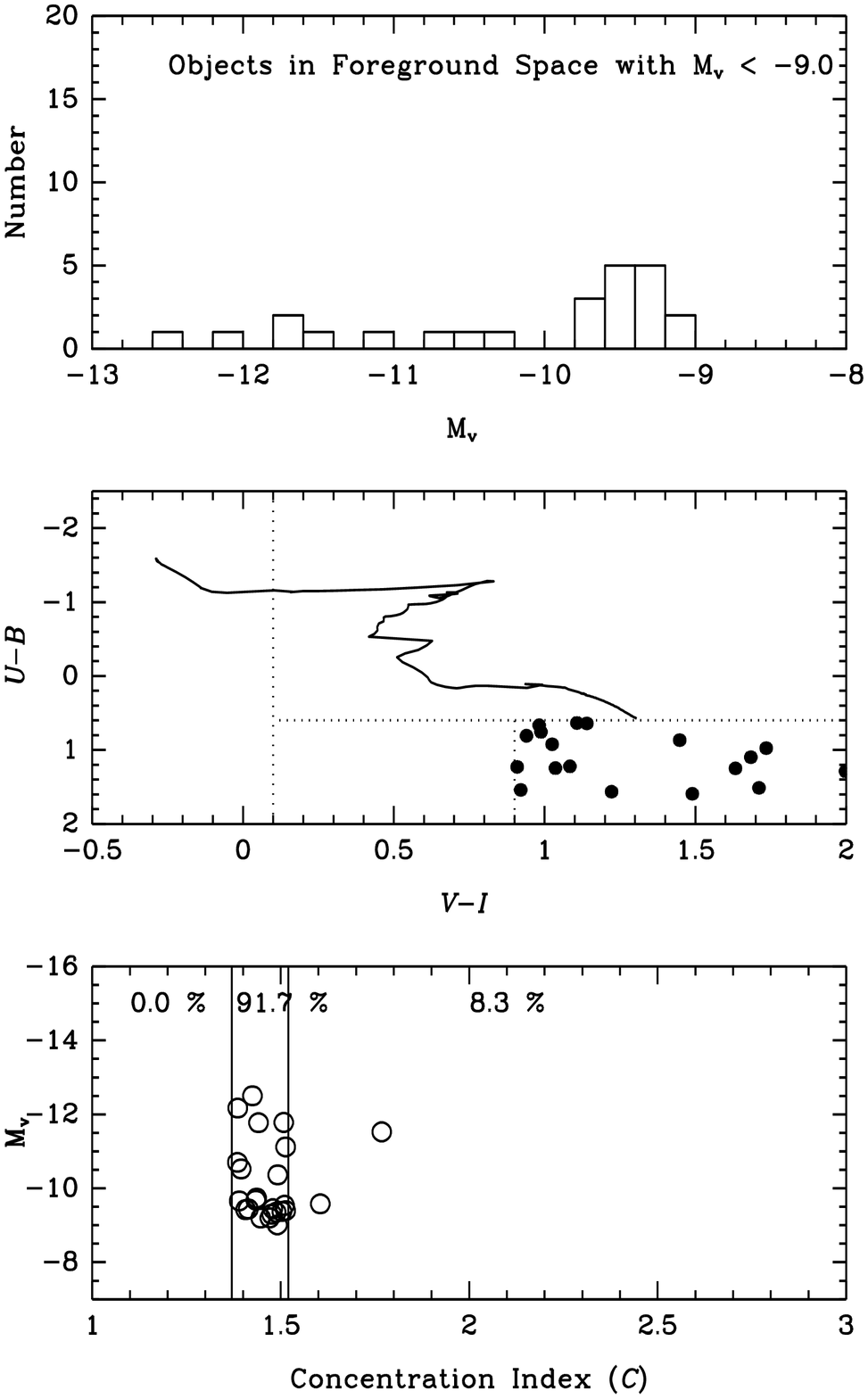}~\includegraphics[scale=0.48]{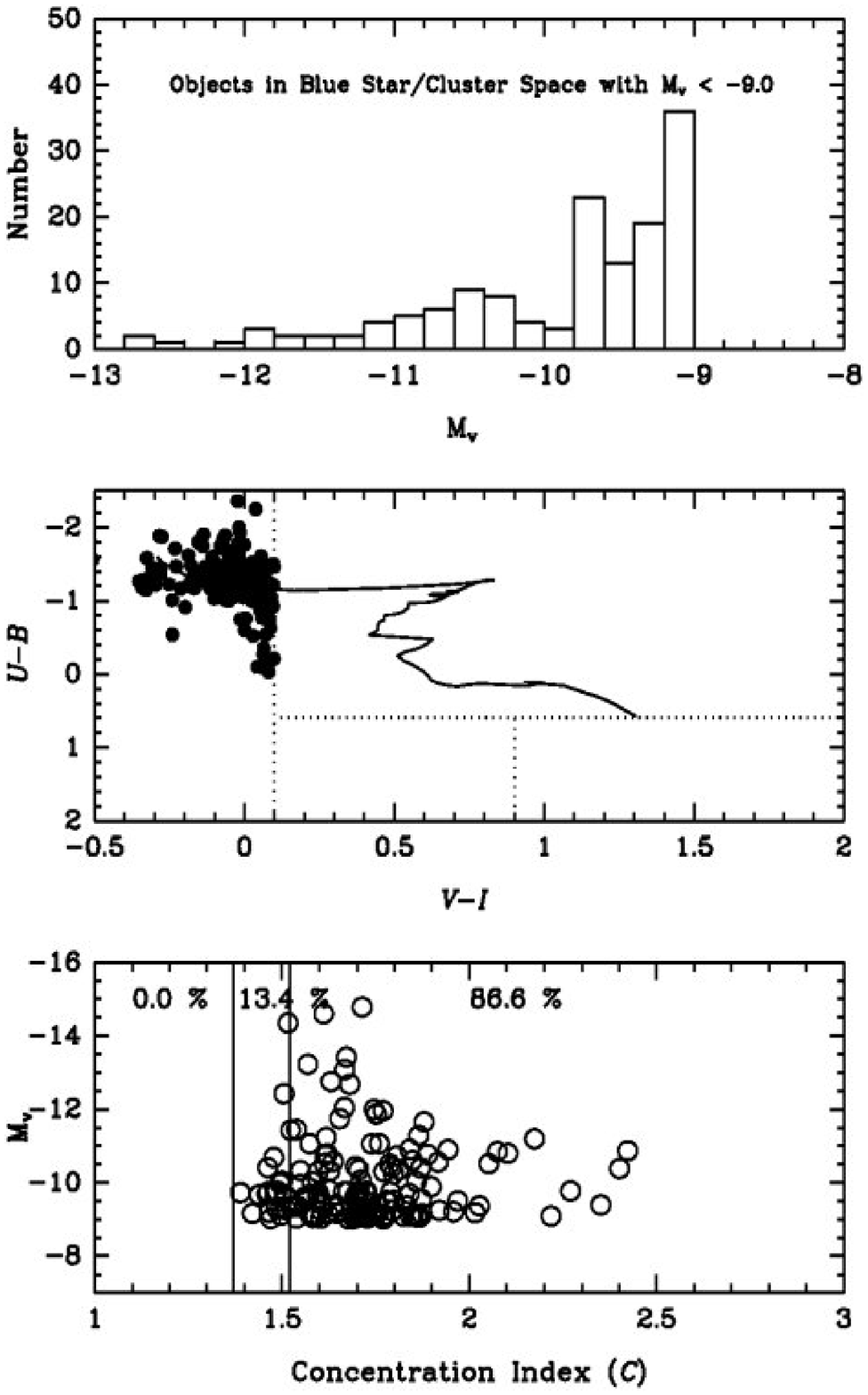}
\caption{(c-left) Same as Figure~31(a), but for objects in ``foreground-star
space.'' (d-right) Same as (a), but for objects in blue-star/cluster space.
}
\label{fig31c}
\end{figure}

%--------
% Figure 14 - foreground and star space
%\setcounter{figure}{30}
%\begin{figure}
%\centering
%\includegraphics[angle=0,scale=0.6]{f31d.eps}
%\caption{(d) - Same as Figure~31(a), but for objects in blue-star/cluster space.
%}
%\label{fig31d}
%\end{figure}

\end{document}